\documentclass[aps,amsmath,amsfonts,amssymb,prb,twocolumn,showpacs,superscriptaddress]{revtex4-1}

\usepackage{graphicx}   % for graphicx
\usepackage{units}      % for nicefrac (and more)
\usepackage{siunitx}    % SI-unit package
\usepackage{stmaryrd}   % for \llbracket
\usepackage{dsfont}     % for \mathds in idmat, e.g.
\usepackage{hyperref}   % for reffering of links
\usepackage{bm}
\usepackage{color}

\allowdisplaybreaks
\newcommand{\figref}[1]{{Fig.~\ref{#1}}}
\newcommand{\tabref}[1]{{Table~\ref{#1}}}
\newcommand{\secref}[1]{{Section~\ref{#1}}}

\newcommand{\coloneqq}{:=}
\newcommand{\eqqcolon}{=:}

\newcommand{\E}{{\bf E}}        % electric field
\newcommand{\D}{{\bf D}}        % electric displacement field
\renewcommand{\H}{{\bf H}}      % magnetic field intensity
\renewcommand{\P}{{\bf P}}      % electric polarization density
\newcommand{\J}{{\bf J}}        % electric current density
\renewcommand{\j}{{\bf j}}      % electric current density (used for surface)

\renewcommand{\k}{{\bf k}}      % k-vector

\newcommand{\br}{{\bf r}}        % location vector x
\newcommand{\normal}{{\bf n}}   % normal vector

\newcommand{\idmatrix}{{\textsf{I}}} % identity matrix % maybe use \newcommand{\idmatrix}{{\mathds 1}}

\newcommand{\nvf}{{N}} % normal vector field: component
\newcommand{\nvfv}{{\bf \nvf}}  % normal vector field: vector

\newcommand{\e}{{\bf e}}        % unit directional vector
\newcommand{\bs}{{\bf s}}       % unit vector s
\newcommand{\bc}{{\bf c}}       % boldfaced c: coefficient vector in generic layer
\newcommand{\ba}{{\bf a}}       % boldfaced a: coefficient vector in layer A
\newcommand{\bb}{{\bf b}}       % boldfaced b: coefficient vector in layer B
\newcommand{\be}{{\bf e}}       % boldfaced e: coefficient vector in evaluation layer E
\newcommand{\bt}{{\bf t}}       % tangential vector

\newcommand{\fmatrix}[1]{{\llbracket #1 \rrbracket}}
\newcommand{\fvec}[1]{{[#1]}}
\newcommand{\rec}[1]{{ {\mathcal R}\left(#1\right)}}    % fourier series reconstruction
\newcommand{\propconst}{{\nu}}  % propagation constant in modal expansion
\newcommand{\propmat}{{\textsf{V}}}  % propagation constant matrix in modal expansion

\renewcommand{\Re}[1]{\mathrm{Re}\left(#1\right)}
\renewcommand{\Im}[1]{\mathrm{Im}\left(#1\right)}

\def\jmo{J.\ Mod.\ Opt.\ }
\def\jap{J.\ Appl.\ Phys.\ }

\def\jo{J.\ Opt.\ }
\def\josaa{J.\ Opt.\ Soc.\ Am.\ A.}
\def\josab{J.\ Opt.\ Soc.\ Am.\ B.}
\def\nat{Nature\ (London)}
\def\natphot{Nat.\ Photon.}
\def\natmat{Nat.\ Mater.}
\def\njp{New\ J.\ Phys.}
\def\nl{Nano\ Lett.}

\def\natnano{Nat. Nanotechnol.}
\def\oe{Opt.\ Express}
\def\ol{Opt.\ Lett.}
\def\ome{Opt.\ Mater.\ Express}
\def\pr{Phys.\ Rev.}
\def\prx{Phys.\ Rev.\ X}
\def\sci{Science}
\def\scirep{Sci.\ Rep.}

\bibpunct{[}{]}{,}{n}{,}{,} % see http://www.andy-roberts.net/writing/latex/bibliographies , under \usepackage{natbib}

\begin{document}
\title{Theoretical and computational analysis of second- and third-harmonic generation in periodically patterned graphene and transition-metal dichalcogenide monolayers}

\author{Martin Weismann}
\affiliation{Department of Electronic and Electrical Engineering, University College London,
Torrington Place, London WC1E 7JE, United Kingdom}
\author{Nicolae C. Panoiu}
\email[Corresponding author:]{n.panoiu@ucl.ac.uk} \affiliation{Department of Electronic and
Electrical Engineering, University College London, Torrington Place, London WC1E 7JE, United
Kingdom}

\date{\today}

% (42.70.-a)  Optical materials,

\begin{abstract}
Remarkable optical and electrical properties of two-dimensional (2D) materials, such as graphene
and transition-metal dichalcogenide (TMDC) monolayers, offer vast technological potential for
novel and improved optoelectronic nanodevices, many of which relying on nonlinear optical effects
in these 2D materials. This article introduces a highly effective numerical method for efficient
and accurate description of linear and nonlinear optical effects in nanostructured 2D materials
embedded in periodic photonic structures containing regular three-dimensional (3D) optical
materials, such as diffraction gratings and periodic metamaterials. The proposed method builds
upon the rigorous coupled-wave analysis and incorporates the nonlinear optical response of 2D
materials by means of modified electromagnetic boundary conditions. This allows one to reduce the
mathematical framework of the numerical method to an inhomogeneous scattering matrix formalism,
which makes it more accurate and efficient than previously used approaches. An overview of linear
and nonlinear optical properties of graphene and TMDC monolayers is given and the various features
of the corresponding optical spectra are explored numerically and discussed. To illustrate the
versatility of our numerical method, we use it to investigate the linear and nonlinear
multiresonant optical response of 2D-3D heteromaterials for enhanced and tunable second- and
third-harmonic generation. In particular, by employing a structured 2D material optically coupled
to a patterned slab waveguide, we study the interplay between geometric resonances associated to
guiding modes of periodically patterned slab waveguides and plasmon or exciton resonances of 2D
materials.
\end{abstract}

\maketitle

\section{Introduction}\label{sec:introduction}
Since its first isolation, preparation, and theoretical description
\cite{novoselov04science,novoselov05nature,novoselov05pnas}, graphene, a monolayer of carbon atoms
distributed in a hexagonal lattice, has attracted a tremendous amount of interest in science and
engineering due primarily to its outstanding physical properties and potential for novel
applications. Graphene was shown to have remarkable mechanical strength
\cite{lee08science,frank07jvacscib,meyer07nature} and extremely high thermal conductivity
\cite{balandin08nl}, making it a particularly appealing materials platform for
nano-electromechanical applications and management of thermal processes in nanoelectronic circuits
\cite{frank07jvacscib,ghosh08apl}. In addition, the high carrier mobility of graphene enhances its
potential for applications to high-frequency electronics
\cite{morozov08prl,lin10science,schwierz10nnano}. These and other remarkable properties of
graphene have spurred new research into and development of new two-dimensional (2D) materials,
such as hexagonal boron nitride (\textit{h}-BN), silicene (a monolayer of silicon), and
transition-metal dichalcogenide (TMDC) monolayers \cite{xu13cr,song10nl,vogt12prl,wang12nnano},
each with their own array of unique physical properties.

One additional compelling aspect of 2D materials (2DMs) is closely related to their optical
properties. Graphene, for example, is nearly transparent at optical frequencies, exhibiting
absorption of only about 2.3\% \cite{nair08science}, which suggests it barely interacts with
light. This optical transparency and the earlier mentioned electro-mechanical properties make
graphene a promising new material for flexible optical devices \cite{bonaccorso10natphot} (e.g.,
touch screens). Moreover, graphene based structures can provide an alternative to conventional
metallo-dielectric structures to spatially confine and guide light at the nanoscale, a research
direction actively pursued in the emerging field of graphene
nanoplasmonics~\cite{cbg12n,fra12n,jablan09prb,ju11natnano,bao12acsnano,jablan13procIEEE,luo13mser,yeung14nl,smirnova15prb,smirnova15prb2}.

In addition to these linear physical properties, the nonlinear optical properties of graphene and
other 2D materials have attracted increased attention. Graphene, as a centrosymmetric material,
exhibits large third-harmonic generation (THG) \cite{hong13prx,cvs14njp}, strong optical Kerr
nonlinearity \cite{zhang12ol}, and induced second-order nonlinearity
\cite{bykov12prb,cheng14oe,an14prb} in a single atomic layer. This allows one to employ graphene
in active photonic devices with improved functionality, including ultra-compact modulators,
optical limiters, frequency converters, and photovoltaic and photoresistive devices
\cite{bonaccorso10natphot,hong13prx,cvs14njp,bykov12prb,zhang12ol,xia09natnano}. In a
complementary fashion, TMDC monolayers are semiconducting materials, which renders them
particularly suitable to be employed in nanoscale transistors and saturable absorbers
\cite{radisavljevic11natnano,zhang14oe}, and have non-centrosymmetric atomic lattice and hence
allow even-order nonlinear optical processes \cite{janish14srep,malard13prb,li13nl,seyler15nn}.
The implementation of these linear and nonlinear optical properties into applications, however,
requires advances in nanofabrication and experimental techniques
\cite{bonaccorso10natphot,kim09nat,li09science}, theoretical models, and numerical methods for
modeling of devices incorporating 2D materials.

There is a multitude of numerical methods for computational study of optical structures and
devices comprising regular, three-dimensional (3D) materials
\cite{taflove2000computational,jin14wiley}, and they can in principle be used for modeling 2D
materials, too. This is customarily done by defining an effective thickness of the material and
incorporating the 2DM into the computational algorithm simply as a very thin layer of 3D material
\cite{gao12acsnano,gosciniak13scirep,nggm12prb}. This computational approach, albeit simple and
easy to implement, has a serious drawback, namely it relies on an obviously ambiguous quantity,
the thickness of the monolayer. Moreover, this thickness is typically orders of magnitude smaller
than the other characteristic lengths of the photonic structure and the operating wavelength,
leading to a large length-scale imbalance that is detrimental to the effectiveness of the spatial
discretization. These issues result in potentially reduced accuracy, increased computational cost,
and numerical artefacts that are difficult to avoid \cite{llatser2012pan,k13ol}. It is hence
desirable to treat the sources of the optical effects pertaining to the 2DM as confined to a 2D
manifold, i.e. as induced by a surface conductivity
\cite{llatser2012pan,nayyeri13ieeetantp,chen11acsnano}.

The numerical method proposed here follows this approach and implements it in the context of the
rigorous coupled-wave analysis (RCWA) method, a modal frequency domain method for modeling
periodic optical structures \cite{mgp95josaa,Li1996josaa,li03josaa}. However, in addition to
previous work on handling 2DMs by means of Fourier series expansion methods
\cite{k13ol,nka14josab}, we also model nonlinear optical effects in 2D materials, in particular
second-harmonic generation (SHG) and THG, and provide the mathematical formulation for arbitrary
nonlinear optical processes. In addition, the method introduced in this work allows one to
describe structured 2D materials, which themselves can be embedded in inhomogeneous 3D optical
structures.

The remainder of this paper introduces the general periodic optical system under consideration in
\secref{sec:geommaterparam} and gives a detailed overview of linear and nonlinear optical
properties of different 2DMs in \secref{sec:linnonlinprop}. Section~\ref{sec:mathformulation}
provides the mathematical formulation for higher-harmonic generation in patterned 2DMs based on
the RCWA and introduces the inhomogeneous scattering matrix formalism for multilayered 3D
structures. This is followed by \secref{sec:convergence}, where the validation and benchmarking of
the numerical method is performed, using as test problems one-dimensional (1D) and 2D periodic
structures. Applications of the proposed numerical method are presented and discussed in
Section~\ref{sec:examples}, where we also investigate different resonant mechanisms to enhance the
nonlinear optical response of 2D-3D heteromaterials containing TDMC monolayers and nanostructured
graphene. Finally, the main conclusions are drawn and an outlook towards future work is given in
\secref{sec:conclusions}.

\section{Photonic system: Geometry and materials parameters}\label{sec:geommaterparam}
The computational method introduced in this article is designed for a very general physical
setting, namely periodically patterned photonic structures that contain both bulk and 2D optical
materials. Our numerical method accurately describes the physics of such photonic structures by
incorporating in the numerical algorithm the relevant linear and nonlinear optical effects
pertaining to 2D materials. Equally important, since the nonlinear optical response of different
2D materials contained by the photonic structure is described via a generic nonlinear
polarization, this computational method can be used to investigate a multitude of nonlinear
optical effects, including SHG, sum- and difference-frequency generation, THG, and four-wave
mixing.

\begin{figure}[t]
    \centering
    \includegraphics[width=\linewidth]{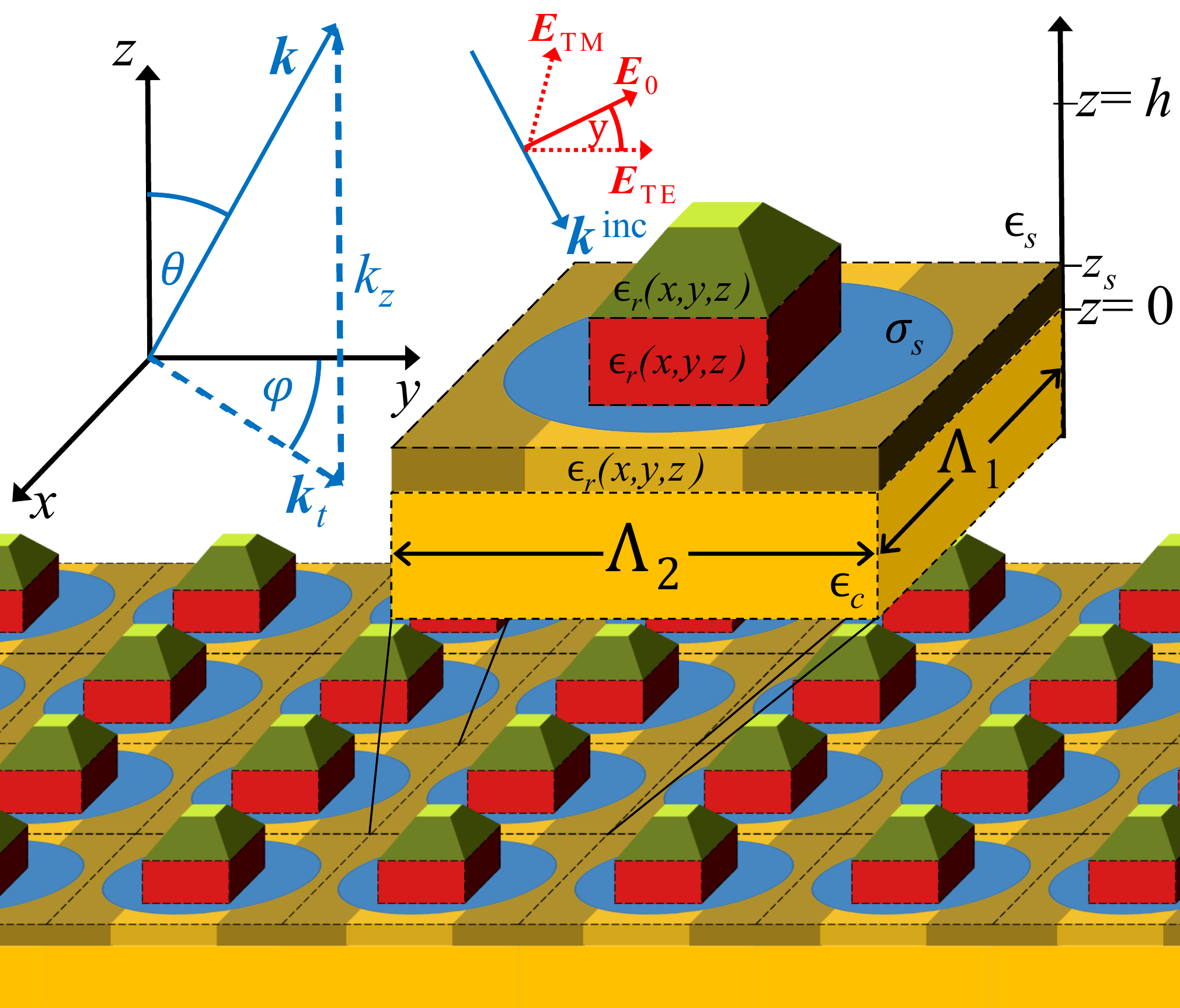}
\caption{Schematic of a generic multilayer structure periodic along the $x$- and $y$-direction
under plane wave incidence, and a close-up of the unit cell. The distribution of the 2D material
(blue) is defined by the position dependent sheet conductance, $\sigma_s(x,y,z_s)$, at $z=z_s$.
The bulk part of the periodic structure, which can consist of slanted (olive green), straight
walled (red), or embedded parts (sand brown), is defined by the permittivity function
$\epsilon_r(x,y,z)$.}
    \label{fig:multilayerstructure}
\end{figure}

The geometric setting and important nomenclature used in the presentation of our numerical method
are introduced in what follows. Thus, consider the generic multilayer, periodic structure
presented in \figref{fig:multilayerstructure}. The bird's-eye view in
\figref{fig:multilayerstructure} depicts the unit cell of a 2D-periodic structure with periods
$\Lambda_1$ and $\Lambda_2$, with the corresponding grating vectors, $\mathbf{\Lambda}_1$ and
$\mathbf{\Lambda}_2$, laying in the $(x,y)$-plane. It consists of several bulky, periodically
structured regions with relative electric permittivity, $\epsilon_r(\br)$, which will be called
bulk-layers, or simply layers. The periodic structure is sandwiched in-between semi-infinite
homogeneous cover and substrate layers with relative permittivity, $\epsilon_c$ and $\epsilon_s$,
respectively. In addition, the structure can comprise 2DM sheets, or simply sheets, located at
$z=z_{s}$, which are assumed to lay in the $(x,y)$-plane between two bulk-layers. Each sheet is
made of homogeneous or periodically patterned 2DMs and is described by its surface conductivity
distribution, $\sigma_s(x,y,z_s)$.

This periodic structure is excited by an incident harmonic plane wave with the electric field
given by:
\begin{equation}\label{eq:incfield}
    \mathcal{\E}^{\mathrm{inc}}(\br,t) = \E_0(\br)e^{-i\omega t} = \E_0e^{i(\k_{t} \cdot \br_{t}-k_{z}z - \omega t)},
\end{equation}
where $\omega$ is the angular frequency, $\br_{t}$ is the in-plane component of the position
vector, $\k_{t} = k_c(\sin\theta\cos\varphi,\sin\theta\sin\varphi)$ and $k_{z}=k_c\cos\theta$ are
the in-plane and normal components of the wave vector of the incident wave in the cover,
respectively, with $k_c = \sqrt{\epsilon_c}k_0=\sqrt{\epsilon_c}\omega/c$ being the wavenumber in
the cover region and $c$ is the speed of light. The polarization state of the incoming plane wave
is described by the field components in the transverse electric (TE) and transverse magnetic (TM)
configurations, namely $\E_\mathrm{TE}$ and $\E_\mathrm{TM}$, respectively, such that
$\E_0=\cos\psi\E_\mathrm{TE} + \sin\psi\E_\mathrm{TM}$ is determined by the polarization angle,
$\psi$.

In the rigorous coupled-wave analysis, the underlying algorithm on which our computational method
is built, oblique structures or devices consisting of several periodic layers are approximated in
a staircasing manner (see Ref.~\cite{popov2002staircase}). To be more specific, the structure is
sliced into thin and $z$-invariant bulk-layers, the optical response of each layer is calculated,
and then the combined response of all layers is determined by properly incorporating the
inter-layers optical coupling described in \secref{sec:inhomSMatrix}. Before this algorithm is
derived thoroughly in \secref{sec:mathformulation}, the optical properties of 2DMs are introduced
and formalized mathematically in the next section.

\section{Linear and nonlinear optical properties of 2D materials}\label{sec:linnonlinprop}
Understanding the physics of 2DMs is a rather recent endeavor and as such a comprehensive
characterization of their linear and nonlinear optical properties is only emerging. In particular,
a complete knowledge of the frequency dependence of the linear and nonlinear optical constants is
a prerequisite for a rigorous computational description of the optical response of these
materials. Theoretical calculations \cite{wss06njp,bhr13prb,cvs14njp,janish14srep} and
experimental investigations \cite{li14prb,mukherjee15ome,hong13prx} are beginning to fill in this
gap, a summary of the results of these studies being briefly outlined in this section.

\subsection{Linear optical properties of 2D materials}\label{sec:2dlinprop}
\begin{figure}[b]
    \centering
    \includegraphics[width=\linewidth]{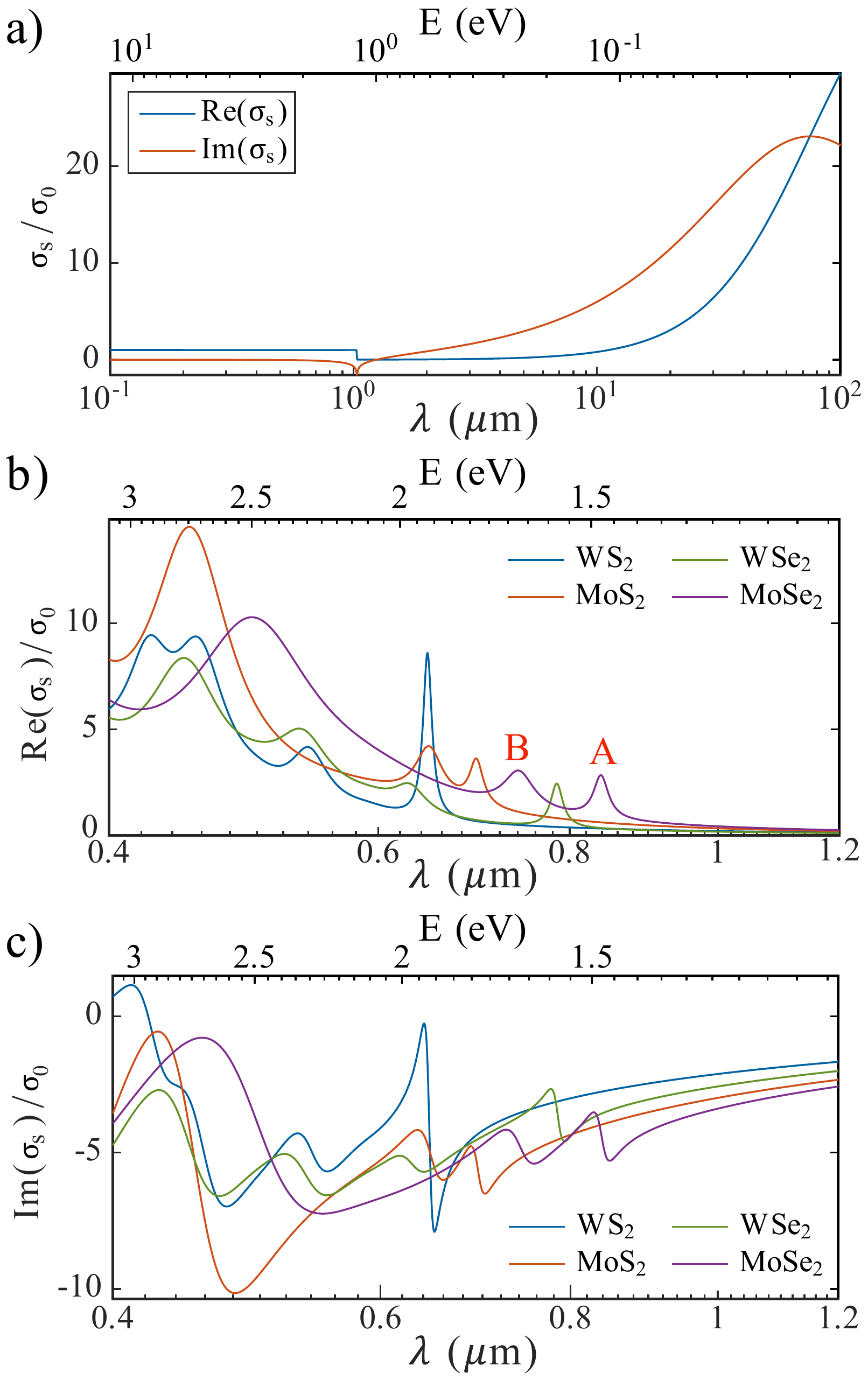}
\caption{a) Dispersion of the complex sheet conductance, $\sigma_s(\omega)$, of graphene in
interband and intraband wavelength range. b), c) Frequency dependence of the real and imaginary
parts of the sheet conductance of several TMDC monolayer materials, respectively.}
    \label{fig:Mat_Linear_Properties}
\end{figure}

Since graphene and other 2DMs are physical systems consisting of a single atomic layer, their
electromagnetic properties are conveniently characterized by surface quantities. For example, in
the case of graphene, the optical properties are described by the sheet conductance (sometimes
simply called conductivity), which in the random-phase approximation and zero temperature
conditions can be expressed as \cite{wss06njp,hd07prb}:
\begin{align}\label{eq:model_sigma}
        &\frac{\sigma_s(\omega)}{\sigma_{0}} = \frac{4\varepsilon_F}{\pi \hbar} \frac{\tau}{1-i\omega\tau}
        + \theta(\hbar\omega - 2\varepsilon_F)+\frac{i}{\pi}
        \ln\left|\frac{\hbar\omega-2\varepsilon_F}{\hbar\omega+2\varepsilon_F}\right|.
%         &\frac{\sigma_s(\omega)}{\sigma_{0}} = \frac{4\varepsilon_F}{\pi \hbar} \frac{i}{\omega+i\tau^{-1}}
%         + \theta(\hbar\omega - 2\varepsilon_F)+\frac{i}{\pi}
%         \ln\left|\frac{\hbar\omega-2\varepsilon_F}{\hbar\omega+2\varepsilon_F}\right|.
\end{align}
Here, $\sigma_{0} = e^2/(4\hbar) = \SI{6.0853e-05}{\ampere\per\volt}$ is the universal dynamic
conductivity of graphene, $e$ denotes the elementary charge, $\varepsilon_F$ the Fermi level,
$\hbar$ the reduced Planck constant, $\tau$ the relaxation time and $\theta(\cdot)$ is the
Heaviside step function. The values of the Fermi level and relaxation time used throughout
this study are $\varepsilon_F=\SI{0.6}{\electronvolt}$ and $\tau=\SI{0.25}{\pico\second}/2\pi$,
 respectively. The dispersion of $\sigma_s(\omega)$ for graphene is shown in
 \figref{fig:Mat_Linear_Properties}(a).

\begin{table*}% [h!]
    {\footnotesize \centering
     \begin{tabular}{ r | c c c | c c c | c c c | c c c }
        & \multicolumn{3}{c|}{$\mathrm{WS_2}$} & \multicolumn{3}{c|}{$\mathrm{WSe_{2}}$} & \multicolumn{3}{c|}{$\mathrm{MoS_{2}}$} & \multicolumn{3}{c}{$\mathrm{MoSe_{2}}$}\\
        \hline
        $h_{\mathrm{eff}}$& \multicolumn{3}{c|}{\SI{6.18}{\angstrom}} & \multicolumn{3}{c|}{\SI{6.49}{\angstrom}} & \multicolumn{3}{c|}{\SI{6.15}{\angstrom}} & \multicolumn{3}{c}{\SI{6.46}{\angstrom}}\\
        \hline
        {\color{white} $|^|_|$} $k$ & $E_k$ (\si{\electronvolt}) & $f^E_k$ (\si{\electronvolt\squared})& $\gamma^E_k$ (\si{\electronvolt})&
                                      $E_k$ (\si{\electronvolt}) & $f^E_k$ (\si{\electronvolt\squared})& $\gamma^E_k$ (\si{\electronvolt})&
                                      $E_k$ (\si{\electronvolt}) & $f^E_k$ (\si{\electronvolt\squared})& $\gamma^E_k$ (\si{\electronvolt})&
                                      $E_k$ (\si{\electronvolt}) & $f^E_k$ (\si{\electronvolt\squared})& $\gamma^E_k$(\si{\electronvolt})  \\
        \hline
       1  & 2.009 & 1.928 & 0.032 & 1.654 & 0.557 & 0.036 & 1.866 & 0.752 & 0.045 & 1.548 & 0.648 & 0.043\\
       2  & 2.204 & 0.197 & 0.250 & 2.426 & 5.683 & 0.243 & 2.005 & 1.883 & 0.097 & 1.751 & 1.302 & 0.097\\
       3  & 2.198 & 0.176 & 0.161 & 2.062 & 1.036 & 0.115 & 2.862 & 36.89 & 0.383 & 2.151 & 4.621 & 0.537\\
       4  & 2.407 & 0.142 & 0.112 & 2.887 & 16.11 & 0.344 & 2.275 & 10.00 & 1.000 & 2.609 & 37.40 & 0.582\\
       5  & 2.400 & 2.980 & 0.167 & 2.200 & 1.500 & 0.300 & 3.745 &  100.0 & 0.533 & 3.959 &  121.4 & 0.896\\
       6  & 2.595 & 0.540 & 0.213 & 2.600 & 1.500 & 0.300 &      &  -   &      &      &  -   &     \\
       7  & 2.644 & 0.050 & 0.171 & 3.800 & 70.00 & 0.700 &      &  -   &      &      &  -   &     \\
       8  & 2.831 & 12.60 & 0.266 & 5.000 & 80.00 & 0.700 &      &  -   &      &      &  -   &     \\
       9  & 3.056 & 8.765 & 0.240 &      &  -   &      &      &  -   &      &      &  -   &     \\
      10  & 3.577 & 29.99 & 1.196 &      &  -   &      &      &  -   &      &      &  -   &     \\
      11  & 5.078 & 49.99 & 1.900 &      &  -   &      &      &  -   &      &      &  -   &     \\
      12  & 5.594 & 79.99 & 2.510 &      &  -   &      &      &  -   &      &      &  -   &     \\
%       1  & 2.009 & 1.948 & 0.031 & 1.654 & 0.557 & 0.036 & 1.866 & 0.752 & 0.045 & 1.548 & 0.648 & 0.043\\
%       2  & 2.126 & 0.422 & 0.294 & 2.426 & 5.683 & 0.243 & 2.005 & 1.883 & 0.097 & 1.751 & 1.302 & 0.097\\
%       3  & 2.221 & 0.318 & 0.193 & 2.062 & 1.036 & 0.115 & 2.862 & 36.89 & 0.383 & 2.151 & 4.621 & 0.537\\
%       4  & 2.395 & 0.234 & 0.048 & 2.887 & 16.11 & 0.344 & 2.275 & 10.00 & 1.000 & 2.609 & 37.40 & 0.582\\
%       5  & 2.401 & 3.265 & 0.188 & 2.200 & 1.500 & 0.300 & 3.745 &  100.0 & 0.533 & 3.959 &  121.4 & 0.896\\
%       6  & 2.583 & 1.377 & 0.247 & 2.600 & 1.500 & 0.300 &      &  -   &      &      &  -   &     \\
%       7  & 2.651 & 0.277 & 0.166 & 3.800 & 70.00 & 0.700 &      &  -   &      &      &  -   &     \\
%       8  & 2.841 & 15.62 & 0.269 & 5.000 & 80.00 & 0.700 &      &  -   &      &      &  -   &     \\
%       9  & 3.056 & 12.78 & 0.242 &      &  -   &      &      &  -   &      &      &  -   &     \\
        \end{tabular}
    }
\caption{Model parameters for the relative permittivity of four TMDC monolayer materials
parameterized by the multi-Lorentzian dispersion relation defined by
Eq.~\eqref{eq:multiLorentzian}. The oscillator strength, $f^E_k=\hbar^2f_k$, the spectral
resonance energy, $E_k = \hbar\omega_k$, and the spectral width, $\gamma^E_k=\hbar\gamma_k$, of
the $k$th oscillator are given in \si{\electronvolt\squared}, \si{\electronvolt}, and
\si{\electronvolt}, respectively.} \label{tab:lorentzianParameters}
\end{table*}

The sheet conductance contains the intraband (Drude) contribution, described by the first term in
the r.h.s. of Eq.~\eqref{eq:model_sigma}, and the interband component given by the next two terms.
A more complete model valid at finite temperature can be used (see
Refs.~\cite{wss06njp,hd07prb,koppens11nl}); however, the model described by
Eq.~\eqref{eq:model_sigma} already captures the main features of graphene conductivity, most
notable being the possibility of tuning $\sigma_s(\omega)$ by changing the Fermi level via
chemical doping or applying a gate voltage.

In many numerical methods pertaining to computational electromagnetics it is more convenient to
work with bulk rather than surface quantities and therefore one often introduces bulk equivalents
of the surface quantities. In particular, instead of the sheet conductance, $\sigma_s$, one uses a
bulk conductivity, $\sigma_{b}=\sigma_s/h_{\mathrm{eff}}$, where $h_{\mathrm{eff}}$ is the
effective thickness of the 2DM. This approach can oftentimes create confusion due to the ambiguity
contained in the definition of the thickness of an atomic monolayer. Moreover, the electromagnetic
properties of 2DMs can alternatively be described by the electric permittivity, $\epsilon$, which
is related to the optical conductivity $\sigma_b$ via the following relation:
\begin{equation}\label{eq:eps_def}
    \epsilon(\omega) = \epsilon_{0}\left(1+\frac{i\sigma_{b}}{\epsilon_{0}\omega}\right)=\epsilon_{0}\left(1+\frac{i\sigma_s}{\epsilon_{0}\omega h_{\mathrm{eff}}}\right).
\end{equation}

In the case of TMDC monolayer materials, we describe their relative electric permittivity,
$\epsilon_r(\omega)=\epsilon(\omega)/\epsilon_{0}$, as a superposition of $N$ Lorentzian
functions:
\begin{align}\label{eq:multiLorentzian}
\epsilon_r(\omega) = \frac{\epsilon(\omega)}{\epsilon_{0}}&= 1 + \sum_{k=1}^{N}
\frac{f_k}{\omega^2_k-\omega^2-i\omega\gamma_k} \nonumber \\
&= 1 + \sum_{k=1}^{N}\frac{f^E_k}{E^2_k-E^2-iE\gamma^E_k},
\end{align}
where $f_k$, $\omega_k$, and $\gamma_k$ are the oscillator strength, resonance frequency, and
spectral width of the $k$th oscillator, respectively. The values of the model parameters for the
four considered TMDC monolayers were determined by fitting Eq.~\eqref{eq:multiLorentzian} to the
experimental data provided in Ref.~\cite{li14prb} and are presented in
\tabref{tab:lorentzianParameters} in terms of $f^E_k=\hbar^2f_k$, $E_k = \hbar\omega_k$, and
$\gamma^E_k=\hbar\gamma_k$. Note that the permittivity given by Eq.~\eqref{eq:multiLorentzian}
fulfils the Kramers-Kronig relation; however, the numerical fitting is less accurate for
$E=\hbar\omega>\SI{3.1}{\electronvolt}$ ($\lambda<\SI{0.4}{\micro\meter}$) due to the limited
range of the available experimental data.

The spectra of $\Re{\sigma_s(\omega)}$, a physical quantity related to the optical absorption in
the material, and $\Im{\sigma_s(\omega)}$ are depicted in Figs.~\ref{fig:Mat_Linear_Properties}(b)
and \ref{fig:Mat_Linear_Properties}(c), respectively. For each TMDC, $\Re{\sigma_s(\omega)}$
exhibits spectral peaks at wavelengths specific to the particular 2DM: the absorption peaks of
$\Re{\sigma_s(\omega)}$ with highest and second highest wavelength for each material (shown for
MoSe$_2$ as $A$ and $B$, respectively) correspond to low-energy interband transitions at the
$K(K^{\prime})$-point of the first Brillouin zone due to splitting of the valence-band by
spin-orbit coupling, whereas peaks at lower wavelengths correspond to higher energy interband
transitions \cite{li14prb}.

\subsection{Nonlinear optical properties of 2D materials}
\label{sec:2dnonlinprop} 
\begin{figure}[b]
    \centering
    \includegraphics[width=\linewidth]{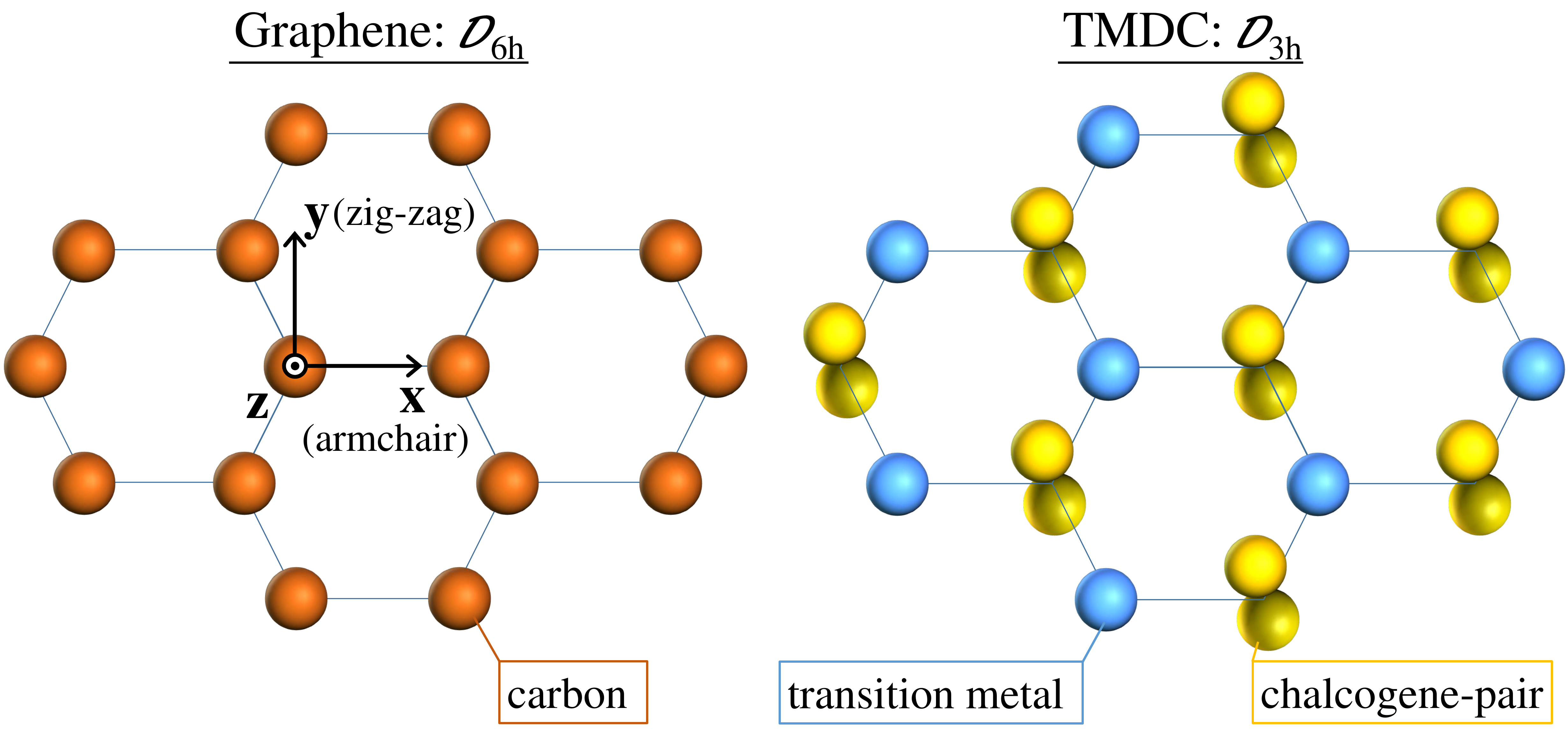}
\caption{Schematic of the atomic structures of graphene (symmetry group $\mathcal{D}_{6h}$, left)
and TMDC monolayers (symmetry group $\mathcal{D}_{3h}$, right). The hexagonal lattice of graphene
is centrosymmetric and thus SHG is forbidden. By contrast, the lattice of TMDC monolayers is
noncentrosymmetric and as such SHG is allowed.}
    \label{fig:groupStructure}
\end{figure}

The lattices of graphene and TMDC monolayers belong to different space
symmetry groups, which means that each of these materials requires a separate treatment. The
graphene lattice belongs to the $\mathcal{D}_{6h}$ space group, as illustrated in
\figref{fig:groupStructure}, which means that graphene is a centrosymmetric material and thus SHG
is a forbidden nonlinear optical process. On the other hand, THG is an allowed, particularly
strong process in graphene \cite{hong13prx,cvs14njp}, which makes it a suitable material for
nonlinear optical applications. By contrast, TMDC monolayers belong to the $\mathcal{D}_{3h}$
space group \cite{soares14prb} so that in this case SHG is the lowest-order nonlinear optical
process.

\begin{figure}[tbp]
    \centering
    \includegraphics[width=\linewidth]{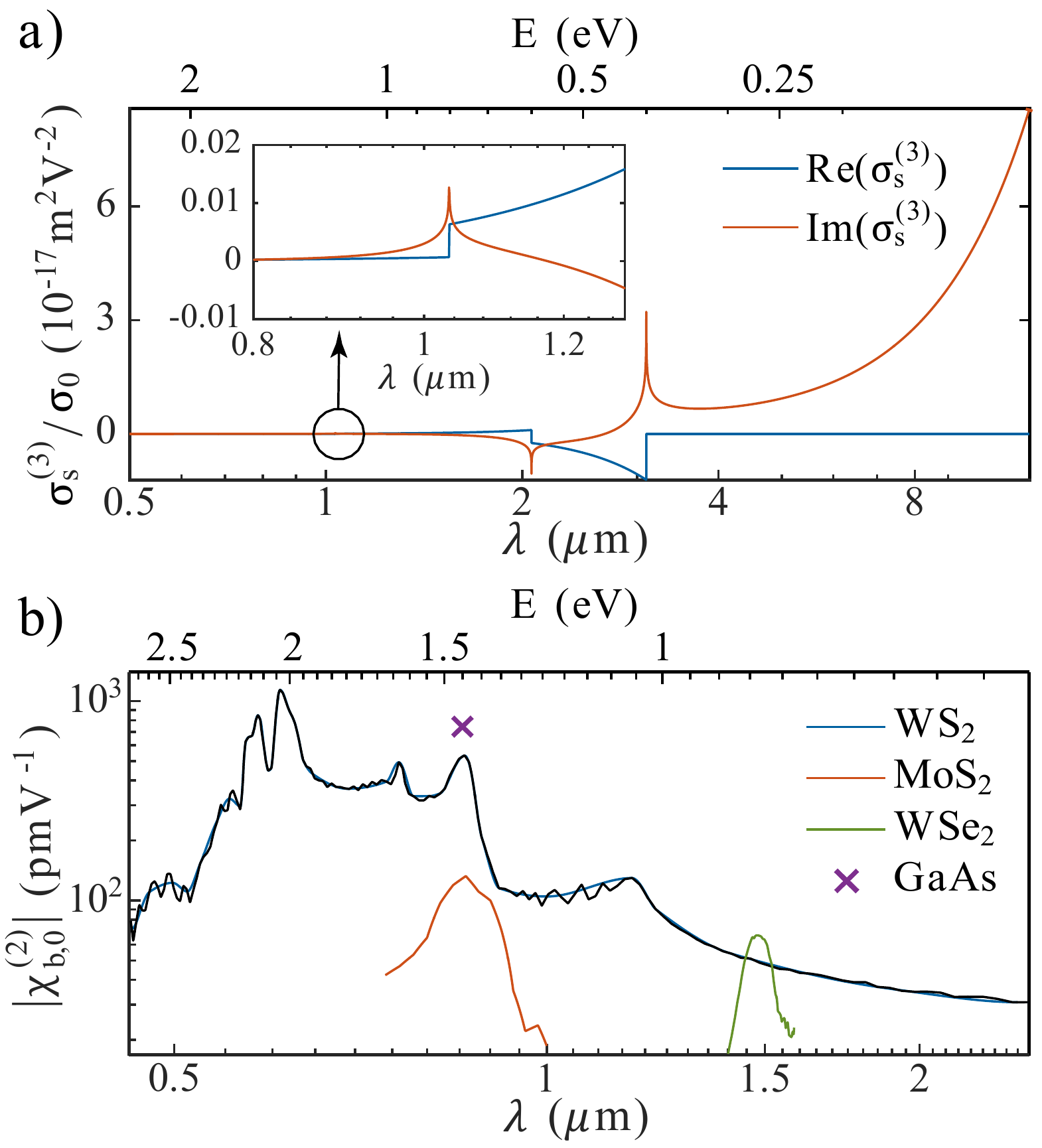}
\caption{a) Real and imaginary part of the third-order nonlinear surface conductivity of graphene,
$\sigma_s^{(3)}$. b) Spectral dependence of the effective bulk quadratic nonlinear susceptibility
$\vert\chi^{(2)}_{b}\vert$ of three TMDC monolayer materials and the value of the dominant
component of $\chi^{(2)}_{b,xyz}$ for GaAs, given for reference. For $\mathrm{WS_{2}}$, the
experimental noisy data (black line) has been smoothed out (blue line).}
    \label{fig:Mat_Nonlinear_Properties}
\end{figure}

The nonlinear optical response of graphene can be described by using the nonlinear optical
conductivity tensor, $\boldsymbol\sigma_s^{(3)}$, which relates the nonlinear surface current
density, $\j^{\mathrm{nl}}$, and the electric field, $\E$, at the fundamental frequency (FF).
Thus, if we assume that the graphene sheet lies in the $(x,y)$-plane at $z$=0, the nonlinear
current density, $\J^{\mathrm{nl}}$, can be written as:
\begin{align}
    \label{eq:defJNL}
    \J^{\mathrm{nl}}(\mathbf{r},t) \!=\! \j^{\mathrm{nl}}(\mathbf{r}_{t},t)\delta(z),
\end{align}
where $\mathbf{r}_{t}$ is the position vector lying in the graphene plane. Then, the nonlinear
surface current density can be written as:
\begin{align}\label{eq:defJNLsheet}
    j_\alpha^{\mathrm{nl}}(\mathbf{r}_{t},t) \!=\! \sigma^{(3)}_{s,\alpha} {E}_\alpha(\mathbf{r}_{t},t) |\E_t(\mathbf{r}_{t},t)|^2,
\end{align}
where $\E_t$ is the electric field component lying in the plane of graphene. In this description
\cite{cvs14njp}, the nonlinear current density lies in the plane of graphene and only depends on
the tangential field components. As a result, $\sigma^{(3)}_{s,x} = \sigma^{(3)}_{s,y} =
\sigma_s^{(3)}$ and $\sigma^{(3)}_{s,z} = 0$. A formula for $\sigma_s^{(3)}$, derived under the
assumptions that electron-electron and electron-phonon scattering as well as thermal effects can
be neglected, has been recently derived in Ref.~\cite{cvs14njp} and reads:
\begin{align}\label{eq:dispnl}
    \sigma_s^{(3)}(\omega) = \frac{i\sigma_{0} (\hbar v_F e)^2}{48\pi(\hbar\omega)^4} T\left(\frac{\hbar
    \omega}{2\varepsilon_F}\right).
\end{align}
In this equation, derived using perturbation theory, $T(x) = 17G(x) - 64G(2x) + 45G(3x)$,
\begin{align*}
    G(x) = \ln\left|\frac{1+x}{1-x}\right| + i\pi \theta(|x|-1),
\end{align*}
and $v_F = 3a_0\gamma_0/(2\hbar)\approx c/300$ is the Fermi velocity, with
$a_0=\SI{1.42}{\angstrom}$ being the nearest-neighbor distance between carbon atoms in graphene
and $\gamma_0 = \SI{2.7}{\electronvolt}$ is the nearest-neighbor coupling constant.
Figure~\ref{fig:Mat_Nonlinear_Properties}(a) depicts $\sigma_s^{(3)}(\omega)$ and additionally
highlights the lowest spectral peak at $\lambda=\SI{1.033}{\micro\meter}$ in the inset of the
figure.

Similarly to the linear case, one can introduce a ``bulk'' nonlinear conductivity,
$\boldsymbol\sigma^{(3)}_{b}=\boldsymbol\sigma_s^{(3)}/h_{\mathrm{eff}}$. This nonlinear
conductivity is particularly useful in experimental investigations of nonlinear optics of graphene
because it is related to an effective bulk third-order nonlinear susceptibility,
$\boldsymbol\chi^{(3)}_{b}$, the physical quantity that is usually measured experimentally. Using
the fact that for harmonic fields
$\J^{\mathrm{nl}}(\mathbf{r},t)=-i\omega\P^{\mathrm{nl}}(\mathbf{r},t)$, where
$\P^{\mathrm{nl}}(\mathbf{r},t)$ is the nonlinear polarization, one can easily show that
\begin{align}\label{eq:chi3tosigma3}
    \boldsymbol\chi^{(3)}_{b} =
    \frac{i}{\epsilon_0\Omega_{t}}\boldsymbol{\sigma}^{(3)}_{b} =
    \frac{i}{\epsilon_0\Omega_{t}h_{\mathrm{eff}}}\boldsymbol{\sigma}_s^{(3)},
\end{align}
where $\Omega_{t}=3\omega_{1}$ is the frequency at the third-harmonic (TH), with $\omega_{1}$
being the fundamental frequency.

In contrast to graphene, TMDC monolayers are non-centrosymmetric (see \figref{fig:groupStructure})
and therefore SHG is allowed \cite{shen03wiley,boyd08ap}. Based on the symmetry properties of
their space group, $\mathcal{D}_{3h}$, it can be shown that the structure of their quadratically
nonlinear susceptibility tensor, $\boldsymbol\chi^{(2)}_{b}$, yields only one independent,
non-vanishing component \cite{shen03wiley,janish14srep,malard13prb}:
%  WS2 janish14srep: P63/mmc
% MoS2: malard13prb: calls it D_3h, gets x,y right
% MoS2: kumar13prb: say D_3h, but they get x'/y' wrong
\begin{align}\label{eq:chi2structure}
    \chi^{(2)}_{b,0} := \chi^{(2)}_{b,xxx} = -\chi^{(2)}_{b,xyy} = -\chi^{(2)}_{b,yxy} = -\chi^{(2)}_{b,yyx},
\end{align}
where $x$ is the armchair direction of the monolayer and $y$ the orthogonal zig-zag-direction. The
nonlinear surface conductivity tensor, ${\boldsymbol\sigma}^{(2)}$, has the same structure and is
related to the nonlinear susceptibility via a relation similar to Eq.~\eqref{eq:chi3tosigma3}:
\begin{align}\label{eq:chi2tosigma2}
    \boldsymbol{\sigma}_s^{(2)} = -i\epsilon_0\Omega_{s}h_{\mathrm{eff}}\boldsymbol\chi^{(2)}_{b},
\end{align}
$\Omega_{s}=2\omega_{1}$ being the second-harmonic (SH) frequency. %for $N_F=1$ in \secref{sec:undepletedPump}.

Numerical values for the effective thickness of the four TMDC monolayer materials are taken from
Ref.~\cite{li14prb} and are presented in \tabref{tab:lorentzianParameters}. Although the
tensorial structure expressed in Eq.~\eqref{eq:chi2structure} already qualitatively determines the
nonlinear surface current,
\begin{align}\label{eq:defJSHG}
    \j_\alpha^{\mathrm{nl}}(\mathbf{r}_{t},t) = \sum_{\beta,\gamma=x,y,z} \boldsymbol\sigma^{(2)}_{s,\alpha\beta\gamma}
    E_\beta(\mathbf{r}_{t},t)E_\gamma(\mathbf{r}_{t},t),
\end{align}
the particular value of $\chi^{(2)}_{b,0}$ (and hence that of ${\boldsymbol\sigma}_s^{(2)}$) is of
practical importance. Reliable values over a certain spectral range exist for $\mathrm{WS_2}$
\cite{janish14srep}, $\mathrm{MoS_2}$ \cite{malard13prb,li13nl}, and $\mathrm{WSe_2}$
\cite{seyler15nn} and are depicted in \figref{fig:Mat_Nonlinear_Properties}(b). Despite these
materials consisting of just a single atomic layer, the largest values of
$\chi^{(2)}_{b}=\SI{1140}{\pico\meter\per\volt}$, $\chi^{(2)}_{b}=\SI{132}{\pico\meter\per\volt}$,
and $\chi^{(2)}_{b}=\SI{67}{\pico\meter\per\volt}$, respectively, have comparable magnitude to
that of GaAs, $\chi^{(2)}_{b,xyz}=\SI{740}{\pico\meter\per\volt}$, a medium with strong bulk
quadratic susceptibility \cite{boyd08ap,bd03prl}.
% at $\lambda=\SI{0.8550}{\micro\meter}$

% kumar13prb: gives a way too high value 10^4 too high.

It is also instructive to compare the strength of the SHG in TMDC monolayer materials to that in
noble metals, as in both cases the SH is generated largely in a single atomic layer. To this end,
we introduce a surface quadratic nonlinear susceptibility tensor,
$\boldsymbol\chi^{(2)}_{s}=h_{\mathrm{eff}}\boldsymbol\chi^{(2)}_{b}$, which we then compare to
the surface quadratic nonlinear susceptibility tensor of Au and Ag. The highest value of
$\boldsymbol\chi^{(2)}_{s}$ of $\mathrm{WS_2}$ is
$\chi^{(2)}_{s}=\SI{7e-19}{\meter\squared\per\volt}$ and is hence comparable to the dominant
component, $\chi^{(2)}_{s,\perp\perp\perp}$, for Ag and Au, i.e.
$\chi^{(2)}_{s,\perp\perp\perp}=\SI{1.59e-18}{\meter\squared\per\volt}$ and
$\chi^{(2)}_{s,\perp\perp\perp}=\SI{1.35e-18}{\meter\squared\per\volt}$ \cite{ktr04jap},
respectively.
% maybe put model for h-BN (li13nl = chapter 6 Yilei Li thesis)

\section{Higher-harmonic generation in periodically patterned 2D materials}\label{sec:mathformulation}
This section will introduce a numerical method for calculating the nonlinear optical interaction
of light with periodic structures consisting of 2DMs embedded in dielectric or metallic patterned
media. Nonlinear optical effects in the bulk part of periodic structures are not considered in
this article, as there already exist numerical formulations that are compatible with
\cite{nty02josaa,bt07josab,prl10josab} or complementary to \cite{wgp15josab} the proposed method
to accurately calculate these bulk nonlinear optical effects.

To this end, the undepleted pump approximation will be used as a means to introduce the nonlinear
optical interactions in the governing Maxwell equations (MEs). Since the basis of the proposed
numerical method is the RCWA, this method will be briefly revisited and the necessary mathematical
formalism derived. Then, a modified boundary condition for interfaces incorporating conductive,
potentially nonlinear 2DMs, is considered and a numerically stable $\mathcal{S}$-matrix algorithm
for propagation of nonlinear fields generated by the monolayers of 2DMs in multilayered periodic
structures is derived to complete the proposed algorithm. Importantly, the thickness of the 2DMs
does not enter in our numerical method, thus removing an ambiguity present in other numerical
methods currently used to describe these materials
\cite{gao12acsnano,gosciniak13scirep,nggm12prb}.

\subsection{Optical higher-harmonic generation in the undepleted pump approximation}\label{sec:undepletedPump}
This section introduces the physical and mathematical model for multi-frequency, nonlinear optical
interaction in the so-called undepleted pump approximation. For a more complete description of
this theoretical approach we refer the reader to Ref.~\cite{boyd08ap}. To this end, let us assume
that the real electric field, $\bm{\mathcal{E}}(\mathbf{r},t)$, as a function of position,
$\mathbf{r}$, and time, $t$, is composed of $N_F+1$ monochromatic waves with pairwise different
optical frequencies, $\omega_n$, $n=0,\ldots,N_F$, that is:
\begin{align}\label{eq:EMultiOmega}
  \bm{\mathcal{E}}(\mathbf{r},t) = \frac{1}{2}\sum_{n=0}^{N_F} \E^{(\omega_n)}(\mathbf{r})\exp(-i\omega_n t) + c.c.,
\end{align}
where $\E^{(\omega_n)}(\mathbf{r})$ is the amplitude of the wave with frequency, $\omega_n$, and
$c.c.$ denotes the complex conjugation operation. The fields with frequencies $\omega_n$,
$n=1,\ldots,N_F$, are assumed to be excitation (pump) fields, whereas the field at
$\Omega\coloneqq\omega_{0}$ is a higher-harmonic, nonlinearly generated field. This nonlinear
optical field depends on the specific nonlinear optical process under investigation. Similar
expressions are assumed for the other electromagnetic quantities. In particular, the polarization,
$\bm{\mathcal{P}}(\mathbf{r},t)$, is expressed as:
\begin{align}\label{eq:PMultiOmega}
\bm{\mathcal{P}}(\mathbf{r},t) = \frac{1}{2}\sum_{n=0}^{N_{F}}
\P^{(\omega_n)}(\bm{\mathcal{E}};\mathbf{r})\exp(-i\omega_n t) + c.c.,
\end{align}
where $\P^{(\omega_n)}(\bm{\mathcal{E}};\mathbf{r})$ encodes a -- possibly nonlinear -- functional
relation between the total, time dependant field $\bm{\mathcal{E}}$ and the polarization with
$\omega_n$-harmonic time dependence at position $\mathbf{r}$.

In most situations of practical interest, depending on particular physical conditions, only
certain nonlinear polarizations are generated with significant strength. Moreover, in the case of
strong excitation fields or when the induced nonlinear polarizations are weak, the following
assumption can be made:
\begin{align*}
\P^{(\omega_n)}(\bm{\mathcal{E}};\mathbf{r}) = \P^{(\omega_n)}(\E^{(\omega_n)};\mathbf{r}) =
\epsilon_0\chi^{(1)}(\mathbf{r};\omega_n)\E^{(\omega_n)}(\mathbf{r}).
\end{align*}
This means that the polarizations at the frequencies of the pump fields, $\omega_n$,
$n=1,\ldots,N_F$, are solely determined by the corresponding linear optical susceptibility,
$\chi^{(1)}(\omega_n)$, of the optical medium and the electric field at frequency $\omega_n$. The
polarization at the nonlinear frequency, $\Omega$, however, consists of both the linear
polarization at $\Omega$, which is proportional to the nonlinear field,
$\E^{(\Omega)}(\mathbf{r})$, and a nonlinear polarization,
$\P^{(\mathrm{nl},\Omega)}(\E^{(\omega_{n\neq0})};\mathbf{r})$, which can incorporate the electric
field from all pump fields with frequencies $\omega_{n}$, $n>0$:
\begin{align*}
  \P^{(\Omega)}(\bm{\mathcal{E};\mathbf{r}}) = \epsilon_0\chi^{(1)}(\Omega)\E^{(\Omega)}(\mathbf{r}) + \P^{(\mathrm{nl},\Omega)}(\E^{(\omega_{n\neq0})};\mathbf{r}).
\end{align*}

Under this assumption, the optical fields at the pump frequencies, $\omega_1,\ldots,\omega_{N_F}$,
are not altered, or in a narrower sense, depleted by the nonlinear processes, hence this
assumption is called the undepleted pump approximation. The particular form of the nonlinear
polarization, $\P^{(\mathrm{nl},\Omega)}(\E^{(\omega_{n\neq0})};\mathbf{r})$, depends on the
nonlinear process under consideration, such as sum- or difference-frequency generation (SFG or
DFG), SHG, and THG. Specific expressions for
$\P^{(\mathrm{nl},\Omega)}(\E^{(\omega_{n\neq0})};\mathbf{r})$ for SHG and THG in 2DMs have been
given in the previous section.

The algorithmic appeal of this approximation is the possibility to obtain a general, one-way
coupled calculation scheme, which only requires the solution of $N_F$ homogeneous linear optical
problems and one affine linear problem with electrical sources. In particular, the numerical
algorithm consists of the following three steps: In the step, \textit{i}) (\textit{linear
calculations}) one calculates the fields $\E^{(\omega_n)}$ at the pump frequencies, $\omega_n$,
$n=1,\ldots,N_F$. In the second step, \textit{ii}) (\textit{polarization evaluation}) one
evaluates the nonlinear polarization,
$\P^{(\mathrm{nl},\Omega)}(\E^{(\omega_{n\neq0})};\mathbf{r})$, for the particular nonlinear
process under consideration. Finally, in the third step, \textit{iii}) (\textit{nonlinear
calculation}), one calculates the generated nonlinear electric field, $\E^{(\Omega)}$.

Given this generic three-step algorithm, in the remaining part of this section we will describe
the numerical implementation of the computational steps \textit{i}) and \textit{iii}) for the
particular case of higher-harmonic generation in multilayered periodic structures, which contain
periodically patterned monolayers of 2DMs that can exhibit quadratic or cubic optical
nonlinearity.

\subsection{RCWA: modal expression for fields in bulk periodic regions}
The proposed method aims at describing nonlinear optical effects in 2D materials, hence, for the
sake of simplicity, we do not consider nonlinear optical effects in the bulky materials involved,
but only in the 2DM sheets. These latter nonlinear effects can be easily incorporated into our
algorithm, as we have recently shown \cite{wgp15josab}. Thus, the electromagnetic fields in the
bulk parts of the structure are governed by the time-harmonic MEs for nonmagnetic media without
sources:
\begin{subequations}\label{eq:ME}
\begin{align}
    &\nabla \cdot [\epsilon_0\epsilon_r^{(\omega)}(\mathbf{r}) \E^{(\omega)}(\mathbf{r})] = 0, \label{eq:ME1} \\
    &\nabla \times \H^{(\omega)}(\mathbf{r}) = - i\omega \epsilon_0\epsilon_r(\mathbf{r}) \E^{(\omega)}(\mathbf{r}), \label{eq:ME2} \\
    &\nabla \times \E^{(\omega)}(\mathbf{r}) = i\omega \mu_0 \H^{(\omega)}(\mathbf{r}), \label{eq:ME3} \\
    &\nabla \cdot \H^{(\omega)}(\mathbf{r}) = 0, \label{eq:ME4}
\end{align}
\end{subequations}
for $\omega = \omega_n$, $n=0,\dots,N_F$. These equations have to be completed by boundary
conditions, which will be the point where both linear and nonlinear interface effects are
incorporated. Before the appropriate boundary conditions are introduced and discussed in
\secref{sec:BC_2DM}, the RCWA is used to describe the solution of Eqs.~\eqref{eq:ME} in each
layer. For now, let us drop the superscript $\omega$, as the description of the modal form of the
electromagnetic fields is independent on whether a pump or a nonlinearly generated frequency is
considered.

The RCWA method is a mature, widely known
algorithm\cite{mgp95josaa,Li1996josaa,l97josaa,srk07josaa}, but in order to make the further
derivation mathematically consistent, the notation and framework used here is described in the
Appendix. The main result that is necessary for the extension of this method to nonlinear 2DMs is
that the solutions in each of the bulk layers are given by a modal expansion, where each mode
profile is given as a Fourier series.

The electromagnetic fields in each layer, denoted with superscript ``$L$``, are expressed in a
linear combination of upward- and downward-propagating modes with coefficients $c_m^{(L,\pm)}$,
where superscripts ``$+$'' and ``$-$'' refer to upward and downward propagation, respectively.
Each mode is fully described by its complex propagation constant, $\propconst^{(L,\pm)}_m$, which
determines the $z$-dependence, and the $z$-independent Fourier series coefficients
$E^{(L,m,\pm)}_{\alpha,n}$ and $H^{(L,m,\pm)}_{\alpha,n}$, for $\alpha=x,y,z$. These coefficients
determine the transverse profile of mode $m$ via the Fourier series reconstruction operator,
$\rec{\cdot}$, defined in the Appendix. This reconstruction of the electric field is hence expressed
as:
\begin{align}
  E_\alpha^{(L)}(x,y,z) = \rec{\fvec{E_\alpha^{(L)}(z)}}(x,y), \label{eq:ModalFieldTotal}
\end{align}
with
\begin{align}
\label{eq:ModalFieldMatrices}
\left(\!\!\begin{array}{c}
\fvec{E_\alpha^{(L)}(z)} \\ \fvec{H_\alpha^{(L)}(z)}
\end{array}\!\!\right) \!&=\! \left[\!\!\begin{array}{cc}
\textsf{E}_{\alpha}^{(L,+)} & \!\!\!\textsf{E}_{\alpha}^{(L,-)} \\
\textsf{H}_{\alpha}^{(L,+)} & \!\!\!\textsf{H}_{\alpha}^{(L,-)}
\end{array}\!\!\!\right]\!\!
\left[\!\!\begin{array}{cc}
    \propmat^{(L,+)}(z) \!\!\!\!\!\!\!\!\!\! & \quad 0  \\
    \!\!\!\!\!\!\!\!0  &\!\!\!\!\!\!\!\!\!\! \propmat^{(L,-)}(z)
\end{array}\!\!\right]\!\!
\left(\!\!\begin{array}{c}
\bc^{(L,+)} \\ \bc^{(L,-)}
\end{array}\!\!\right).\!\!
\end{align}
Here, the $m$th column of $\textsf{E}_{\alpha}^{(L,\pm)}$ and $\textsf{H}_{\alpha}^{(L,\pm)}$ is
the vector of Fourier coefficients $\fvec{E_\alpha^{(L,m,\pm)}}$ and
$\fvec{H_\alpha^{(L,m,\pm)}}$, respectively, of the $\alpha$-component of the $m$th mode. The
propagation matrix, $\propmat^{(L,\pm)}(z)$, is diagonal with entries $\propmat^{(L,\pm)}_{mm}(z)
= e^{ik_0\propconst_m^{(L,\pm)}(z-z^{(L,\mp)})}$, where $z^{(L,\pm)}$ is the bottom/top
$z$-coordinate of layer $L$. This means that the electromagnetic fields in each layer are fully
determined by their coefficients $\bc^{(L,\pm)}$.

\begin{figure*}[t]
    \centering
    \includegraphics[width=\linewidth]{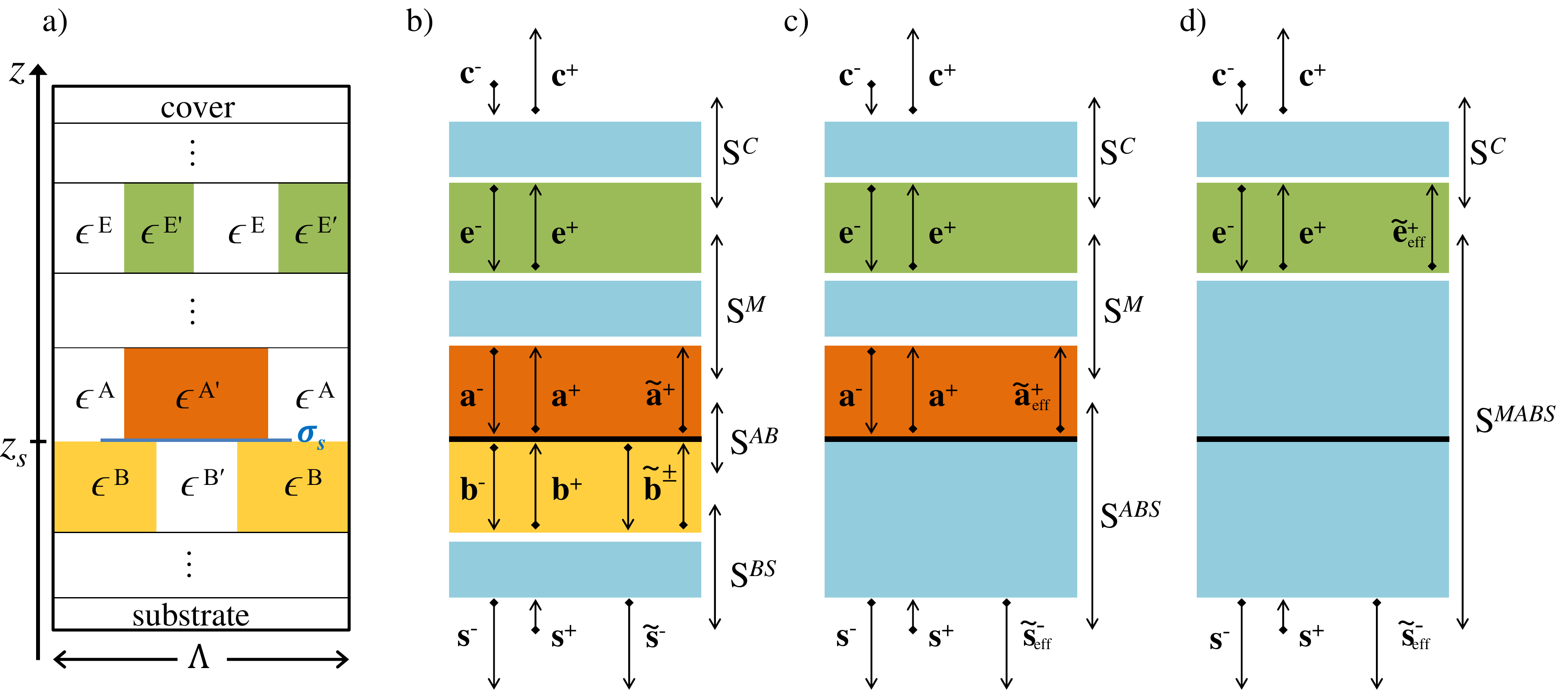}
\caption{Schematic of a multilayer structure located between cover and substrate and containing a
2DM sheet with surface conductance $\sigma_s$, the neighboring layers $A$ and $B$, and the
evaluation layer, $E$, in which the electromagnetic field is to be determined. b) Linear
($\bc^\pm,\be^\pm,\ldots$) and nonlinear ($\tilde\ba^+, \tilde\bb^\pm, \tilde\bs^-$) mode
coefficients in the respective layers and the $\mathcal{S}$-matrices that connect them. c)
Combination of layers $B$ and $S$ yields the combined $\mathcal S$-matrix $\textsf{S}^{ABS}$ and
effective nonlinear coefficients $\tilde \ba^{+}_\mathrm{eff}$ and $\tilde \bs^-_\mathrm{eff}$. d)
Repeated combination of $\mathcal S$-matrices allows the calculation of the field coefficients,
$\be$, from the effective nonlinear mode coefficients, $\tilde\be^{+}_\mathrm{eff}$.}
    \label{fig:SMatrixCollapse}
\end{figure*}
\subsection{Modeling 2D materials via RCWA by means of boundary conditions}\label{sec:BC_2DM}
The optical response of 2DMs can chiefly be described in two different ways. One direct approach
in RCWA and other numerical methods is to model the monolayer material as a periodic, very thin
film, similar to the layers in which the periodic bulk components are decomposed. To this end, a
physical effective thickness, $h_{\mathrm{eff}}$, of the 2DM under consideration has to be chosen
as well as its permittivity. This is a rather questionable and inefficient approach, for three
main reasons. First, the choice of $h_{\mathrm{eff}}$ is somewhat ambiguous, both because the
thickness of an atomic monolayer does not have a clear meaning in the classical physics context
and due to the fact that the experimentally measured value of this thickness varies considerably,
for all 2DMs. Second, this approach suffers from numerical artefacts and slow convergence, as
reported in recent works~\cite{k13ol,nka14josab}. This is understandable as the thickness of the
monolayer is much smaller than the optical wavelength, $h_{\mathrm{eff}}\ll\lambda$. Finally, it
is computationally costly to find RCWA modes of the 2DMs since this requires the numerical
solution of an eigenvalue problem (see the Appendix).

An alternative procedure for describing sheets of structured 2DMs is introduced in this article
and consists of modeling such 2DM-sheets by means of a surface conductance, which enters the
algorithm in via electromagnetic boundary conditions between adjacent bulk layers. Thus, consider
the schematic of the multilayer structure presented in \figref{fig:SMatrixCollapse}(a). At a
horizontal plane with $z=z_s$, which is located between two adjacent bulk layers, the following
relations of the tangential $\E$ and $\H$ fields in the top region (superscript $A$) and the
bottom region (superscript $B$) have to be fulfilled \cite{jackson99book}:
\begin{subequations}\label{eq:BC}
\begin{align}
    &\hat{\normal} \times \left( \E^{(\omega,A)}(x,y,z_s) - \E^{(\omega,B)}(x,y,z_s)\right) = 0, \label{eq:BC_electric}\\
    &\hat{\normal} \times \left( \H^{(\omega,A)}(x,y,z_s) - \H^{(\omega,B)}(x,y,z_s)\right) = \j^{(\omega)}(x,y),  \label{eq:BC_magnetic}
\end{align}
\end{subequations}
where $\hat{\normal} = (0,0,1)$ denotes the unit vector along the $z$-direction, pointing towards
region $A$. The first of these equations expresses the continuity of the tangential components of
$\E$ at the interface. The second equation warrants further discussion. Thus, in the presence of a
2DM at $z=z_s$, the tangential component of $\H$ is discontinuous, its variation across the
\textit{i}th interface being given by the surface current, $\j^{(\omega)}(x,y) =
\j^{(\omega,\mathrm{lin})}(x,y) + \j^{(\omega,\mathrm{nl})}(x,y)$. In particular, the total
surface current contains a linear component given by,
\begin{align}\label{eq:linSurfaceCurrent}
    \j^{(\omega,\mathrm{lin})}(x,y) = \sigma_s^{(\omega)}(x,y)\E^{(\omega,s)}(x,y),
\end{align}
and a nonlinear surface current, $\j^{(\omega,\mathrm{nl})}(x,y)$. The linear surface current
depends only on the electric field at the interface, $\E^{(\omega,s)}(x,y,z_{s})$, at the same
frequency, $\omega$, and the sheet conductance distribution at the interface. The nonlinear
surface current, on the other hand, is assumed to be different from zero only at the frequency
$\omega=\omega_0$ and generally depends on the electric field at all the other frequencies,
$\omega=\omega_n$, $n=1\ldots,N_F$.

One can easily see that $\j^{(\omega)}(x,y)$ is a pseudo-periodic function of $x$- and
$y$-coordinate, hence it is determined by its Fourier vector coefficient, $\fvec{\j^{(\omega)}}$.
Special attention is, however, necessary when one calculates the Fourier coefficients of the
linear current, $\fvec{\j^{(\omega,\mathrm{lin})}}$, as in the real space it is given by a product
of two periodic functions, $\sigma_s^{(\omega)}(x,y)$ and $\E^{(\omega,s)}(x,y)$. This issue,
known as the fast Fourier factorization problem, must be properly addressed in order to achieve
high accuracy and fast convergence of methods relying on Fourier series representation
\cite{lm96josaa,l97josaa,li03josaa,srk07josaa}.

\begin{figure}[t]
\centering
\includegraphics[width=\linewidth]{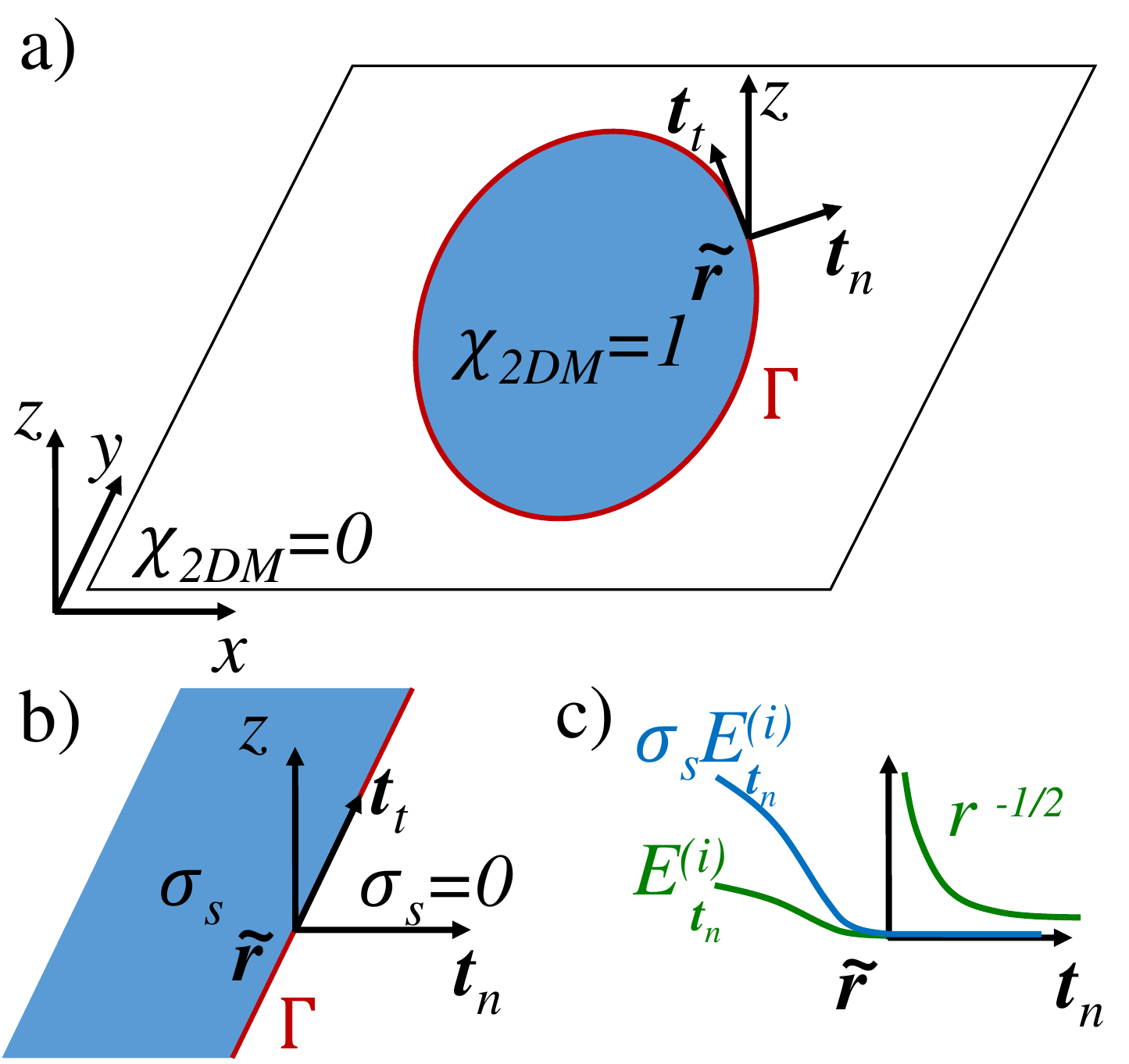}
\caption{a) Cross-section of the periodic structure through the plane $z=z_{s}$ containing the 2DM
distribution $\chi_\mathrm{2DM}(x,y)$ with the boundary contour, $\Gamma$, showing a unit cell and
the local coordinate system at a generic location, $\tilde{\mathbf{r}}$. b) Local coordinate
system at $\tilde{\mathbf{r}}$. c) Qualitative behavior of the surface quantities
$E_{\bt_n}^{(i)}$ and $j_{\bt_n}^{(\omega,\mathrm{lin})}=\sigma^{(\omega)}_s
E_{\bt_n}^{(\omega,s)}$.} \label{fig:localCS}
\end{figure}
To find the correct factorization rule for Eq.~\eqref{eq:linSurfaceCurrent}, the continuity
property of the two factors in the r.h.s. of this equation as well as that of the product have to
be investigated. To this end, consider \figref{fig:localCS}, which depicts the $z=z_s$ plane that
comprises a surface conductivity,
$\sigma^{(\omega)}_s(x,y)=\chi_\mathrm{2DM}(x,y)\sigma_s(\omega)$. Hereby,
$\chi_\mathrm{2DM}(x,y)$ denotes the characteristic function of the 2DM distribution and
$\sigma_s(\omega)$ is the sheet conductance of the 2DM, e.g. Eq.~\eqref{eq:model_sigma} for
graphene. At each point, $\tilde{\br}$, of the 1D boundary $\Gamma$ of the 2DM we introduce a
local coordinate system defined by the orthogonal vectors $\bt_n$, $\bt_t$, and $\e_z$, where
$\bt_n$ and $\bt_t$ are the in-plane unit vectors normal and tangent to the contour $\Gamma$,
respectively, at the point $\tilde{\mathbf{r}}$ and $\e_z$ is the unit vector along the $z$-axis.
One can assume that the sheet conductance function is smooth along the $\bt_t$ direction, yet it
is discontinuous along the $\bt_n$ direction, as per \figref{fig:localCS}(b). The continuity
relation, Eq.~\eqref{eq:BC_electric}, of the tangential electric field at $z=z_s$ allows one to
define a tangential surface electric field, $E_{\bt_n}^{(\omega,s)}(x,y)$, as the limit of the
volumetric electric fields, $E^{(\omega)}_{\bt_n}(x,y,z)$, from either side of $z=z_{s}$ plane:
\begin{align}\label{eq:defInterfaceTangentialField}
    E_{\bt_n}^{(\omega,s)}(x,y) \coloneqq \lim_{z\to z_s} E^{(\omega)}_{\bt_n}(x,y,z).
\end{align}

The electromagnetic near-field in the vicinity of such a conductive sheet can be calculated
analytically \cite{born00cap} and here we only summarize the main results relevant to our
numerical method: \textit{i}) The $\bt_n$-component of the surface current vanishes at
$\tilde{\mathbf{r}}$ as
\begin{align*}
    \lim_{\rho\to 0^{+}} j_{\bt_n}^{(\omega,\mathrm{lin})}(\tilde{\mathbf{r}} - \rho\bt_n) = \lim_{\rho\to 0^{+}} C\sqrt{k\rho}e^{ik\rho}=
    0.
\end{align*} % eq.48 Chapter 11.5.4
Since $j_{\bt_n}^{(\omega,\mathrm{lin})}(\tilde{\mathbf{r}} + \rho\bt_n)=0$ for $\rho>0$,
$j_{\bt_n}^{(\omega,\mathrm{lin})}(\tilde{\mathbf{r}})$ is continuous around $\tilde{\mathbf{r}}$.
\textit{ii}) The $\bt_n$-component of the surface electric field, $E_{\bt_n}^{(\omega,
s)}(\tilde{\mathbf{r}})$, is discontinuous at the boundary $\Gamma$. More specifically,
\begin{align*}
    \lim_{\rho\to 0^{+}} E_{\bt_n}^{(\omega,s)}(\tilde{\mathbf{r}} - \rho\bt_n) =0,
\end{align*}
that is $E_{\bt_n}^{(\omega,s)}(\tilde{\mathbf{r}})$ inside the 2DM vanishes near the boundary,
whereas it diverges when approaching the interface from outside the 2DM,
\begin{align*}
    \lim_{\rho\to 0^{+}} E_{\bt_n}^{(\omega,s)}(\tilde{\mathbf{r}} + \rho\bt_n) = \lim_{\rho\to 0^{+}} C (k\rho)^{-1/2} = \infty.
\end{align*}
Finally, \textit{iii}) the in-plane tangential component
$E_{\bt_t}^{(\omega,s)}(\tilde{\mathbf{r}})$ of the interfacial electric field is continuous and
bounded at the boundary $\Gamma$. Findings \textit{i}) and \textit{ii}) are schematically depicted
in \figref{fig:localCS}(c).

These results imply that $j_{\bt_n}^{(\omega,\mathrm{lin})}(x,y)$ is a continuous function but
both factors, $\sigma_s^{(\omega)}(x,y)$ and $E_{\bt_n}^{(\omega,s)}(x,y)$, in
Eq.~\eqref{eq:linSurfaceCurrent} are discontinuous. Therefore, Laurent's product rule cannot be
applied, as it would result in spurious field oscillations near the boundary $\Gamma$ and slow
convergence of the algorithm. The inverse rule is also not applicable, as the first factor,
$\sigma_s^{(\omega)}(x,y)$, vanishes in some regions of the unit cell, outside of the domains
occupied by the 2DM. To overcome this problem, a small but nonzero surface conductivity,
$\sigma_{s,\mathrm{add}}\neq0$, is added at the interface regions where there is no 2DM with
finite conductivity. As a result, this modified sheet conductance distribution in the unit cell is
defined as:
\begin{align}\label{eq:modSigma}
  \tilde \sigma^{(\omega)}_s(x,y) \coloneqq \chi_\mathrm{2DM}(x,y)\sigma_s(\omega)  +
        \left[1-\chi_\mathrm{2DM}(x,y)\right]\sigma_{s,\mathrm{add}}.
\end{align}

As it has been seen in \secref{sec:2dlinprop}, $\sigma_{s}(\omega)$ is a function of frequency and
can vary by orders of magnitude. Therefore, it is natural to scale the added conductivity,
$\sigma_{s,\mathrm{add}}$, relative to the absolute value of the physical conductivity of the 2DM,
that is $\sigma_{s,\mathrm{add}}(\omega, \eta) = -i\eta\vert\sigma_s(\omega)\vert$, where $\eta$
is a small scaling constant. This choice ensures in addition that $\sigma_{s,\mathrm{add}}$ is a
negative imaginary quantity, so that one does not introduce artificial losses in the system but
only a vanishingly small phase change at the interface. However small this effect is, it
introduces an error and hence its quantitative contribution to the optical response of the
photonic structure must be investigated. The influence of $\sigma_{s,\mathrm{add}}$, and
implicitly that of $\eta$, on the calculated near- and far-fields by the proposed method will be
investigated in \secref{sec:convergence}.

A further justification of our nonzero conductivity model is provided by the application of
Ampere's law for a loop around the interface, an approach that has been recently introduced
elsewhere \cite{k13ol,nka14josab}. This alternative model also amounts to using a modified
conductivity at the interface similar to that in Eq.~\eqref{eq:modSigma}.

In order to complete the derivation of the correct Fourier factorization of the surface current
defined by Eq.~\eqref{eq:linSurfaceCurrent} for $\tilde \sigma^{(\omega)}_s(x,y)\neq 0$, let
$\nvfv(x,y)$ denote a normal vector field (NVF), namely a vector field that is normal to the 1D
contour $\Gamma$ at any point of the interface and is analytically continued into the regions away
from $\Gamma$. This can be done analytically for certain cross-sections and ways to automatically
generate a NVF for arbitrary cross-sections are readily available \cite{gotz08oe}. This allows one
to express the factorization of $\fvec{\j^{(\omega,\mathrm{lin})}}$ in terms of the conductivity
distribution coefficients, $\fmatrix{\tilde\sigma^{(\omega)}_s}$ and
$\fmatrix{1/\tilde\sigma^{(\omega)}_s}$, and tangential field coefficients,
$\fvec{\E^{(\omega,s)}}$, in such a way that the normal $\bt_n$-component of
$\j^{(\omega,\mathrm{lin})}(x,y)$ is decomposed using the inverse rule, and its tangential
$\bt_t$-component is decomposed using the regular product rule:
\begin{align}\label{eq:FacRuleJ}
    \fvec{j^{(\mathrm{lin})}_\alpha} = & \sum_{\beta=x,y}
    \Delta\textsf{\nvf}_{\alpha,\beta}\fvec{E_\beta^{(\omega,s)}},
\end{align}
where the conductivity difference matrix is given by
\begin{align}\label{eq:DefDeltaNVF}
      \Delta\textsf{\nvf}_{\alpha,\beta} = &\ \delta_{\alpha\beta} \fmatrix{\tilde \sigma^{(\omega)}} + \frac{1}{2}\fmatrix{\nvf_\alpha\nvf_\beta} \left(\fmatrix{1/ \tilde \sigma^{(\omega)}_s}^{-1} - \fmatrix{\tilde \sigma^{(\omega)}_s}\right) \nonumber \\
                   &\ + \frac{1}{2}\left(\fmatrix{1/ \tilde \sigma^{(\omega)}_s}^{-1} - \fmatrix{\tilde \sigma^{(\omega)}_s}\right)\fmatrix{\nvf_\alpha\nvf_\beta}.
\end{align}
This procedure is similar to the factorization rule for the displacement field used in the regular
RCWA for bulk materials employing the NVF approach for fast Fourier factorization given in
\eqref{eq:ConstRelationNVF}; see also Refs.~\cite{srk07josaa,wgp15jo}.

The calculation of $\j_{\bt_n}^{(\omega_0,\mathrm{nl})}(x,y)$ as a function of $\E$ requires the
electric near-field distribution at the interface. This near-field is difficult to calculate
accurately in the RCWA even when the correct Fourier factorization rules are employed, as was
noted in Refs.~\cite{popov2002staircase,lj98jmo,wgp15jo}. A revealing insight into the accuracy of
near-field calculations is that RCWA relies on the accurate description of continuous quantities,
as they can be readily expanded in Fourier series. This can be exploited to achieve an accurate
interface-field description: knowing that the $\bt_t$-component of the interface electric field,
$\E^{(\omega,s)}_{\bt_t}$, is continuous, it is convenient to evaluate this component directly by
Fourier series reconstruction. Since, however, its $\bt_n$-component is discontinuous (and in fact
singular), it can only be poorly represented by its Fourier series, so that the reconstructed
field $\E^{(\omega,s)}$ will experience unphysical oscillations and the Gibbs phenomenon.

A more well-behaved quantity is the surface current, which is continuous and hence properly
described by its Fourier series. With the definition of the Fourier coefficients of the vectorial
normal surface current and tangential surface fields as:
\begin{subequations}\label{eq:improvedEval}
\begin{align}
    \fvec{\E_{\bt_t}^{(\omega,s)}} =& \fmatrix{\textsf{I}-\nvfv\nvfv^T}\fvec{\E_{\bt}^{(s)}}, \label{eq:improvedEval_t} \\
    \fvec{\j_{\bt_n}^{(\omega)}} =& \frac 12\left(\fmatrix{\nvfv\nvfv^T} \fmatrix{1/ \tilde \sigma^{(\omega)}_s}^{-1} \right. \nonumber \\
    &+ \left.\fmatrix{1/ \tilde \sigma^{(\omega)}_s}^{-1} \fmatrix{\nvfv\nvfv^T}\right) \fvec{\E_{\bt}^{(\omega,s)}}, \label{eq:improvedEval_n}
\end{align}
\end{subequations}
one obtains the reconstructed field at the interface as:
\begin{align}\label{eq:defReconInterfaceField}
  \E^{(\omega,s)}_{\bt}(x,y) =\rec{\fvec{\E_{\bt_t}^{(\omega,s)}} }(x,y) +
  \frac{1}{\sigma_s(\omega)}\rec{\fvec{\j_{\bt_n}^{(\omega)}}}(x,y).
\end{align}
Here, all Fourier series represent functions continuous at the in-plane boundaries of the 2DM. The
fact that this approach only allows a reconstruction of the near-field at the 2DM interface, where
$\chi_\mathrm{2DM}\sigma_s\neq 0$, is not particularly concerning since the nonlinear surface
current is \textit{a priori} nonvanishing there only.

\subsection{Inhomogeneous $\mathcal{S}$-matrix formalism}\label{sec:inhomSMatrix}
In he previous section we have introduced the modal expansion of the electromagnetic fields inside
and around the diffraction grating structure and cast them in a concise matrix form, as per
Eq.~\eqref{eq:ModalFieldMatrices}. It remains now to determine the coefficients $\bc^\pm$ in every
computational layer in order to obtain the electromagnetic fields. This is achieved by fulfilling
the boundary conditions given by Eq.~\eqref{eq:BC_magnetic} between all computational layers, the
results of these calculations being cast in a versatile inhomogeneous $\mathcal{S}$-matrix
formalism.

To this end, let us consider again the multilayer structure and the computational variables
defining the corresponding electromagnetic field schematically illustrated in
Figs.~\ref{fig:SMatrixCollapse}(a) and \ref{fig:SMatrixCollapse}(b), respectively. The optical
structure consists of exactly one 2DM sheet located at $z=z_s$, defined by its distribution of
conductivity, and an arbitrary number of bulk layers, defined by their respective electrical
permittivity distribution and identified by their superscript. Three of the layers are of
particular interest, namely the two computational layers directly enclosing the 2DM sheet,
identified by ``$A$'' and ``$B$'', and the evaluation layer, ``$E$'', which can be any
computational bulk layer in the grating structure. In addition, the semi-infinite layers ``$C$''
and ``$S$'' identify the cover and substrate, respectively. These capital letters are used to
label the mode shape matrices, $\textsf{E}_{\alpha}^{(L,\pm)}$ and
$\textsf{H}_{\alpha}^{(L,\pm)}$, and the propagation matrix, $\propmat^{(L,\pm)}(z)$, in each
layer, $L\in\{A,B,E,C,S\}$. The vector of mode coefficients in each layer is denoted by the
corresponding bold lower-case letters, $\ba^{\pm}$, $\bb^{\pm}$, $\be^{\pm}$, $\bc^{\pm}$, and
$\bs^{\pm}$.

At each of the pump frequencies, plane wave incidence is assumed, i.e. the incoming cover and
substrate coefficients, $\bc^{-}$ and $\bs^{+}$, respectively, are given and at least one of their
entries is nonzero. At the generated frequency, no field is assumed to be incident, hence $\bs^{-}
= \bc^{+} = 0$. Moreover, the higher-harmonics optical field is generated due to the nonlinear
surface current, $\j^{(\omega,\mathrm{nl})}(x,y)$. With this set-up, the goal of the remaining
derivation is to determine the mode coefficients, $\be^\pm$, in the evaluation layer, $E$. Since
the layer $E$ was arbitrarily chosen, this suffices to determine the mode coefficients, and hence
the electromagnetic field, in any layer.

Since the fields in layers $A$ and $B$ are expressed using their Fourier series according to
Eq.~\eqref{eq:ModalFieldTotal}, the boundary conditions given by Eq.~\eqref{eq:BC} can be
expressed in terms of the Fourier vector coefficients of the fields and read:
\begin{subequations}\label{eq:BC_FS}
    \begin{align}
        &\fvec{E_{\bt}^{(A)}(z_s)} = \fvec{E_{\bt}^{(B)}(z_s)}, \label{eq:BC_electricFS}\\
        &\fvec{H_{\bt}^{(A)}(z_s)} - \fvec{H_{\bt}^{(B)}(z_s)} = \fvec{\delta H_{\bt}^{(s)}}.  \label{eq:BC_magneticFS}
    \end{align}
\end{subequations}
Hereby, $\delta H_{\bt}^{(s)} = \left(j_{y}, -j_{x}\right)^T$ denotes the variation of the
tangential components of $\H$ across the surface $z=z_s$ and the superscript, $\omega$, is dropped
to simplify the notation. The symbol ``$T$'' means matrix transpose operation.

The vector Fourier coefficients, $\fvec{E_{\bt}^{(A)}(z_s)}$ and $\fvec{H_{\bt}^{(A)}(z_s)}$, can
be expressed in terms of their mode coefficients, $\ba^\pm$ and $\bb^\pm$, via
Eq.~\eqref{eq:ModalFieldMatrices}. This same procedure is performed with the linear surface
current according to the factorization rule expressed by Eq.~\eqref{eq:FacRuleJ}, which relies on
the tangential field at $z=z_s$. Due to the continuity of the tangential surface field, in
principle, the field values from either side $A$ or $B$, or their average, could be taken, as they
are all equal. The latter is chosen in our approach, thus yielding $\fvec{E_{\bt}^{(s)}}
=(\fvec{E_{\bt}^{(A)}(z_s)}+\fvec{E_{\bt}^{(B)}(z_s)})/2$.

Using the factorization rule given in Eq.~\eqref{eq:FacRuleJ}, one obtains for the
$\alpha$-component of the linear surface current:
\begin{align}
    &\fvec{j_{\alpha}^{(\mathrm{lin})}} = \frac 12 \sum_{\beta=x,y} \Delta\nvf_{\alpha\beta} \nonumber \\
    & ~~~\times\left\{ \left[\textsf{E}_{\beta}^{(A,+)} \ \textsf{E}_{\beta}^{(B,-)}\right]
    \left[\!\!\begin{array}{cc}
        \propmat^{(A,+)}(z_s) \!\!\!\!\!\!\!\!\!\! & \quad 0  \\
        \!\!\!\!\!\!\!\! 0 &\!\!\!\!\!\!\!\!\!\! \propmat^{(B,-)}(z_s)
    \end{array}\!\!\right] \left(\!\!\begin{array}{c}
    \ba^{+} \\ \bb^{-}
\end{array}\!\!\right) \right. \nonumber\\
& \left. \ \ \ \ \ \, +\left[\textsf{E}_{\beta}^{(A,-)} \ \textsf{E}_{\beta}^{(B,+)}\right]
\left[\!\!\begin{array}{cc}
    \propmat^{(A,-)}(z_s) \!\!\!\!\!\!\!\!\!\! & \quad 0  \\
    \!\!\!\!\!\!\!\! 0 &\!\!\!\!\!\!\!\!\!\! \propmat^{(B,+)}(z_s)
\end{array}\!\!\right] \left(\!\!\begin{array}{c}
\ba^{-} \\ \bb^{+}
\end{array}\!\!\right) \right\}.
\end{align}

Defining the block matrix $\textsf{G}_{\bt}^{(L,\pm)}$ for layers $A$ and $B$ as
\begin{align}
    \textsf{G}_{\bt}^{(L,\pm)} =\frac{1}{2}
    \left[\!\!\begin{array}{c}
        \Delta\nvf_{yx}\textsf{E}_{x}^{(L,\pm)} + \Delta\nvf_{yy}\textsf{E}_{y}^{(L,\pm)} \\
        -\Delta\nvf_{xx}\textsf{E}_{x}^{(L,\pm)} - \Delta\nvf_{xy}\textsf{E}_{y}^{(L,\pm)}
    \end{array} \!\!\right],
\end{align}
the surface variation $\fvec{\delta H_{\bt}^{(s)}}$ is found to be:
\begin{align}\label{eq:deltaHinab}
    & \fvec{\delta H_{\bt}^{(s)}}= \fvec{\delta H_{\bt}^{(s,\mathrm{nl})} }  \nonumber \\
    &  ~~~+\left[\!\!\begin{array}{cc} \textsf{G}_{\bt}^{(A,+)} & \textsf{G}_{\bt}^{(B,-)} \end{array}\!\!\right]
    \left[\!\!\begin{array}{cc}
        \propmat^{(A,+)}(z_s) \!\!\!\!\!\!\!\! & \quad 0  \\
        \!\!\!\!\!\! 0 &\!\!\!\!\!\!\!\! \propmat^{(B,-)}(z_s)
    \end{array}\!\!\right] \left(\!\!\begin{array}{c}
    \ba^{+} \\ \bb^{-}
\end{array}\!\!\right) \nonumber \\
& ~~~+\left[\!\!\begin{array}{cc} \textsf{G}_{\bt}^{(A,-)} & \textsf{G}_{\bt}^{(B,+)} \end{array}\!\!\right]
\left[\!\!\begin{array}{cc}
    \propmat^{(A,-)}(z_s) \!\!\!\!\!\!\!\! & \quad 0  \\
    \!\!\!\!\!\! 0 &\!\!\!\!\!\!\!\! \propmat^{(B,+)}(z_s)
\end{array}\!\!\right] \left(\!\!\begin{array}{c}
\ba^{-} \\ \bb^{+}
\end{array}\!\!\right).
\end{align}
Inserting this equation and Eq.~\eqref{eq:ModalFieldMatrices} into Eqs.~\eqref{eq:BC_FS}, the
following matrix relation is derived:
\begin{align}\label{eq:BC_modecoeffs}
\textsf{L}^{AB} \left(\!\!\begin{array}{c} \ba^{+} \\ \bb^{-}
\end{array}\!\!\right) =
\left(\!\!\begin{array}{c} 0 \\ \fvec{\delta H_{\beta}^{(s,\mathrm{nl})}}\end{array} \!\!\right) +
\textsf{R}^{AB} \left(\!\!\begin{array}{c} \bb^{+} \\ \ba^{-}
\end{array}\!\!\right),
% mode-coefficients
\end{align}
where  % a matrix relation between the incoming coefficients
{\small
\begin{align*}
\textsf{L}^{AB} \!\!&=\!\! \left[\!\!\begin{array}{cc}
        \textsf{E}_{\bt}^{(A,+)}                  &\!\!\!\! -\textsf{E}_{\bt}^{(B,-)} \\
        \textsf{H}_{\bt}^{(A,+)} + \textsf{G}_{\bt}^{(A,+)} &\!\!\!\! -\textsf{H}_{\bt}^{(B,-)}-\textsf{G}_{\bt}^{(B,-)}
    \end{array} \!\!\right]\!\!
    \left[\!\!\begin{array}{cc}
        \propmat^{(A,+)}(z_s) \!\!\!\!\!\!\!\!\!\! & \quad 0  \\
        \!\!\!\!\!\!\!\!0  &\!\!\!\!\!\!\!\!\!\! \propmat^{(B,-)}(z_s)
    \end{array}\!\!\right],\\
\textsf{R}^{AB} \!\!&= \!\!
    \left[\!\!\begin{array}{cc}
        \textsf{E}_{\bt}^{(B,+)}                  &\!\!\!\! -\textsf{E}_{\bt}^{(A,-)} \\
        \textsf{H}_{\bt}^{(B,+)} + \textsf{G}_{\bt}^{(B,+)} &\!\!\!\! - \textsf{H}_{\bt}^{(A,-)} - \textsf{G}_{\bt}^{(A,-)}
    \end{array} \!\!\right]\!\!
    \left[\!\!\begin{array}{cc}
        \propmat^{(B,+)}(z_s) \!\!\!\!\!\!\!\!\!\! & \quad 0  \\
        \!\!\!\!\!\!\!\! 0 &\!\!\!\!\!\!\!\!\!\! \propmat^{(A,-)}(z_s)
    \end{array}\!\!\right].
\end{align*}}
Here, $\textsf{E}_\bt^{(L,\pm)} = \left[\textsf{E}^{(L,\pm)}_x; \textsf{E}^{(L,\pm)}_y\right]$
denotes the $2N_0\times 2N_0$ matrix of all tangential Fourier components of the modes in layer
$L$. Note that the vectors of Fourier coefficients have already been arranged in a way suitable
for the $\mathcal{S}$-matrix formalism, which is explained in what follows.

By multiplying from the left both sides of Eq.~\eqref{eq:BC_modecoeffs} with the inverse of matrix
$\textsf{L}^{AB}$, the Fourier coefficients of the outgoing modes, $\ba^{+}$ and $\bb^{-}$, can be
determined in terms of the coefficients of the incoming modes, $\ba^{-}$ and $\bb^{+}$, and the
nonlinear surface current:
\begin{align}\label{eq:inhomSMatrix}
    \left(\!\!\begin{array}{c}
        \ba^{+} \\ \bb^{-}
    \end{array}\!\!\right) &= \left(\textsf{L}^{AB}\right)^{-1}\left\{\left(\!\!\begin{array}{c} 0 \\ \fvec{\delta H_{\bt}^{(s,\mathrm{nl})}}\end{array} \!\!\right)
    + \textsf{R}^{AB} \left(\begin{array}{c}
        \bb^{+} \\ \ba^{-}
    \end{array}\right)\right\} \nonumber \\
   & = \left(\!\!\begin{array}{c}
    \tilde{\ba}^{+} \\ \tilde{\bb}^{-}
\end{array}\!\!\right)
       + \textsf{S}^{AB} \left(\!\!\begin{array}{c}
        \bb^{+} \\ \ba^{-}
    \end{array}\!\!\right)
    = \left(\!\!\begin{array}{c}
        \tilde{\ba}^{+} \\ \tilde{\bb}^{-}
    \end{array}\!\!\right)
    + \left(\!\!\begin{array}{c}
        \overline{\ba}^{+} \\ \overline{\bb}^{-}
    \end{array}\!\!\right),
\end{align}
where $\textsf{S}^{AB}={(\textsf{L}^{AB})}^{-1}\textsf{R}^{AB}$ is the scattering matrix
($\mathcal{S}$-matrix) of the interface system. This equation shows that the outgoing coefficients
are comprised of two parts: $\overline{\ba}^{+}$ and $\overline{\bb}^{-}$ are the contributions to
the total coefficients, $\ba^{+}$ and $\bb^{-}$, respectively, given by linear scattering at the
interface of the incident modes described by the coefficients $\ba^{-}$ and $\bb^{+}$, whereas
$\tilde{\ba}^{+}$ and $\tilde{\bb}^{-}$ are the contributions of the nonlinear surface current to
the total coefficients $\ba^{+}$ and $\bb^{-}$, respectively, and only enter if the generated
frequency $\omega=\omega_0$ is considered.

If the considered structure only consists of cover, substrate, and one periodically patterned 2DM
sheet, Eq.~\eqref{eq:inhomSMatrix} is sufficient to fully describe its optical response. In
multilayer structures, however, the contributions of $\mathcal{S}$-matrices of different layers
and interfaces have to be properly incorporated. To this end, consider three layers, $A$, $B$, and
$S$, with coefficients $\ba^{\pm}$, $\bb^{\pm}$, and $\bs^{\pm}$, respectively, depicted in
\figref{fig:SMatrixCollapse}(b). We assume that the $\mathcal{S}$-matrix, which connects the mode
coefficients in layers $B$ and $S$, is known and fulfills the relation
\begin{align}\label{eq:SMatBS}
    \left(\!\!\begin{array}{c}
        \bb^{+} \\ \bs^{-}
    \end{array}\!\!\right)
    = \textsf{S}^{BS} \left(\!\!\begin{array}{c}
        \bb^{-} \\ \bs^{+}
    \end{array}\!\!\right) +
    \left(\!\!\begin{array}{c}
        \tilde{\bb}^{+} \\ \tilde{\bs}^{-}
    \end{array}\!\!\right).
\end{align}
Combining this equation and Eq.~\eqref{eq:inhomSMatrix}, one determines the scattering matrix
relation between the coefficients associated to the top and bottom layers, $\ba^{\pm}$ and
$\bs^{\pm}$, respectively, by eliminating the coefficients $\bb^{\pm}$:
\begin{align}\label{eq:def_inhomS-Matrix}
    \left(\!\!\begin{array}{c}
        \ba^{+} \\ \bs^{-}
    \end{array}\!\!\right)
    = \textsf{S}^{ABS} \left(\!\!\begin{array}{c}
        \bs^{+} \\ \ba^{-}
    \end{array}\!\!\right)
    + \textsf{T}^{ABS} \left(\!\!\begin{array}{c}
        \tilde{\bb}^{+} \\ \tilde{\bb}^{-}
    \end{array}\!\!\right) +
    \left(\!\!\begin{array}{c}
        \tilde{\ba}^{+} \\ \tilde{\bs}^{-}
    \end{array}\!\!\right),
\end{align}
where the four $N_0\times N_0$ sub-blocks of the combined $\mathcal{S}$-matrix,
$\textsf{S}^{ABS}$, are given by:
\begin{subequations}\label{eq:DefRedheffer}
\begin{align}
    &\textsf{S}^{ABS}_{11} = \textsf{S}^{AB}_{11} + \textsf{S}^{AB}_{12} \left(\idmatrix-\textsf{S}^{BS}_{11}\textsf{S}^{AB}_{22}\right)^{-1}\textsf{S}^{BS}_{11}\textsf{S}^{AB}_{21},\\
    &\textsf{S}^{ABS}_{12} =               \textsf{S}^{AB}_{12} \left(\idmatrix-\textsf{S}^{BS}_{11}\textsf{S}^{AB}_{22}\right)^{-1}S^{BS}_{12},\\
    &\textsf{S}^{ABS}_{21} =               \textsf{S}^{BS}_{21} \left(\idmatrix-\textsf{S}^{AB}_{22}\textsf{S}^{BS}_{11}\right)^{-1}\textsf{S}^{AB}_{21},\\
    &\textsf{S}^{ABS}_{22} = \textsf{S}^{BS}_{22} + \textsf{S}^{BS}_{21}
    \left(\idmatrix-\textsf{S}^{AB}_{22}\textsf{S}^{BS}_{11}\right)^{-1}\textsf{S}^{AB}_{22}\textsf{S}^{BS}_{12},
\end{align}
\end{subequations}
and the four sub-blocks of the combined matrix $\textsf{T}^{ABS}$ are expressed as: % this is NOT a T-matrix
\begin{subequations}\label{eq:DefRedhefferT}
    \begin{align}
        &\textsf{T}^{ABS}_{11} = \textsf{S}^{AB}_{12} \left(\idmatrix-\textsf{S}^{BC}_{11}\textsf{S}^{AB}_{22}\right)^{-1},\\
        &\textsf{T}^{ABS}_{12} = \textsf{S}^{AB}_{12} \left(\idmatrix-\textsf{S}^{BS}_{11}\textsf{S}^{AB}_{22}\right)^{-1}\textsf{S}^{BS}_{11},\\
        &\textsf{T}^{ABS}_{21} = \textsf{S}^{BS}_{21} \left(\idmatrix-\textsf{S}^{AB}_{22}\textsf{S}^{BS}_{11}\right)^{-1}\textsf{S}^{AB}_{22},\\
        &\textsf{T}^{ABS}_{22} = \textsf{S}^{BS}_{21} \left(\idmatrix-\textsf{S}^{AB}_{22}\textsf{S}^{BS}_{11}\right)^{-1}.
    \end{align}
\end{subequations}
These relations are found by straightforward matrix calculations, where one of the intermediate
steps yields the coefficients of the middle layer, $B$:
\begin{subequations}\label{eq:DefRedhefferMiddle}    % reduced problem
    \begin{align}
        \bb^{-} =& \left(\idmatrix-\textsf{S}^{AB}_{22}\textsf{S}^{BS}_{11}\right)^{-1} \nonumber \\
        & ~\times\left[ \textsf{S}^{AB}_{21}\ba^{-} + \textsf{S}^{AB}_{22}\textsf{S}^{BS}_{12}\bs^{+} + \tilde{\bb}^{-} + \textsf{S}^{AB}_{22}\tilde{\bb}^{+}\right], \\
        \bb^{+} =& \left(\idmatrix-\textsf{S}^{BS}_{11}\textsf{S}^{AB}_{22}\right)^{-1} \nonumber \\
        &~\times\left[\textsf{S}^{BS}_{11}\textsf{S}^{AB}_{21}\ba^{-} + \textsf{S}^{BS}_{12}\bs^{+} + \textsf{S}^{BS}_{11}\tilde{\bb}^{-} +
        \tilde{\bb}^{+}\right],
    \end{align}
\end{subequations}
expressed solely in terms of incoming and known coefficients $\ba^-,$ $\bs^+$, and
$\tilde{\bb}^\pm$.

The matrix operation, $\textsf{S}^{ABS}=\textsf{S}^{AB}\otimes \textsf{S}^{BS}$, is known as the
Redheffer star-product \cite{li96josaaSMatrix}. It is associative, noncommutative, and has the
neutral element $\idmatrix^\otimes=[\textsf{0}, \idmatrix; \idmatrix, \textsf{0}]$. It can be
applied repeatedly and hence at all pump frequencies, where all nonlinear coefficients
($\tilde{\bc}^{\pm},\ \tilde{\be}^{\pm}, \ldots$) vanish, it enables the calculation of the
outgoing mode coefficients $\bc^{+}$ and $\bs^{-}$ from the incident mode coefficients, $\bc^{-}$
and $\bs^{+}$.

Note that the term
\begin{align}\label{eq:def_inhomEffCoefficient}
    \textsf{T}^{ABS} \left(\!\!\begin{array}{c}
        \tilde{\bb}^{+} \\ \tilde{\bb}^{-}
    \end{array}\!\!\right) +
    \left(\!\!\begin{array}{c}
        \tilde{\ba}^{+} \\ \tilde{\bs}^{-}
    \end{array}\!\!\right) \eqqcolon
    \left(\!\!\begin{array}{c}
    \tilde{\ba}^{+}_\mathrm{eff} \\ \tilde{\bs}_\mathrm{eff}^{-}
\end{array}\!\!\right)
\end{align}
in Eq.~\eqref{eq:def_inhomS-Matrix} can be viewed as the effective coefficients of the modes that
are radiated by the combined multilayer-interface system, $ABS$. More specifically, this term
accounts for the linear propagation of the internally radiated modes at the generated frequency,
with coefficients $\tilde{\bb}^{\pm}$ in bulk layer $B$, and it accounts for linear optical
interaction (reflection, transmission, and absorption) with the 2DM located at the $AB$-interface
or the $BS$-layer-interface system.

Equipped with these matrix-vector relations, the calculation of all solution coefficients at the
generated frequency can now be completed. In order to evaluate the mode coefficients $\be^{\pm}$
in the evaluation layer $E$, one calculates the combined $\mathcal{S}$-matrix, $\textsf{S}^C$, of
all layers and interfaces above the evaluation layer $E$, and the combined $\mathcal{S}$-matrix,
$\textsf{S}^{M}$, of all interfaces and layers between the evaluation layer $E$ and up to but
excluding the 2DM sheet. This is depicted in \figref{fig:SMatrixCollapse}(c). Note that this
procedure allows for $\textsf{S}^C$, $\textsf{S}^{M}$, or $\textsf{S}^{BS}$ to be equal to
$\idmatrix^\otimes$, i.e. the evaluation layer can be any layer above the nonlinear 2DM sheet,
which itself can be located at any interface, including just above the substrate.

Under these circumstances, the governing $\mathcal{S}$-matrix relations read as follows:
\begin{subequations}\label{eq:InhomSMatrixRelations}
    \begin{align}
    &\left(\!\!\begin{array}{c}
            \bc^{+} \\ \be^{-}
        \end{array}\!\!\right)
        = \textsf{S}^{C} \left(\!\!\begin{array}{c}
            \bc^{-} \\ \be^{+}
        \end{array}\!\!\right), \\
    &\left(\!\!\begin{array}{c}
            \be^{+} \\ \ba^{-}
        \end{array}\!\!\right)
        = \textsf{S}^{M} \left(\!\!\begin{array}{c}
            \be^{-} \\ \ba^{+}
        \end{array}\!\!\right), \\
        &\left(\!\!\begin{array}{c}
            \ba^{+} \\ \bs^{-}
        \end{array}\!\!\right)
        = \textsf{S}^{ABS} \left(\!\!\begin{array}{c}
            \ba^{-} \\ \bs^{+}
        \end{array}\!\!\right) + \left(\!\!\begin{array}{c}
        \tilde{\ba}^{+}_\mathrm{eff} \\ \tilde{\bs}_\mathrm{eff}^{-}
    \end{array}\!\!\right).
\end{align}
\end{subequations}

%\begin{subequations}\label{eq:InhomSMatrixRelations}
%    \begin{align}
%     &\left(\!\!\begin{array}{c}
%         \bc^{(0,+)} \\ \be^{(-)}
%        \end{array}\!\!\right)
%        = S^{A} \left(\!\!\begin{array}{c}
%            \bc^{(0,-)} \\ \be^{(+)}
%     \end{array}\!\!\right), \\
%     &\left(\!\!\begin{array}{c}
%         \be^{(+)} \\ \ba^{(-)}
%        \end{array}\!\!\right)
%        = S^{M} \left(\!\!\begin{array}{c}
%            \be^{(0,-)} \\ \ba^{(+)}
%     \end{array}\!\!\right), \\
%     &\left(\!\!\begin{array}{c}
%         \ba^{(0,+)} \\ \bb^{(-)}
%        \end{array}\!\!\right)
%        = S^{I} \left(\!\!\begin{array}{c}
%            \ba^{(0,-)} \\ \bb^{(+)}
%    \end{array}\!\!\right) + \left(\!\!\begin{array}{c}
%    \doverline{\ba}^{(0,+)} \\ \doverline{\bb}^{(-)}
%   \end{array}\!\!\right), \\
%    &\left(\!\!\begin{array}{c}
%            \bb^{(+)} \\ \bc^{(L,-)}
%        \end{array}\!\!\right)
%        = S^{B} \left(\!\!\begin{array}{c}
%            \bb^{(-)} \\ \bc^{(L,+)}
%    \end{array}\!\!\right).
%    \end{align}
%\end{subequations}
By applying Eq.~\eqref{eq:def_inhomS-Matrix} to the matrix relations for $\textsf{S}^{M}$ and
$\textsf{S}^{ABS}$, one obtains $\textsf{S}^{MABS}=\textsf{S}^{M}\otimes \textsf{S}^{ABS}$ and
effective affine coefficients, $\tilde \be^{+}_\mathrm{eff} \coloneqq \textsf{T}^{MABS}_{11}
\tilde{\ba}^{+}_\mathrm{eff}$ and $\tilde \bs^{-}_\mathrm{eff} \coloneqq \textsf{T}^{MABS}_{21}
\tilde{\ba}^{+}_\mathrm{eff} +\tilde{\bs}^{-}_\mathrm{eff}$, as per
Eq.~\eqref{eq:def_inhomEffCoefficient}.

The final constellation of the remaining two $\mathcal{S}$-matrices, $\textsf{S}^{C}$ and
$\textsf{S}^{MABS}$, and corresponding coefficients is depicted in
\figref{fig:SMatrixCollapse}(d). This configuration is similar to the initial system of
$\mathcal{S}$-matrices, $\textsf{S}^{AB}$ in Eq.~\eqref{eq:inhomSMatrix} and $\textsf{S}^{BS}$ in
Eq.~\eqref{eq:SMatBS}. Hence, it allows the calculation of the evaluation coefficients $\be^\pm$
by means of Eq.~\eqref{eq:DefRedhefferMiddle}, with the following replacements $\bb^{\pm} \to
\be^{\pm}$, $\textsf{S}^{AB} \to \textsf{S}^{C}$, $\textsf{S}^{BS} \to \textsf{S}^{MABS}$,
$\ba^{-} \to 0$, $\bs^{+} \to 0$, $\tilde{\bb}^{-}\to 0$, and $\tilde{\ba}^{+} \to
\tilde{\be}^{+}_\mathrm{eff}$, yielding
\begin{align*}
    \be^{-} =& \left(\idmatrix-\textsf{S}^{C}_{22}S^{MABS}_{11}\right)^{-1} \textsf{S}^{C}_{22}\tilde{\be}^{+}_\mathrm{eff}, \\
    \be^{+} =& \left(\idmatrix-\textsf{S}^{MABS}_{11}\textsf{S}^{C}_{22}\right)^{-1}
    \tilde{\be}^{+}_\mathrm{eff}.
\end{align*}

The treatment of an evaluation layer below the nonlinear layer can be performed in a similar
manner. Moreover, if the structure contains more than one nonlinear 2DM sheet, the algorithm we
just described can be repeated independently for each interface where nonlinear 2DM is located and
then sum the individually obtained solution coefficients. It should be noted that in addition to
the nonlinear optical response of 2DMs, the linear scattering effects at interfaces are naturally
incorporated in our algorithm. This overall approach to linear and nonlinear light scattering in
layered photonic structures containing 2DMs is possible because of the affine linearity of the
total system.

\section{Validation of the numerical method and convergence analysis}\label{sec:convergence}
In this section, aiming to validate our numerical method, we consider a series of generic test
cases of diffraction gratings containing 2DMs and analyze the corresponding convergence
characteristics of the numerical method. We consider both 1D- and 2D-periodic structures made of
graphene, which has cubic nonlinearity. No additional test cases for the TMDC monolayers are shown
here because they are far less challenging than graphene from a computational point of view: TMDC
monolayers are poorly conductive materials and thus do not affect the optical field as strongly as
graphene does.

In our analysis, we investigate physical quantities describing the near- and far-field, so as to
fully assess the stability and convergence properties of our modal method, as was discussed in
Ref.~\cite{wgp15jo}. To characterize the far-field, we used the optical absorption, $A$, at the
FF, which is given by $A=1-R-T$, where $R$ and $T$ denote the fraction of the intensity of the
incident light that is reflected and transmitted, respectively. At the TH, the total radiation,
$R^{\prime}+T^{\prime}$, is chosen as a far-field quantity suitable to validate our method, where
$R^{\prime}$ and $T^{\prime}$ denote the intensity at the TH radiated in the direction of
reflection and transmission, respectively.
\begin{figure}[b]
    \centering
    \includegraphics[width=\linewidth]{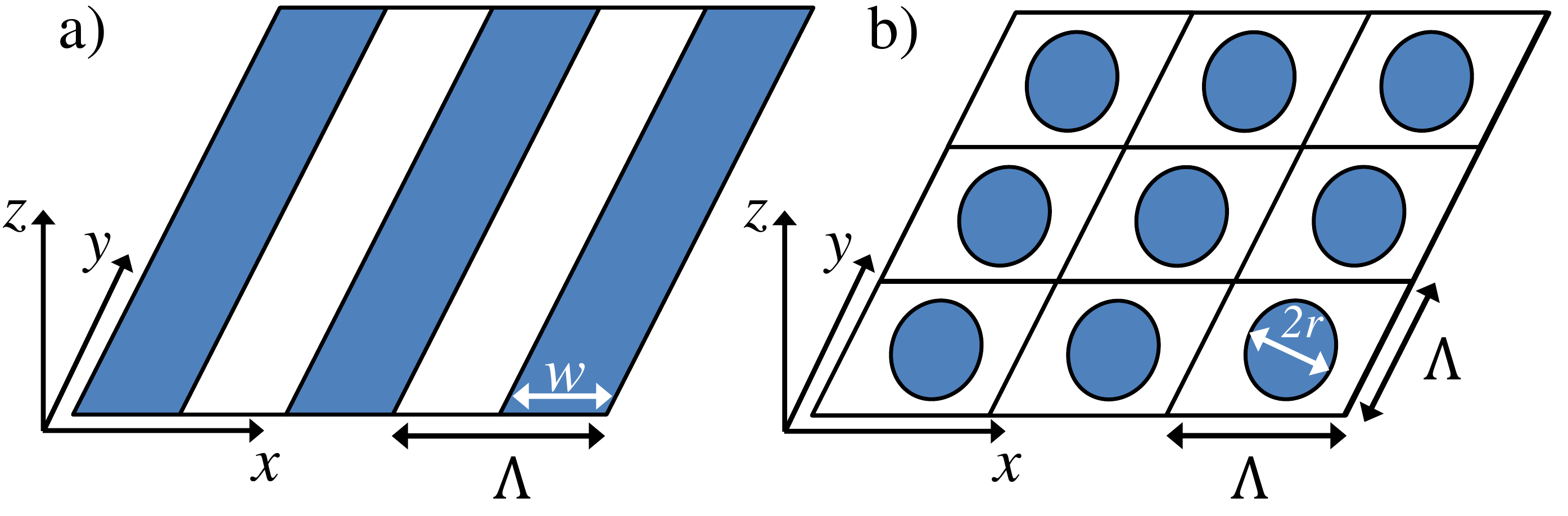}
\caption{Generic structures for numerical convergence analysis. a) 1D periodic array of graphene
ribbons with period, $\Lambda$, and width, $w$. b) 2D array of circular graphene disks with
radius, $r$, and periods, $\Lambda_{1}=\Lambda_{2}=\Lambda$.}
    \label{fig:nDStructures}
\end{figure}

\subsection{One-dimensional binary graphene gratings}\label{sec:convergence:1D}
As a first example of a periodic diffraction grating containing nonlinear 2DMs, consider the 1D
periodic array of graphene ribbons depicted in \figref{fig:nDStructures}(a), sandwiched in-between
homogeneous cover and substrate materials with electric permittivity, $\epsilon_c=3$ and
$\epsilon_s=4$, respectively. The period of the grating is $\Lambda=\SI{8}{\micro\meter}$ and the
spacing between adjacent ribbons is $w=\Lambda/2=\SI{4}{\micro\meter}$. The incident light is
normally impinging onto this binary graphene grating and is TM-polarized.

In order to illustrate the effectiveness and benefits of our algorithm with added conductivity,
Eq.~\eqref{eq:modSigma} with $\sigma_{s,\mathrm{add}}=-i \eta|\sigma_{s}(\omega)|$, and correct
Fourier factorization, Eq.~\eqref{eq:FacRuleJ}, we compare it with two other versions of the
proposed method. The first of these two algorithms employs a zero conductivity,
$\sigma_{s,\mathrm{add}}=0$, in the regions without graphene and only uses the product
factorization rule for Eq.~\eqref{eq:linSurfaceCurrent}, i.e. an incorrect factorization rule. In
the second algorithm we assume that graphene has a finite thickness,
$h_{\mathrm{eff}}=\SI{0.33}{\nano\meter}$, i.e. we model the array of graphene ribbons as a
periodic bulk layer with relative permittivity, $\epsilon_r=1+i\sigma_s/(\epsilon_{0}\omega
h_{\mathrm{eff}})$. This can be done using a standard RCWA implementation. We stress, however,
that this is computationally more costly as it involves the determination of the RCWA modes in
this bulk layer representing the array of graphene ribbons. In addition, for the sake of
completeness, computational results from Ref.~\cite{k13ol} are included, too.

\begin{figure}[b]
    \centering
    \includegraphics[width=\linewidth]{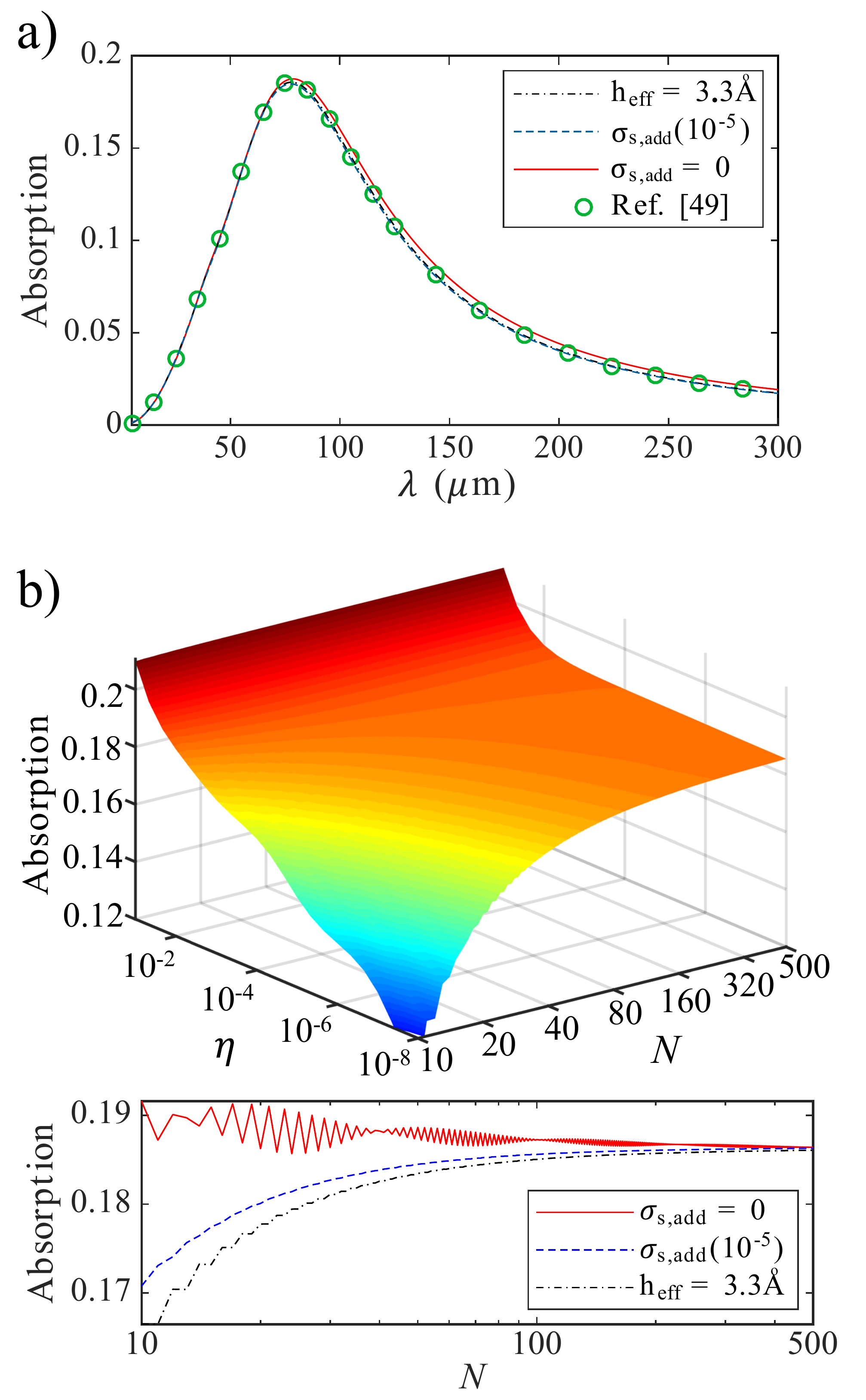}
\caption{a) Linear absorption spectra obtained by setting the added conductivity to zero,
$\sigma_{s,\mathrm{add}}=0$ (solid line), by using
$\sigma_{s,\mathrm{add}}=\sigma_{s,\mathrm{add}}(10^{-5})$ (dashed line), using a standard RCWA
with $h_{\mathrm{eff}}=\SI{.33}{\nano\meter}$ (dashed-dotted line) and results from \cite{k13ol}
(green circles). b) Top panel shows the dispersion map of the absorption vs. $N$ and $\eta$,
determined using the algorithm with added conductivity, whereas the bottom panel shows the
dependence of absorption on $N$, calculated using a standard RCWA (dashed-dotted line), the the
algorithm with $\sigma_{s,\mathrm{add}}=0$ (solid line), and the algorithm with added
conductivity, $\sigma_{s,\mathrm{add}}=\sigma_{s,\mathrm{add}}(10^{-5})$ (dashed line).}
    \label{fig:1DLinearPlots}
\end{figure}

The linear absorption spectra calculated using a moderate number of harmonics, $N=100$, are shown
in \figref{fig:1DLinearPlots}(a). The device absorption presents a broad resonance peak with
maximum of about $\SI{18.5}{\percent}$ at the wavelength, $\lambda=\SI{80}{\micro\meter}$, with
negligible absorption being observed at both shorter and longer wavelengths. The results obtained
using the algorithm with added conductivity with $\eta=10^{-5}$ shows very good agreement with
those found using conventional bulk RCWA calculations and with the results taken from
Ref.~\cite{k13ol}. The model without added conductivity slightly overestimates the absorption, in
the region $\lambda\gtrsim\SI{70}{\micro\meter}$.

Keeping in mind that the sheet conductance of graphene, $\sigma_{s,\mathrm{add}}(\eta) = -i
\eta|\sigma^{s}(\omega)|$, has been introduced somewhat artificially to facilitate the use of the
correct (inverse) factorization rule, the influence of $\sigma_{s,\mathrm{add}}(\eta)$ on the
accuracy of the computed results needs to be carefully investigated. The asymptotic behavior of
both near- and far-field physical quantities are suitable tools for performing this analysis. To
this end, we have determined the dependence of the absorption at $\lambda=\SI{18.5}{\micro\meter}$
on $N$ and $\eta$, as depicted in \figref{fig:1DLinearPlots}(b). Among other things, this figure
shows that the convergence with respect to $N$ is faster for larger $\eta$. This dependence is not
surprising, because the interface containing the structured graphene sheet becomes more optically
homogeneous as $\eta$ increases, i.e. the description of $1/\tilde{\sigma}_s(x,y)$ as a Fourier
series becomes more accurate. However, this does not prove the accuracy of the method just yet:
$\sigma_{s,\mathrm{add}}$ was introduced at a purely mathematical level and as such it should
vanish in physical diffraction gratings. The accuracy of the method is demonstrated by the fact
that as $\eta\rightarrow0$, convergence of the absorption is reached; for example, as
$\eta\rightarrow0$ the maximum absorption converges to $A=\SI{18.63}{\percent}$ for increasing
value of $N$.

In the bottom panel of \figref{fig:1DLinearPlots}(b) we contrast the convergence characteristics
of the three algorithms we just described. As it can be seen in this figure, all three approaches
converge to the same value but, importantly, the convergence speed in the case of finite added
conductivity is the fastest among the three cases. It is also instructive to remark that the
slowest convergence is observed in the case of zero conductivity, an added drawback in this case
being the oscillatory dependence of the absorption on $N$, which confirms that this formulation is
incorrect as was theoretically argued in \secref{sec:BC_2DM}.
\begin{figure}[t]
    \centering
    \includegraphics[width=\linewidth]{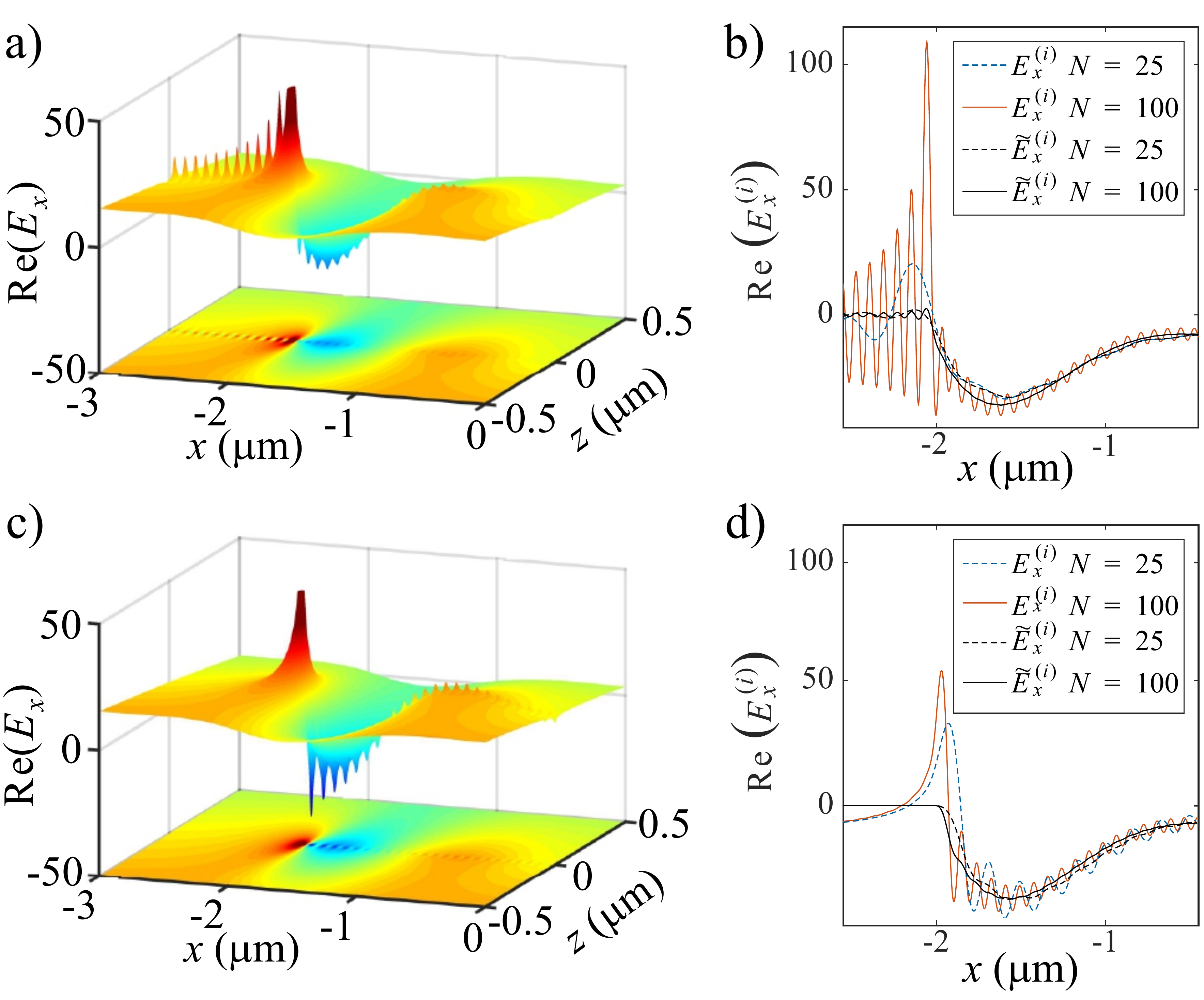}
\caption{a), b) Electric field component $E_x$ around the graphene ribbon calculated for $N=100$
and the fields at the surface of the graphene ribbon corresponding to the improper (colored lines)
and correct (black lines) field evaluation, respectively. In both cases,
$\sigma_{s,\mathrm{add}}=0$. c), d) The same quantities as in a) and b), respectively, but
determined for $\sigma_{s,\mathrm{add}}=\sigma_{s,\mathrm{add}}(10^{-5})$. Only a part of the left
half of the unit cell, defined by $x\in\left[-4\si{\micro\meter},0\right]$, is shown, as the
results are symmetric in $x$ and rather featureless far away from the graphene ribbon located at
$|x|<\SI{2}{\micro\meter}$.}
    \label{fig:1DLinearFields}
\end{figure}

The spatial profile of the electric near-field deserves special attention as well because it
reveals new important features pertaining to the convergence of the numerical method. To
illustrate this idea, the $x$-component of the electric field for the cases in which
$\sigma_{s,\mathrm{add}}=0$ and when the conductivity is finite,
$\sigma_{s,\mathrm{add}}=\sigma_{s,\mathrm{add}}(10^{-5})$, are shown in
Figs.~\ref{fig:1DLinearFields}(a) and \ref{fig:1DLinearFields}(c), respectively. The operating
wavelength is $\lambda=\SI{33}{\micro\meter}$ and $N=100$ in both cases. In these plots, the
boundary of the graphene ribbon in the unit cell is located at $x=\SI{-2}{\micro\meter}$. It can
be seen from these plots that without the added conductivity the field exhibits very strong,
unphysical oscillations near the boundary of the graphene ribbons, i.e. at $z=z_s=0$, which spread
over the whole unit cell along the $x$-direction. The spatial frequency of these oscillations is
equal to the highest spatial discretization frequency, namely to $2\pi N/\Lambda$. In the case of
finite $\sigma_{s,\mathrm{add}}$, on the other hand, there are no such spurious field oscillations
outside graphene regions and only weak oscillations are seen inside the graphene ribbon, as per
\figref{fig:1DLinearFields}(c). In both cases, these field oscillations only occur very close to
the interface where the graphene sheet is located and can merely be observed at distances $\vert
z-z_{s}\vert\gtrapprox\SI{5}{\nano\meter}$. These oscillations are due to the fact that the
Fourier series decomposition does not resolve the singularity of the electric field at the edges
of the graphene ribbons, as we discussed in \secref{sec:BC_2DM}.

One of the aims of the accurate near-field formulation introduced by Eqs.~\eqref{eq:improvedEval}
is to overcome this shortcoming of Fourier series decompositions and thus to allow the accurate
field evaluation exactly at the location of the 2DM monolayers, i.e. at $z=z_s$. To illustrate how
our method achieves this, we depict in Figs.~\ref{fig:1DLinearFields}(b) and
\ref{fig:1DLinearFields}(d) the interface field, $E_x^{(i)}(x,y,z=z_{s})$, for the cases in which
$\sigma_{s,\mathrm{add}}=0$ and when the conductivity is finite,
$\sigma_{s,\mathrm{add}}=\sigma_{s,\mathrm{add}}(10^{-5})$, respectively. Without using an added
conductivity, the improper field evaluation $\mathcal{R}\big(\fvec{E_{x}^{(i)}}\big)(x,y)$ of the
surface field leads to a strongly oscillatory spatial dependence, which in addition changes
significantly with $N$, as shown by the blue and red lines in \figref{fig:1DLinearFields}(b) for
$N=25$ and $N=100$, respectively. By contrast, the interface field obtained by using
Eqs.~\eqref{eq:improvedEval} with $\fmatrix{1/\tilde{\sigma_s}}^{-1}$ replaced by
$\fmatrix{\tilde\sigma_s}$ displays hardly any oscillations, as per the black lines in this
figure, yet it does not vanish at the edge of the graphene ribbon as required by the theoretical
analysis presented in \secref{sec:BC_2DM}. These remaining oscillations are merely due to the fact
that the incorrect factorization rule was used and not because of the unresolved singularity of
the field.

This description changes significantly if one uses a small, finite value for
$\sigma_{s,\mathrm{add}}$, see \figref{fig:1DLinearFields}(d). The improper evaluation using a
finite value for $\sigma_{s,\mathrm{add}}$ yields a smooth field outside the graphene region,
$|x|>\SI{2}{\micro\meter}$, but still displays unphysical oscillations inside the graphene region,
$|x|\leq\SI{2}{\micro\meter}$, and does not vanish at the graphene boundary,
$|x|=\SI{2}{\micro\meter}$, as required. On the other hand, by employing the correct evaluation of
the surface field and surface current given by Eqs.~\eqref{eq:improvedEval}, one obtains a surface
field that vanishes at $|x|=\SI{2}{\micro\meter}$, is free of spurious oscillations, and converges
rapidly with $N$.

We stress that since the nonlinear polarization generated in our photonic structures is determined
by the optical near-field at the location of 2DMs, an accurate evaluation of these near-fields is
a paramount prerequisite to a rigorous description of the nonlinear optical response of these
nonlinear photonic devices. To this end, as we just showed, this can be readily achieved by using
the correct Fourier factorization expressed as Eq.~\eqref{eq:FacRuleJ} in the model with finite
sheet conductance, the correct field evaluation given by Eqs.~\eqref{eq:improvedEval}, and a
properly chosen number of harmonics.
\begin{figure}[t]
    \centering
    \includegraphics[width=\linewidth]{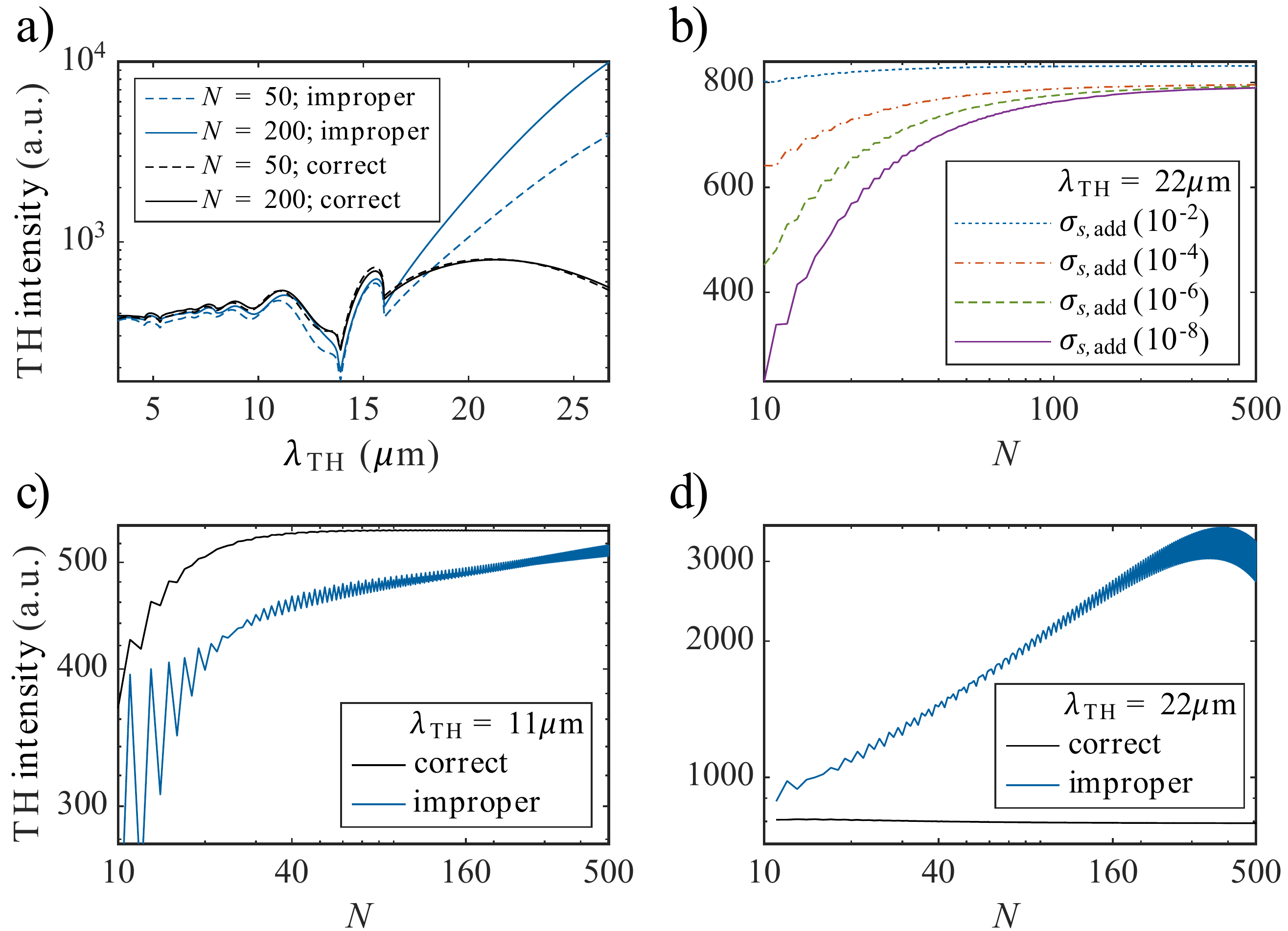}
\caption{a) Nonlinear radiation spectrum for $N=50$ (dashed lines) and $N=200$ (solid lines)
calculated using the improper field evaluation (blue lines) and correct field evaluation (black
lines) at the FF. b) Intensity of TH radiation at $\lambda_\mathrm{TH}=\SI{22}{\micro\meter}$ vs.
$N$ determined for different values of added conductivity $\sigma_{s,\mathrm{add}}$. c), d)
Intensity of TH radiation at $\lambda_\mathrm{TH}=\SI{11}{\micro\meter}$ and
$\lambda_\mathrm{TH}=\SI{22}{\micro\meter}$, respectively, vs. $N$ determined using improper field
evaluation (blue lines) and correct field evaluation (black lines) at the FF. In both cases
$\sigma_{s,\mathrm{add}}=\sigma_{s,\mathrm{add}}(10^{-5})$.}
    \label{fig:1DNonlinearPlots}
\end{figure}

The total intensity of radiated TH is considered as the first quantity characterizing the
nonlinear optical response of our graphene gratings, the corresponding computational results being
summarized in \figref{fig:1DNonlinearPlots}. Thus, \figref{fig:1DNonlinearPlots}(a) depicts the
total radiated power at the TH, determined for the case of finite added conductivity,
$\sigma_{s,\mathrm{add}}=\sigma_{s,\mathrm{add}}(10^{-5})$, and number of harmonics, $N=50$ and
$N=200$. The nonlinear source current given by Eq.~\eqref{eq:defJNLsheet} was calculated using
both the improperly evaluated interface field, $\E^{(i)}$, and the correctly calculated interface
field, $\tilde\E^{(i)}$. Interestingly, in the wavelength range
$\lambda_\mathrm{TH}\lessapprox\SI{15}{\micro\meter}$, both calculation methods yield
qualitatively similar spectra. For longer wavelengths, however, results differ considerably and in
fact only the results obtained using $\tilde\E^{(i)}$ field converge.

Before investigating in more detail this behavior, we consider first the convergence
characteristics of the far-field at the TH with respect to the value of the added conductivity,
$\sigma_{s,\mathrm{add}}$. The main features of this dependence, illustrated by the data plotted
in \figref{fig:1DNonlinearPlots}(b), which corresponds to
$\lambda_\mathrm{TH}=\SI{22}{\micro\meter}$, are similar to those observed in the case of linear
calculations. More specifically, the larger $\sigma_{s,\mathrm{add}}$ is the faster
self-convergence with respect to $N$ is observed and the computational results converge for
vanishingly small $\sigma_{s,\mathrm{add}}$.

The difference in convergence behavior for increasing $N$ of the two approaches used to evaluate
the surface field is illustrated in Figs.~\ref{fig:1DNonlinearPlots}(c) and
\ref{fig:1DNonlinearPlots}(d), for two representative wavelengths,
$\lambda_\mathrm{TH}=\SI{11}{\micro\meter}$ and $\lambda_\mathrm{TH}=\SI{22}{\micro\meter}$,
respectively. These figures show that the correct field evaluation leads to rapid convergence at
both wavelengths, whereas the improper approach yields slow and oscillatory convergence at
$\lambda_\mathrm{TH}=\SI{11}{\micro\meter}$ and completely fails to converge at
$\lambda_\mathrm{TH}=\SI{22}{\micro\meter}$.
\begin{figure}[tp]
    \centering
    \includegraphics[width=\linewidth]{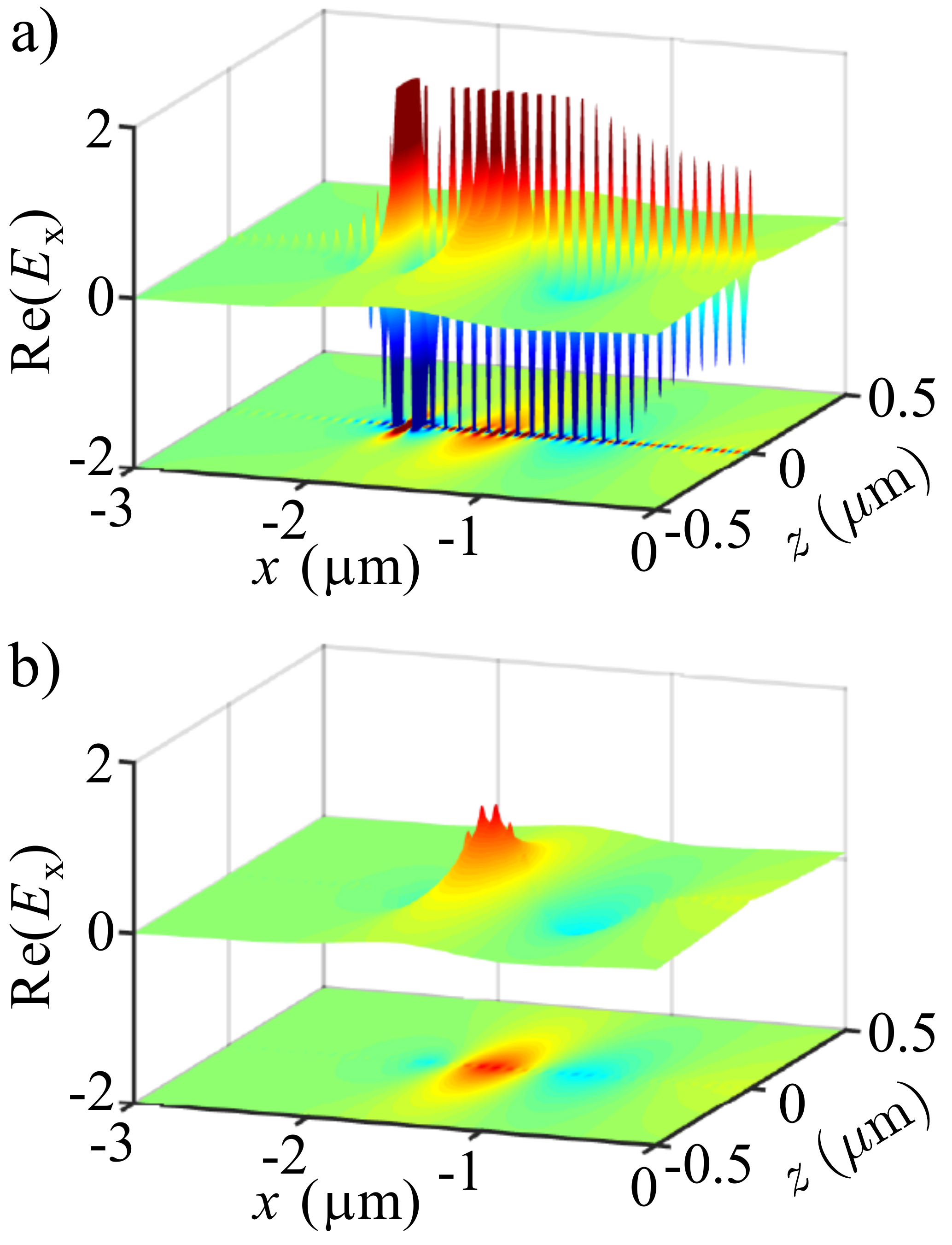}
\caption{a) Third-harmonic near-field around a graphene ribbon determined using the improperly
evaluated interface field at the FF, $E^{(i)}_x$, to calculate the nonlinear polarization,
$\P^\mathrm{nl}$. b) The same as in a) but using the correctly evaluated field at the FF, $\tilde
E^{(i)}_x$, to compute $\P^\mathrm{nl}$.}
    \label{fig:1DNonlinearPlotsB}
\end{figure}

The electromagnetic near-field, $\E^{(\Omega)}$, at the TH wavelength,
$\lambda_\mathrm{TH}=\SI{11}{\micro\meter}$, determined using the two algorithmic choices and
$N=100$ spatial harmonics, is plotted in \figref{fig:1DNonlinearPlotsB}. Thus, the TH $E_x$ field
derived from the nonlinear source polarization, $\P^{\mathrm{nl}}(\E^{(i)})$, comprising the
incorrectly evaluated field at the FF, $\E^{(i)}$, is completely swamped by unphysical
oscillations, i.e. numerical artefacts, at the interface where the graphene ribbon is located, as
per \figref{fig:1DNonlinearPlotsB}(a). On the other hand, the $x$-component of the TH electric
field, $E^{(\Omega)}_x$, which has the nonlinear polarization $\P^{\mathrm{nl}}(\tilde \E^{(i)})$
as its source, shows almost no oscillatory behavior and thus has the expected spatial profile, as
shown in \figref{fig:1DNonlinearPlotsB}(b). Only further away from the interface, some agreement
between the data in Figs.~\ref{fig:1DNonlinearPlotsB}(a) and \ref{fig:1DNonlinearPlotsB}(b) can be
observed.

The faster convergence, the more physically correct near-fields at both FF and TH, combined with
the expected electric field behavior at graphene boundaries allow us to conclude that the properly
evaluated field, $\tilde\E^{(i)}$, at the FF yields the correct results at the TH, whereas the TH
optical response obtained using the improperly evaluated field, $\E^{(i)}$, at the FF is plagued
by numerical artefacts and unphysical behavior.

\subsection{Two-dimensional graphene diffraction gratings}\label{sec:convergence:2D}
Let us now consider 2D diffraction gratings that contain 2DMs, namely the periodically arranged
graphene disks depicted in \figref{fig:nDStructures}(b). We assume that the two periods are the
same, $\Lambda_{1}=\Lambda_{2}=\Lambda=\SI{250}{\nano\meter}$ and the diameter of the graphene
disks is $D=2r=\SI{175}{\nano\meter}=0.7\Lambda$. For the sake of simplicity, we consider that the
incoming light is normally impinging onto the grating, the cover and substrate media being air
($\epsilon_c=1$) and glass ($\epsilon_s = 2.0852$), respectively.
\begin{figure}[t]
    \centering
    \includegraphics[width=\linewidth]{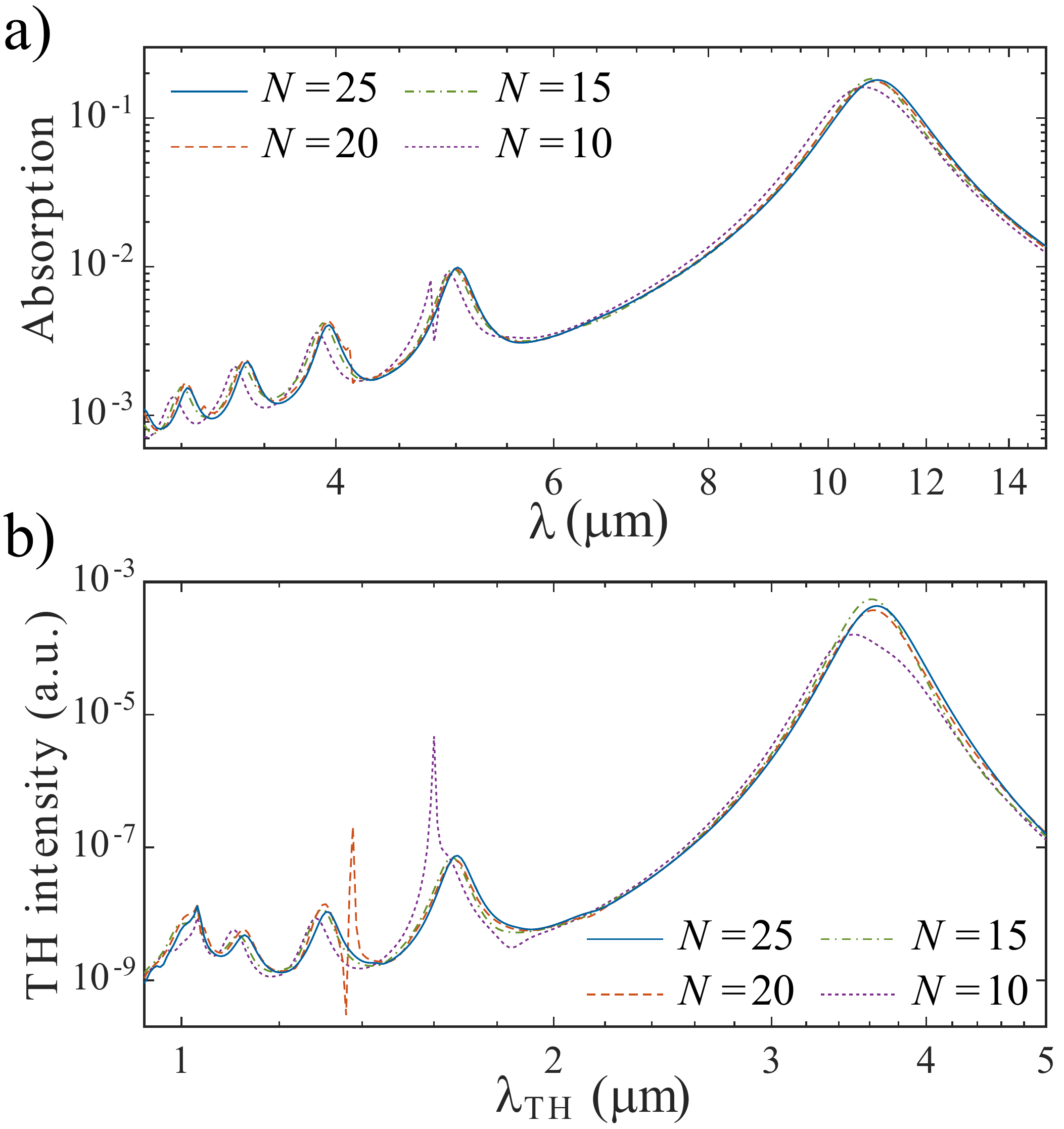}
\caption{a), b) Linear absorption and intensity of TH radiation spectra calculated using a finite
added conductivity, $\sigma_{s,\mathrm{add}}=\sigma_{s,\mathrm{add}}(10^{-3})$, and number of
harmonics $N=$ \numlist[list-final-separator={, and }]{10;15;20;25}.}
    \label{fig:2DBothPlot}
\end{figure}

The absorption spectrum for $x$-polarized incoming light, calculated for $N=$
\numlist[list-final-separator={, and }]{10;15;20;25} harmonics, is presented in
\figref{fig:2DBothPlot}(a). All results are obtained using the finite conductivity model defined
by Eq.~\eqref{eq:modSigma}, the scaling parameter being $\eta=10^{-3}$. The zero-conductivity
model or the improper interface field evaluation yield highly oscillatory near-field profiles at
FF and TH and fails to deliver converging TH far-field results. Thus, it can be seen that the
spectra calculated using different values of $N$ agree well and exhibit similar features. In
particular, all spectra show a series of resonances whose amplitude and spectral width increase
with the wavelength. Figure~\ref{fig:2DBothPlot}(b) depicts the dependence of the intensity of the
total radiated TH from the array of graphene disks, $R^{\prime}_{N}$, on the number of harmonics,
$N$. The nonlinear radiation spectra corresponding to the largest values $N=25$ and $N=20$ already
show good agreement, which suggests that convergence has been achieved.

\begin{figure}[t]
    \centering
    \includegraphics[width=\linewidth]{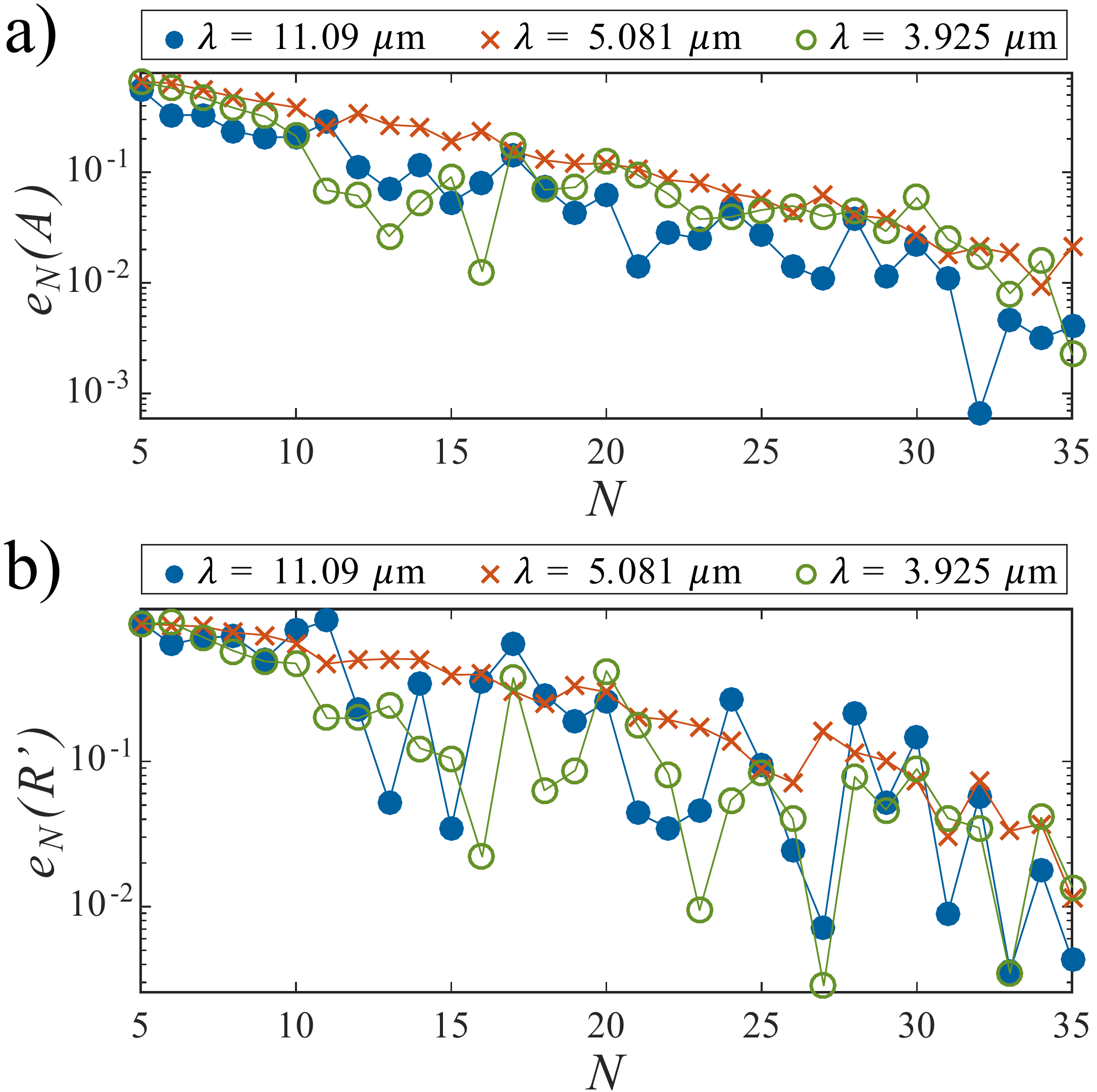}
\caption{a), b) Relative self-error for linear absorption and intensity of TH radiation,
respectively, determined at the plasmon resonance wavelengths $\lambda=\SI{11.09}{\micro\meter}$
(dots), $\lambda=\SI{5.081}{\micro\meter}$ (crosses) and $\lambda=\SI{3.925}{\micro\meter}$
(circles).}
    \label{fig:2DConvergence}
\end{figure}

We now investigate in more detail the self-convergence characteristics of linear calculations, the
main conclusions of this analysis being summarized in \figref{fig:2DConvergence}(a). We considered
the wavelengths of the first three absorption resonances seen in \figref{fig:2DBothPlot}(a), that
is $\lambda=\SI{11.09}{\micro\meter}$, $\lambda=\SI{5.081}{\micro\meter}$, and
$\lambda=\SI{3.925}{\micro\meter}$. Setting as converged values the results obtained by using a
large number of harmonics, $N=40$, specifically $\bar{A}=A_{40}$, the relative self-error
corresponding to a number of harmonics, $N$, is then defined as:
\begin{align}\label{eq:def_ref_error}
    e_N(A) = \frac{A_{N}-\bar{A}}{\bar{A}}.
\end{align}
This self-error function can be used as a reliable measure of the convergence of the method as
long as the value of $N$ for which the reference absorption is defined is chosen to be
sufficiently large.

At all three resonance wavelengths, relative errors of $e_N(A)\approx\SI{1}{\percent}$ are
achieved for $N\geq35$. It should be noted that in terms of computational effort, using
$N_{2D}=35$ in 2D simulations is comparable to using $N_{1D}=2520$ in 1D simulations. Moreover,
the accuracy of the 2D simulation with $N_{2D}=35$ is still correlated to and limited by the
highest spatial frequency, $2\pi N_{2D}/\Lambda$.

To asses the convergence of the TH simulations more rigorously, the relative error,
$e_N(R^{\prime})$, of the TH radiation intensity, $R^{\prime}_{N}$, for a given number of
harmonics $N$, which is defined similarly to $e_N(A)$ in Eq.~\eqref{eq:def_ref_error} with
$\bar{R}^{\prime}=R^{\prime}_{40}$, is depicted in \figref{fig:2DConvergence}(b). A relative self
error of $e_N(R^{\prime})\approx2\%$ can be achieved at $N=35$ for all three resonance
wavelengths. Note also that the intensity of the TH radiation varies over six orders of magnitude
and has maxima at the locations of the spectral resonances of the linear absorption, as per
\figref{fig:2DBothPlot}(a).

\begin{figure}[t]
    \centering
    \includegraphics[width=\linewidth]{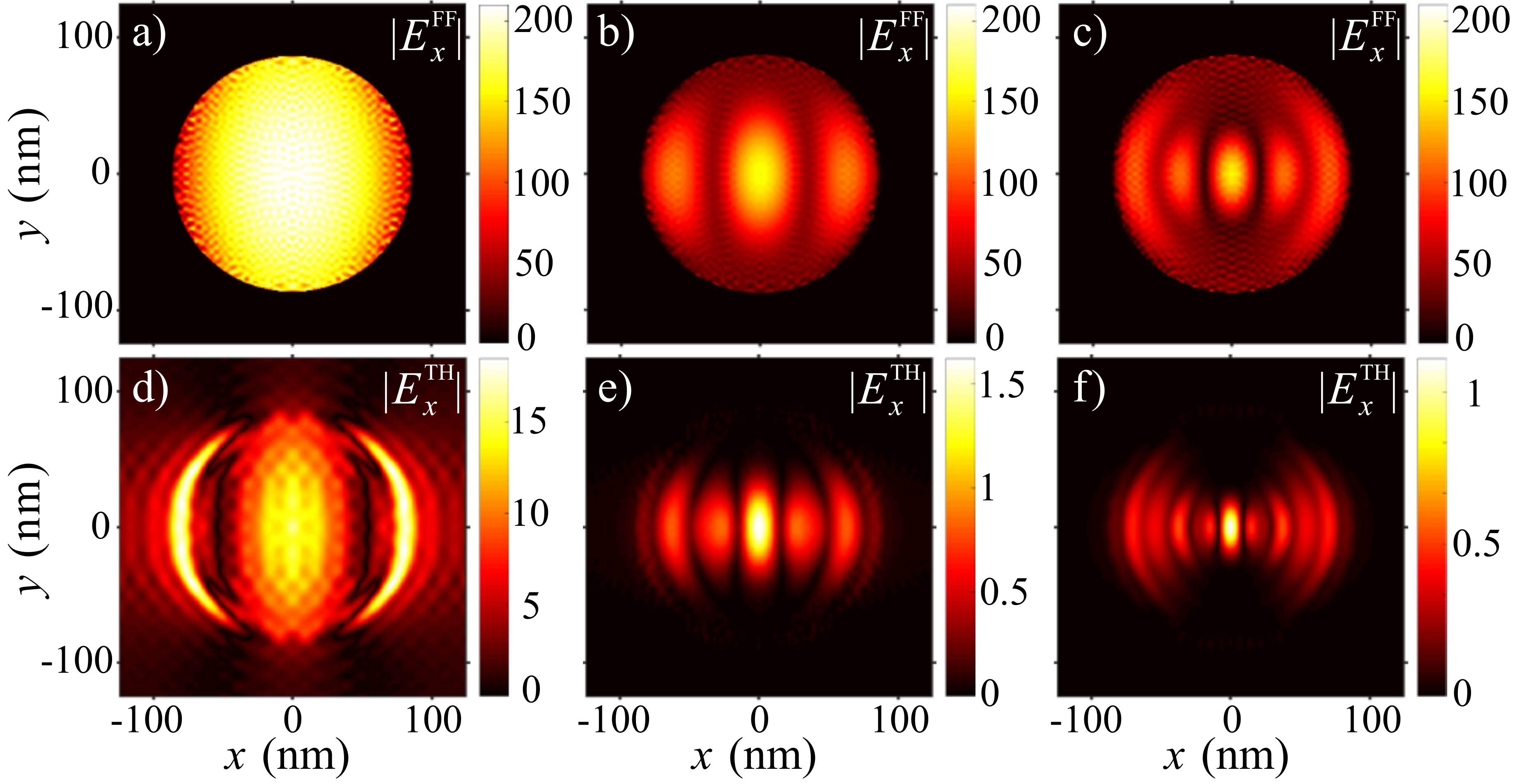}
\caption{Dominant electric field component, $|E_x|$, at the FF (top panels) and TH (bottom panels)
at the surface of a graphene disk for the resonance wavelengths, from left to right,
$\lambda_{FF}=\SI{11.09}{\micro\meter}$, $\lambda_{FF}=\SI{5.081}{\micro\meter}$, and
$\lambda_{FF}=\SI{3.925}{\micro\meter}$. The number of harmonics used in the simulations was
$N=40$ at the FF and $N=30$ at the TH.}
    \label{fig:2DFieldProfiles}
\end{figure}

Graphene is a lossy conductor in the spectral range considered and therefore allows the excitation
of surface waves \cite{koppens11nl,nggm12prb,gao12acsnano}. This is the case with each of the
absorption maxima seen in the linear spectrum, as illustrated by
Figs.~\ref{fig:2DFieldProfiles}(a)--\ref{fig:2DFieldProfiles}(c). Thus, we show in these figures
the dominant field component, $|E_x|$, of the linear electric field at the first three resonance
wavelengths, $\lambda=\SI{11.09}{\micro\meter}$ in Fig.~\ref{fig:2DFieldProfiles}(a),
$\lambda=\SI{5.081}{\micro\meter}$ in Fig.~\ref{fig:2DFieldProfiles}(b), and
$\lambda=\SI{3.925}{\micro\meter}$ in Fig.~\ref{fig:2DFieldProfiles}(c). These field profiles
exhibit distinct mode shapes with one, three and five maxima, which demonstrates that the
corresponding resonances represent the first three plasmon modes of the graphene disks.

Strongly enhanced and largely confined optical field resulting from the excitation of localized
plasmon modes gives rise to enhanced THG. This fact is supported by the field profiles plotted in
the bottom panels of \figref{fig:2DFieldProfiles}, where the dominant electric field component at
the TH, $|E_x|$, is shown. Indeed, regions of strong field enhancement can be seen, with field
profiles with three, seven, and eleven maxima being observed at
$\lambda_\mathrm{TH}=\SI{11.09}{\micro\meter}/3$ in Fig.~\ref{fig:2DFieldProfiles}(d),
$\lambda_\mathrm{TH}=\SI{5.081}{\micro\meter}/3$ in Fig.~\ref{fig:2DFieldProfiles}(e), and
$\lambda_\mathrm{TH}=\SI{3.925}{\micro\meter}/3$ in Fig.~\ref{fig:2DFieldProfiles}(f),
respectively.

\section{Diffraction in nonlinear gratings containing 2D materials}\label{sec:examples}
In this section the linear and nonlinear optical response of diffraction gratings incorporating
TMDC monolayer materials or graphene is investigated and the application of the inhomogeneous
$\mathcal{S}$-matrix formulation introduced in \secref{sec:inhomSMatrix} is illustrated in several
specific cases.

Throughout this section, the intensity of the incident beam is chosen to be
$I_0=\SI{e12}{\watt\per\square\meter}$, which is a moderately high peak intensity generated by a
pulsed laser. Changing the incident intensity in the undepleted pump approximation does not alter
the numerical results at any of the incident frequencies, $n=1,\ldots,N_F$ but it substantially
changes the magnitude of the electromagnetic field at the generated frequency, $n=0$. More
specifically, for SHG and THG the intensity of the generated waves behaves as
$I_\mathrm{SH}\propto I_0^2$ and $I_\mathrm{TH}\propto I_0^3$, respectively, and as such they can
increase to significant values. A problem that might arise in the undepleted pump approximation is
that if the intensity of the generated waves becomes comparable to the intensity of the incident
wave, the use of this approximation would become questionable. This is not the case in most
practical situations and certainly not the case here, as can be seen by the intensity of generated
optical fields in the examples hereafter.

\subsection{SHG from TMDC monolayer ribbons}\label{sec:example1}
\begin{figure}[t]
    \centering
    \includegraphics[width=\linewidth]{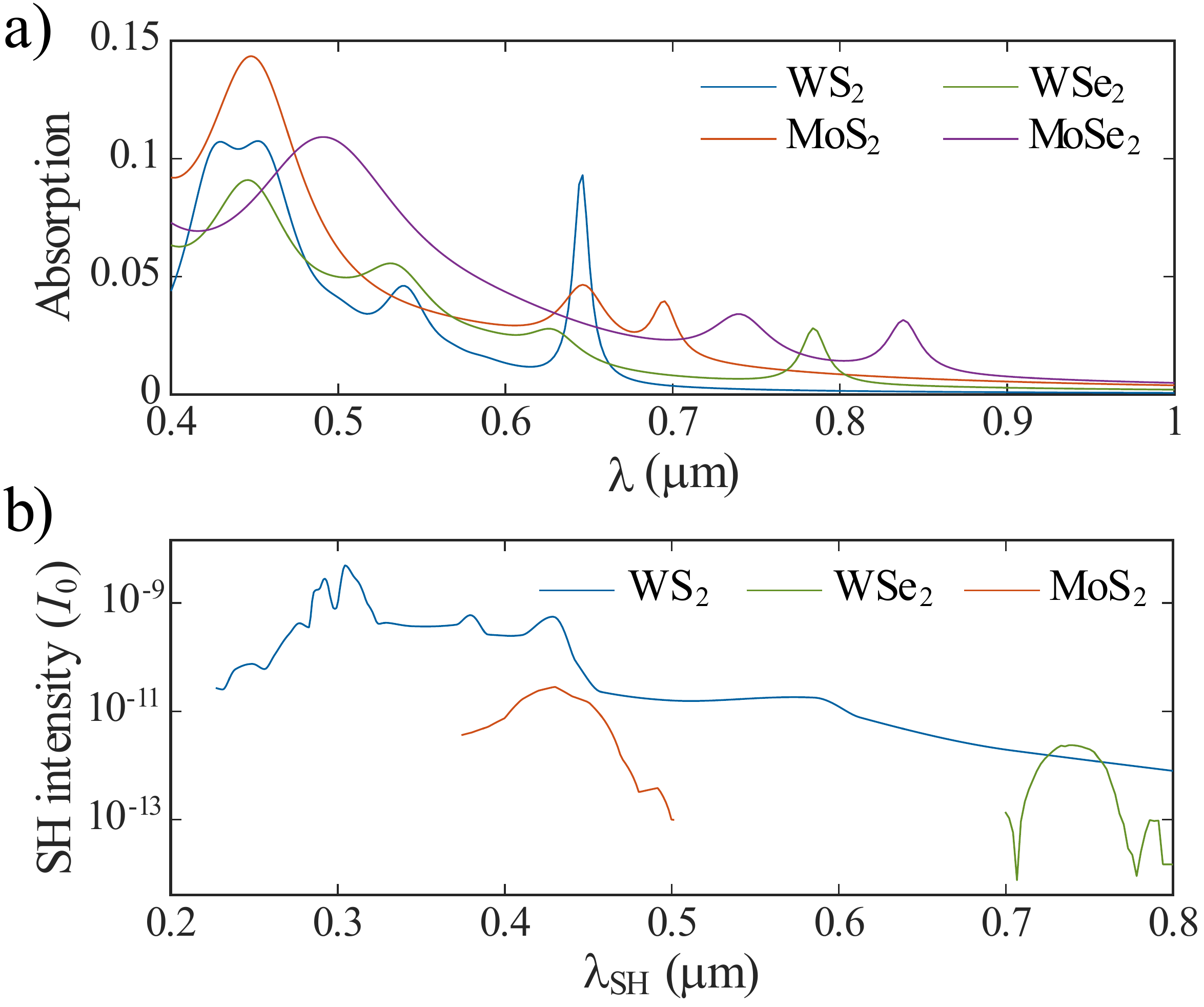}
\caption{a) The FF absorption spectra for ribbons made of $\mathrm{WS_2}$, $\mathrm{MoS_2}$,
$\mathrm{WSe_2}$, and $\mathrm{MoSe_2}$. b) The SH radiation spectra for ribbons made of
$\mathrm{WS_2}$, $\mathrm{MoS_2}$, and $\mathrm{WSe_2}$.}\label{fig:TMDC_ribbons}
\end{figure}

To begin with, we consider a 1D binary TMDC grating placed on top of a glass substrate with
$\epsilon_s = 1.44$. Its period is $\Lambda=\SI{100}{\nano\meter}$ and the filling factor is
$0.9$. An $x$-polarized plane wave is normally incident onto the grating and a spectral range of
\SIrange{0.4}{4}{\micro\meter} for the incident wavelength $\lambda$ is considered. The
computations were performed using $N=200$ harmonics and an added sheet conductance of
$\sigma_{s,\mathrm{add}}=-i 10^{-5} |\sigma_s(\omega)|$, the results being presented in
\figref{fig:TMDC_ribbons}. The convergence of these calculations has been assured as diligently as
for the binary graphene diffraction gratings discussed in \secref{sec:convergence:1D}.

The plots presented in \figref{fig:TMDC_ribbons}(a) reveal that the linear absorption spectra are
primarily determined by the linear material properties $\sigma_s(\omega)$ of the TMDC materials.
The absence of any additional resonant features in the spectra has two main reasons: the
dielectric nature of the TMDCs does not allow the formation of plasmons, as in the case of
graphene, whereas the vanishingly small thickness of the TMDC monolayers precludes the existence
of geometric, Mie-type resonances. These explanations are supported also by the fact that linear
and nonlinear spectra are qualitatively similar if one varies the grating period and the feature
size of the TMDC ribbons.

In order to obtain the SH radiation of the 2DM grating, we used the nonlinear conductivity,
${\boldsymbol \sigma}^{(2)}$, discussed in \secref{sec:2dnonlinprop}. Since no values of
${\boldsymbol \sigma}^{(2)}$ for $\mathrm{MoSe_2}$ were available, we determined the nonlinear
optical response of the grating only in the case of three TMDC materials, $\mathrm{WS_2}$,
$\mathrm{MoS_2}$, and $\mathrm{WSe_2}$. As our calculations showed that there is no resonant field
enhancement at the TMDC ribbons at the FF, no strong enhancement of generated SH is expected. The
nonlinear SH radiation spectra presented in \figref{fig:TMDC_ribbons}(b) suggest that indeed the
intensity of radiated SH closely follows the magnitude of $\boldsymbol \sigma^{(2)}$, as can be
found by comparison with the plots in \figref{fig:Mat_Nonlinear_Properties}(b). The maximal
intensity of generated SH is $I_\mathrm{SH} = \num{2e-9}I_0$, rendering valid the undepleted pump
approximation.

\subsection{Nonlinear efficiency enhancement for TMDC monolayers on a slab waveguide}\label{sec:example2}
\begin{figure}[t]
    \centering
    \includegraphics[width=\linewidth]{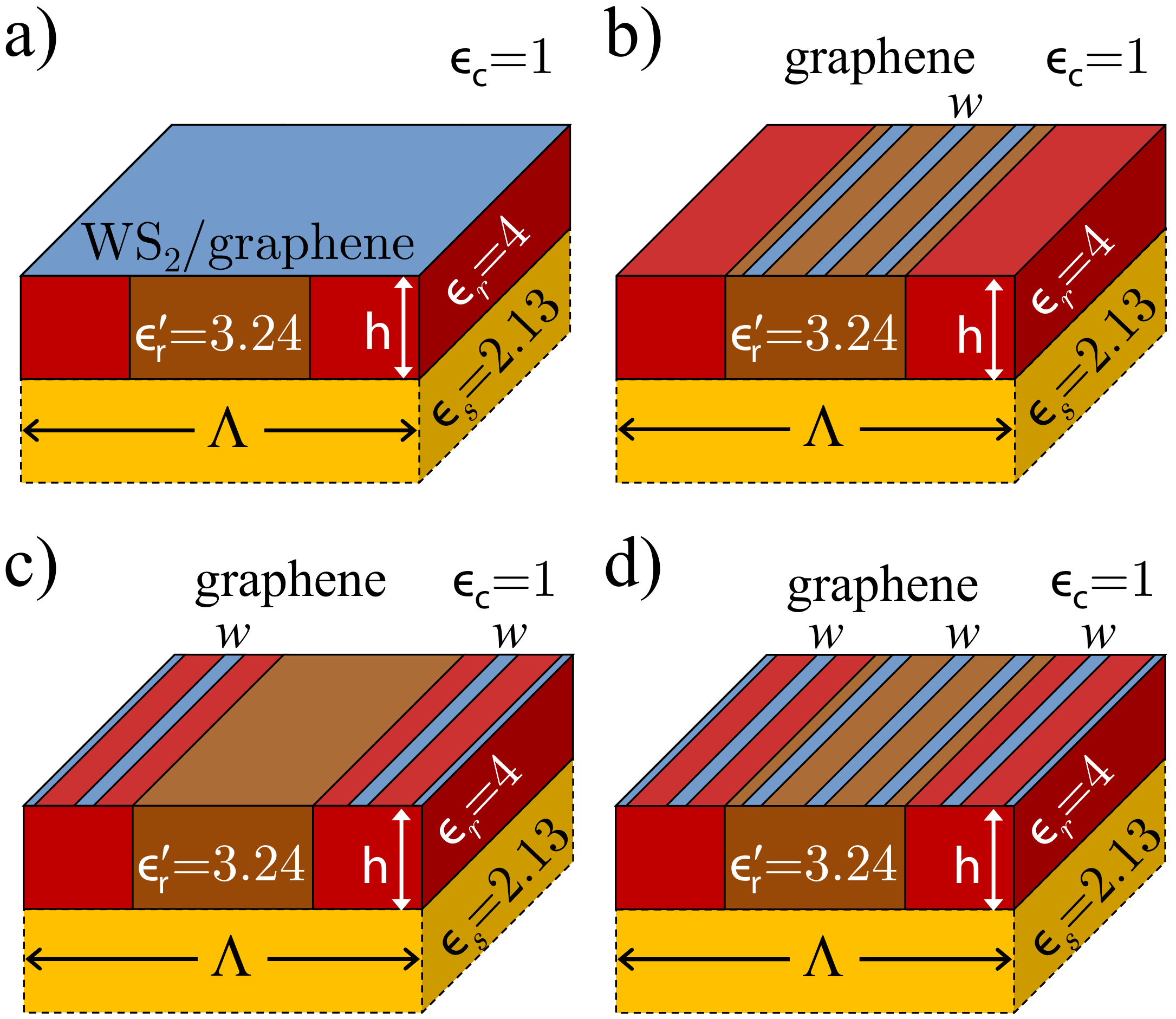}
\caption{Waveguide structures comprising a periodically patterned slab waveguide with
permittivities $\epsilon_{r}$ and $\epsilon_r^{\prime}<\epsilon_{r}$ covered by air and placed on
a dielectric substrate with $\sqrt{\epsilon_s}=1.46$. On top of the slab waveguide different 2DMs
are placed: a) A uniform monolayer of $\mathrm{WS_2}$ or graphene. b), c) Graphene ribbons with
width $w$ distributed over the slab material with lower and higher permittivity, respectively. d)
Graphene ribbons with width $w$ distributed over both regions of the slab waveguide.}
    \label{fig:WG_setting}
\end{figure}

Since TMDC monolayers themselves do not possess optical modes, we use a different approach to
achieve enhanced nonlinear optical interactions in these 2D materials. Thus, we combine a TMDC
monolayer with a bulk structure that possesses waveguide modes whose excitation leads to strong
local field enhancement. A very effective structure for this purpose is a periodically patterned
slab waveguide, covered by a TMDC monolayer, as depicted in \figref{fig:WG_setting}(a). Dielectric
slab waveguides exhibit very narrow spectral resonances, due to the resonant excitation of guiding
slab modes, a phenomenon that can be used to enhance linear and nonlinear optical response of
certain devices \cite{pm96josaa,po07ol,wgp15josab,cy02prb}.

The waveguide structure under consideration consists of a slab of height, $h$, placed between a
monolayer WS$_2$ and substrate with relative permittivity $\epsilon_s=1.46^2$ (index of
refraction, $n_s=1.46$). The 2DM monolayer is adjacent to the cover region, which is assumed to be
air, $\epsilon_c=1$. The slab itself is periodically patterned, that is it consists of alternating
regions with permittivity $\epsilon_r = 4$ ($n=2$) and $\epsilon_r^{\prime}=3.24$
($n^{\prime}=1.8$) with a period $\Lambda=\SI{400}{\nano\meter}$. This particular choice of
parameters was inspired by Ref.~\cite{pm96josaa}.

For now, consider the unperturbed waveguide consisting of a material with relative permittivity
$\bar{\epsilon}_r = (\epsilon_r + \epsilon_r^{\prime})/2$. A slab waveguide supports optical
guided modes and their excitation strongly affects its reflective and transmissive
characteristics. A mode of order $\nu$ with in-plane propagation constant $\beta_\nu$ can be
excited, when the in-plane wavenumber, $k_\parallel$, of an incident wave coincides with
$\beta_\nu$. This is not possible for a homogeneous waveguide due to the particular dispersion
properties of $k_\parallel$ and $\beta_\nu$, but can be achieved if the waveguide permittivity is
periodically modulated, as illustrated in \figref{fig:WG_setting}(a). This periodic perturbation
effectively folds the propagation constant $\beta_\nu$ into the first Brillouin-zone of the
$k$-space of the modes of the periodic waveguide and enables phase matching between $k_\parallel$
and $\beta_\nu$, i.e. the excitation of the mode $\nu$. It should be stressed that, albeit less
effectively, free-space photons couple to waveguide modes even if the periodic slab waveguide is
covered by a thin, optically homogeneous layer, such as a 2DM, because the effective refractive
index of the combined structure is periodic in this case, too.

In what follows, we demonstrate the resonant excitation of modes in the TMDC-waveguide structure
for a fixed height of $h=\SI{0.18}{\micro\meter}$. Subsequently we optimize the height of the slab
waveguide to obtain maximal generated SH and investigate the interplay between various resonant
mechanisms in the combined waveguide-2DM device that lead to enhanced nonlinear optical response.
Only monolayer WS$_2$ is considered as covering 2DM in this example, because for this TMDC
monolayer the dispersion of the nonlinear conductivity, $\boldsymbol{\sigma}^{(2)}$, is known over
the broadest spectral domain. The WS$_2$ monolayer is oriented such that the arm-chair direction
of the atomic lattice is aligned with the $x$-axis of the structure. Due to the particular
tensorial structure of $\boldsymbol{\sigma}^{(2)}$, this configuration only yields TM-polarized SH
for a TM-polarized fundamental field.
\begin{figure}[t]
    \centering
    \includegraphics[width=1\linewidth]{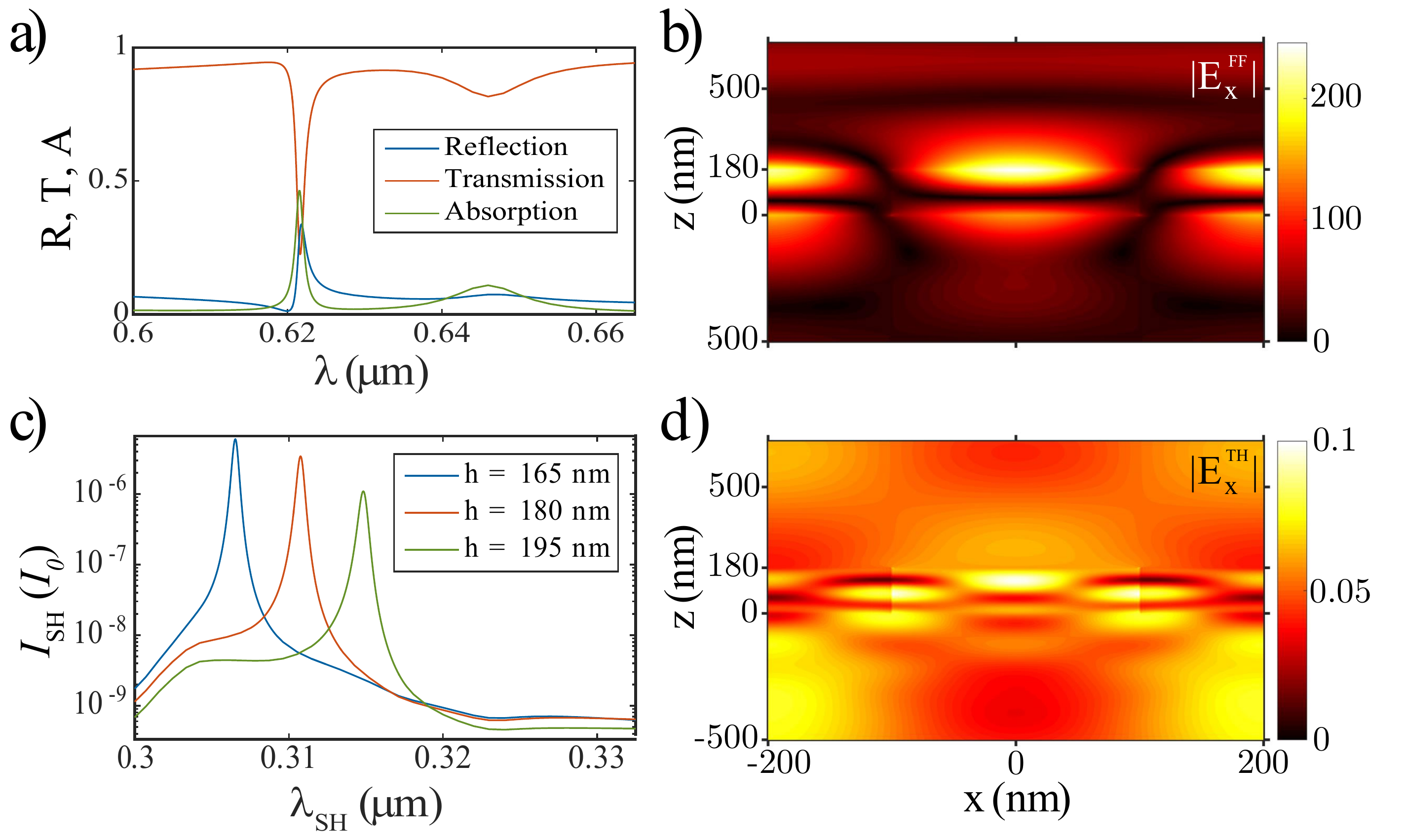}
\caption{a) Linear reflection, absorption, and transmission for waveguide height,
$h=\SI{0.18}{\micro\meter}$, demonstrate the effect of the intrinsic material absorption at
$\lambda\approx$ \SI{0.645}{\micro\meter} and the Fano resonance due to a waveguide resonance at
$\lambda=$ \SI{0.6215}{\micro\meter}. b) Electric near-field in and around the waveguide for
$\lambda=$ \SI{0.6215}{\micro\meter} exhibits the spatial profile of the TM$_0$ mode, with strong
field enhancement at waveguide interfaces. c) Nonlinear radiation intensity spectra near the
fundamental frequency corresponding to the TM$_0$ mode, determined for three values of the
waveguide height, $h$. d) Electric near-field at SH wavelength, $\lambda_\mathrm{SH}=$
\SI{0.311}{\micro\meter}, for $h=\SI{180}{\nano\meter}$.}
    \label{fig:WGR_TMDC_closeupNearField}
\end{figure}

In order to illustrate the excitation of a waveguide mode, we considered a TM-polarized, normally
incident plane wave in a wavelength range at FF of \SIrange{0.6}{0.67}{\micro\meter}. The
corresponding reflection, transmission, and absorption spectra are depicted in
\figref{fig:WGR_TMDC_closeupNearField}(a). A steep increase of the absorption is observed, from
less than \SI{10}{\percent} to a maximum of \SI{45}{\percent} at $\lambda =
\SI{0.6215}{\micro\meter}$. Moreover, the transmission and reflection have their minimum and
maximum at this wavelength, respectively. This is due to the excitation of the TM$_0$ waveguide
mode, as can be confirmed by the inspection of the electric near-field profile in
\figref{fig:WGR_TMDC_closeupNearField}(b): $|E_x|$ has maxima at the top and bottom facets of the
waveguide, which also implies a maximum of $|H_y|$ at its center, as one expects for a TM$_0$
waveguide mode. Another local maximum of the absorption can be seen at
$\lambda=\SI{0.645}{\micro\meter}$ and is due to one of the exciton absorption peaks of monolayer
WS$_2$ [c.f. \figref{fig:Mat_Linear_Properties}(b)].

The enhancement of the fundamental field at the top of the waveguide, where the WS$_2$ monolayer
is located, yields a strongly increased intensity of SH radiation with a maximal value of
$I_\mathrm{SH}=\SI{4e-6}{}I_0$, as shown in \figref{fig:WGR_TMDC_closeupNearField}(c). Comparison
of the radiation spectra for different waveguide heights, $h=\SI{165}{\nano\meter}$,
$h=\SI{180}{\nano\meter}$, and $h=\SI{195}{\nano\meter}$, already shows the large sensitivity of
the spectral location of waveguide modes to changing height. This is the basis for the parameter
study in the remainder of this section. Before that, let us inspect the electric near-field at the
SH wavelength, $\lambda_\mathrm{SH}= \SI{0.3107}{\micro\meter}$, presented in
\figref{fig:WGR_TMDC_closeupNearField}(d). This profile is markedly different from that of the
linear near-field, namely it is spatially more inhomogeneous and has a different distribution of
local minima and maxima. Finally, we point out that the absorption peak at
$\lambda=\SI{0.645}{\micro\meter}$ does no translate to a notable increase of SH radiation, which
is similar to the findings reported in \secref{sec:example1}.

\begin{figure}[t]
    \centering
    \includegraphics[width=\linewidth]{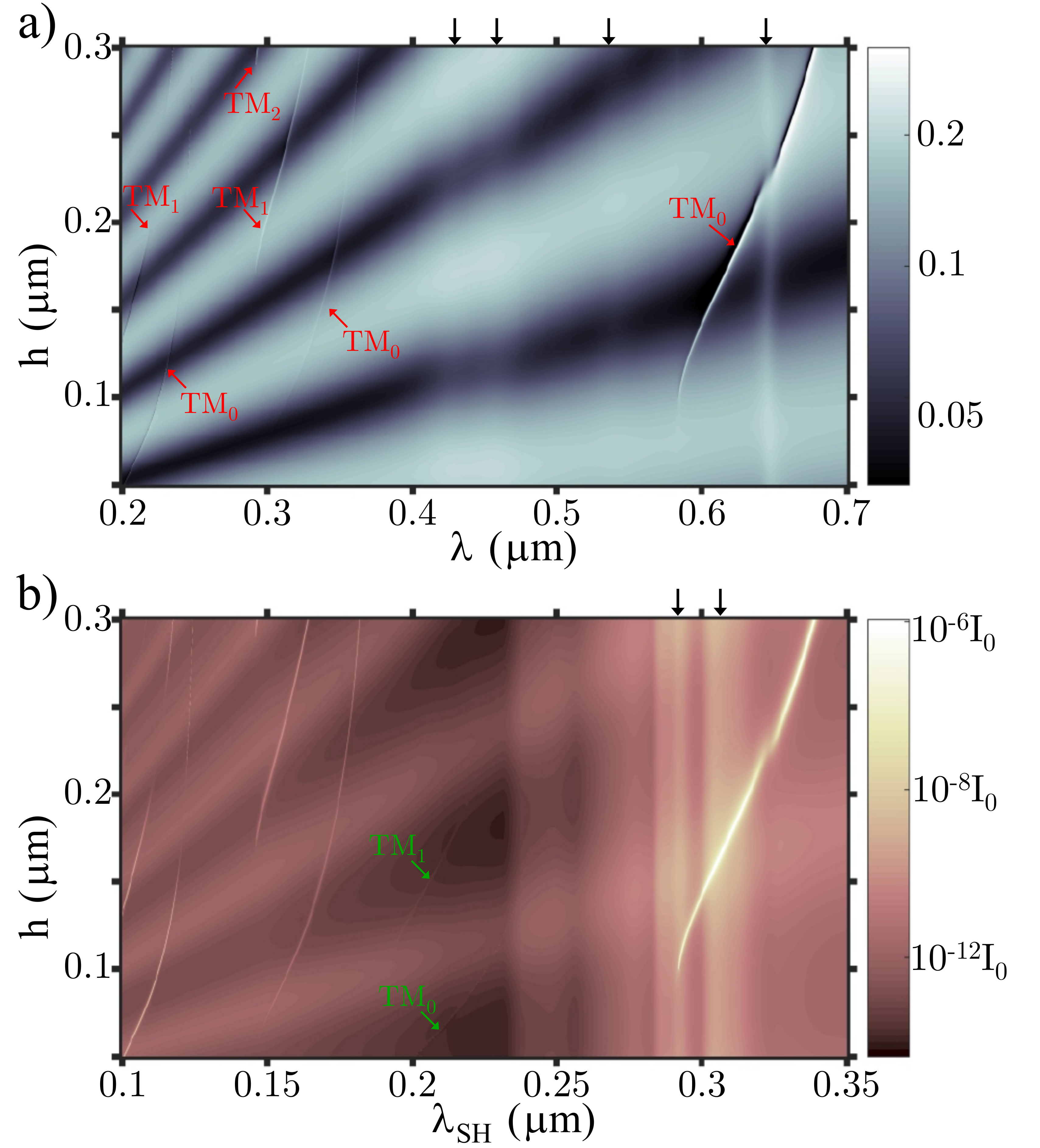}
\caption{a) Map of linear reflection spectra vs. waveguide height $h$ exhibits Fabry-Perot
resonances, resonances due to exciton generation ($\boldsymbol\downarrow$), and resonances due to
excitation of waveguide modes ({\color{red}{$\boldsymbol\searrow$}}). Interaction between the
TM$_0$ waveguide mode and excitons of WS$_2$ monolayer as well as Fano resonances can be observed.
b) Map of nonlinear radiation spectra determined for different waveguide height, $h$.}
    \label{fig:WGR_TMDC}
\end{figure}

In the rest of this section we explore the interplay among different resonant mechanisms in the
combined waveguide-2DM structure that lead to enhanced nonlinear optical response and investigate
how one can exploit them to increase the intensity of SH generated from this optical device. To
this end, TM-polarized incident light in a wavelength range at the FF of
\SIrange{0.2}{0.7}{\micro\meter} is considered and simulations for waveguide height ranging from
\SIrange{50}{300}{\nano\meter}, using $N=25$ harmonics, are performed. The results of these
calculations, corresponding to the FF, are shown in \figref{fig:WGR_TMDC}(a) in terms of the
reflection spectra maps, and will be discussed now. The nonlinear part of the device study is
summarized in \figref{fig:WGR_TMDC}(b) in terms of maps of its outgoing radiation at the SH and
will be investigated subsequently.

The linear optical characteristics of the TMDC-covered waveguide determined for different heights
are depicted as a 2D map of reflectivity values. This 2D map of reflection spectra exhibits a
smooth dependence on $\lambda$ and $h$ with minimal and maximal values of $R=\num{4.9e-06}$ and
$R=0.5428$, respectively, except when one of several mechanisms leads to resonant enhancement of
the reflectivity. \textit{i})~The most evident and spectrally broadest features are due to the
Fabry-Perot interference mechanism, which yields maximal (minimal) reflectivity if the multiple
reflections inside the slab waveguide are in (out-of) phase \cite{hernandez1988cambridge}. These
Fabry-Perot resonances appear in the reflectivity map as spectrally broad variations from
reflection minima to maxima. \textit{ii)}~The second kind of spectral feature is due to the
resonant increase of intrinsic optical absorption of WS$_2$, which occurs at wavelengths at which
excitons are generated in the WS$_2$ monolayer. The strongest of these absorption peaks is at
$\lambda=\SI{0.645}{\micro\meter}$. Their spectral location is determined by the dispersion of
$\Re{\sigma_s(\omega)}$, given in \figref{fig:Mat_Linear_Properties}(b), and is largely
independent of the electromagnetic environment and hence does not depend on $h$. The intrinsic
optical absorption mostly increases the absorption in the combined waveguide-2DM device, but also
leads to increasing reflection and decreasing transmission, as was already shown for
$h=\SI{180}{\nano\meter}$ around $\lambda=\SI{0.65}{\micro\meter}$ in
\figref{fig:WGR_TMDC_closeupNearField}(a). \textit{iii)}~The third kind of resonance is due to the
excitation of TM$_{\nu}$ waveguide modes and manifests itself as a spectrally narrow, asymmetric,
and steep variation of the reflectivity of the device. A detailed analysis of the resonant
excitation of the TM$_{0}$ mode for $h=\SI{180}{\nano\meter}$ and
$\lambda=\SI{0.6215}{\micro\meter}$ was already presented in relation to
\figref{fig:WGR_TMDC_closeupNearField}. By varying the waveguide height and wavelength, the
excitation of the TM$_1$ and TM$_2$ modes was found, too.

As the map in \figref{fig:WGR_TMDC}(a) suggests, there is a mutual interaction among resonances of
the TMDC-covered waveguide, giving rise to several interesting phenomena. First, one can observe
the generation of Fano resonances, which generally result from the interference between a discrete
state and a broad continuum and are characterized by an asymmetric spectral profile
\cite{fano1961pr,mirosh2010rmp,luk2010nm}. Fano resonances arise via different scenarios in the
considered structure, most notably due to the interference of the TM$_0$ waveguide mode (the
discrete state) and the Fabry-Perot resonance (the broad continuum). For example, for a device
height of $h=\SI{180}{\nano\meter}$ and for increasing wavelengths around $\lambda =
\SI{0.62}{\micro\meter}$, the reflection decreases to a minimum value of $R=0.01$, then steeply
increases to $R=0.33$, as shown in \figref{fig:WGR_TMDC_closeupNearField}(a). Similar behavior can
be observed when TM$_1$ and TM$_2$ waveguide modes are excited. The absorption and transmission
spectra exhibit similar features, but the spectral asymmetry, a characteristic feature of Fano
resonances, is not as well pronounced in these cases. Hence, only the reflection is shown here.
The second phenomenon revealed by \figref{fig:WGR_TMDC_closeupNearField}(a) is the crossing of the
TM$_0$ waveguide mode with the spectrally highest exciton absorption peak of monolayer WS$_2$. In
particular, the two resonances exhibit an anti-crossing behavior at the wavelength of their
strongest interaction, which is a well-known phenomenon in photonics and other physical systems
\cite{weisbuch1992prl,yoshie2004nature,novotny2010amjp}.

Having understood the key features of the linear response of the combined waveguide-2DM optical
system, we now explore its nonlinear optical properties. To this end, consider
\figref{fig:WGR_TMDC}(b), which depicts the map of the intensity of the total generated SH at
wavelengths, $\lambda_\mathrm{SH}$, ranging from \SIrange{0.1}{0.35}{\micro\meter} and for the
same values of the waveguide height as in \figref{fig:WGR_TMDC}(a). This intensity varies over
almost 6 orders of magnitude, from a minimum of $I_\mathrm{SH}=\SI{9}{\watt\per\square\meter}$ at
$h=\SI{300}{\nano\meter}$ and $\lambda_\mathrm{SH}=\SI{0.228}{\micro\meter}$ to a maximum of
$I_\mathrm{SH}= \SI{6.3e+06}{\watt\per\square\meter}\approx\num{6e-6}I_0$ for
$h=\SI{161.5}{\nano\meter}$ and $\lambda_\mathrm{SH}=\SI{0.306}{\micro\meter}$. It exhibits a
smooth dependence on the system parameters, except for the excitation of certain resonances via
mechanisms similar to those examined in the linear case.

In general, the nonlinear radiation is affected by two factors, which we call \textit{inherited}
and \textit{intrinsic} effects. Inherited effects are due to the enhancement via certain
mechanisms of the optical field at the FF, at the location of the WS$_2$ monolayer, which
increases the nonlinear source current and consequently the intensity of the generated SH.
Intrinsic effects, on the other hand, are resonant effects at the SH wavelength, which can also
influence the intensity of radiated waves at the SH.

The most important inherited effects leading to resonant enhancement of nonlinear radiation, seen
in the map plotted in \figref{fig:WGR_TMDC}(b), are as follows: \textit{i)}~The inherited
Fabry-Perot reflection minima lead to a moderate increase of the SH over broad wavelength ranges.
\textit{ii)}~The excitation of waveguide modes at the FF leads to particularly strong enhancement
of the fundamental field and yields the highest intensity of SH radiation, most notably when the
TM$_0$ mode is excited. In particular, SH radiation with intensity $I_\mathrm{SH}>10^{-6}I_0$ is
consistently achieved when this mode is excited, except for fundamental wavelengths near the
exciton absorption maximum at $\lambda=\SI{0.645}{\micro\meter}$. \textit{iii)}~Finally, the
interaction of the fundamental TM$_0$ mode and the WS$_2$ exciton leads to a reduced enhancement
of the fundamental field, as compared to the case of the sole excitation of the TM$_0$ mode, and
results in a decrease of the SH intensity to $I_\mathrm{SH}=\num{8.1e-8}I_0$.

Amongst the intrinsic effects, two important mechanisms that lead to enhancement of SH intensity
were identified: \textit{i)}~The generated electric field at the SH wavelength acts as excitation
waves for the TMDC monolayer-waveguide system and resonantly excites waveguide modes existing at
the SH, namely the TM$_0$ and TM$_1$ modes. This leads to a relatively small increase in the SH
intensity. \textit{ii)}~The frequency dispersion of the nonlinear optical conductivity of
monolayer WS$_2$ is apparent in the SH spectra: maxima of $\sigma_s^{(2)}$, which are naturally
independent of the waveguide height, $h$, correspond to maxima in the SH radiation spectrum.
Moreover, the SHG due to the combined inherited TM$_0$ mode and the intrinsic maximum of
$\sigma_s^{(2)}$ for three values of $h$ is presented in \figref{fig:WGR_TMDC_closeupNearField}(c)
and shows that the two effects constructively add to increase the intensity of the SHG.

Note that the reflection, transmission, and absorption spectra, the nonlinear radiation spectra,
as well as the resonance wavelengths of the slab waveguide were accurately calculated even for the
moderate number of $N=25$ harmonics, as was ensured with a convergence check for fixed height $h$
and with the rigorous procedures described in \secref{sec:convergence:1D}. A total of $385297$
simulations for pairs of $(h,\lambda)$ were performed, where a higher spectral resolution was used
near the resonance wavelengths of the device in order to accurately resolve the spectrally narrow
effect of the waveguide resonances.

\subsection{Nonlinear interaction between waveguide modes and graphene plasmons}\label{sec:example3}
In the preceding section, we combined a 2D material, WS$_2$, which does not support localized
optical modes, with an optical device consisting of a periodic slab waveguide, and achieved a
strong enhancement of the nonlinear efficiency of the combined device. In this section we follow a
similar approach and combine a similar slab waveguide with graphene structures, which we have
already shown that support localized surface plasmon modes, in order to achieve a multiresonant,
highly nonlinear optical device. Thus, the structure under consideration is schematically depicted
in \figref{fig:WG_setting}(d). It consists of a periodically patterned slab waveguide with the
same optical parameters as the one in the preceding section, which now is covered by graphene
ribbons with width $w=\SI{230}{\nano\meter}$. We will call the three graphene ribbons centered
on top of the material with permittivity $\epsilon_r^{\prime}$ and $\epsilon_r$ the inner and
outer ribbons, respectively, which is natural given the definition of the unit cell in
\figref{fig:WG_setting}(d). The center-to-center distance of the inner ribbons (and outer
ribbons) is $3w=\SI{0.69}{\micro\meter}$ and the center-to-center distance between a inner
ribbon to a neighboring outer ribbon is $\SI{1.37}{\micro\meter}$. In this case, the height
of the slab waveguide is $h=\SI{1.5}{\micro\meter}$ and the period is
$\Lambda=\SI{5.5}{\micro\meter}$.

The small feature size of the graphene ribbons, $w = 0.0418\Lambda$, is required to excite
graphene plasmons at moderately small wavelengths, at which waveguide modes exist, too. This is
not a conceptual drawback of this particular structure, but it is computationally costly to
accurately resolve graphene ribbons with very small width. Thus, $N=251$ harmonics were used
throughout this computational analysis and an added conductivity of
$\sigma_{s,\mathrm{add}}(10^{-3})$ was introduced.

To fully understand the linear and nonlinear optical properties of the graphene-waveguide
structure, let us first investigate two less complex, complementary structures where the graphene
ribbons are located on the waveguide sections with either low or high index of refraction, as per
Figs.~\ref{fig:WG_setting}(b) and \ref{fig:WG_setting}(c), respectively, and also the waveguide
covered with an unstructured, uniform graphene sheet, as per \figref{fig:WG_setting}(a).
\begin{figure}[t]
    \centering
    \includegraphics[width=\linewidth]{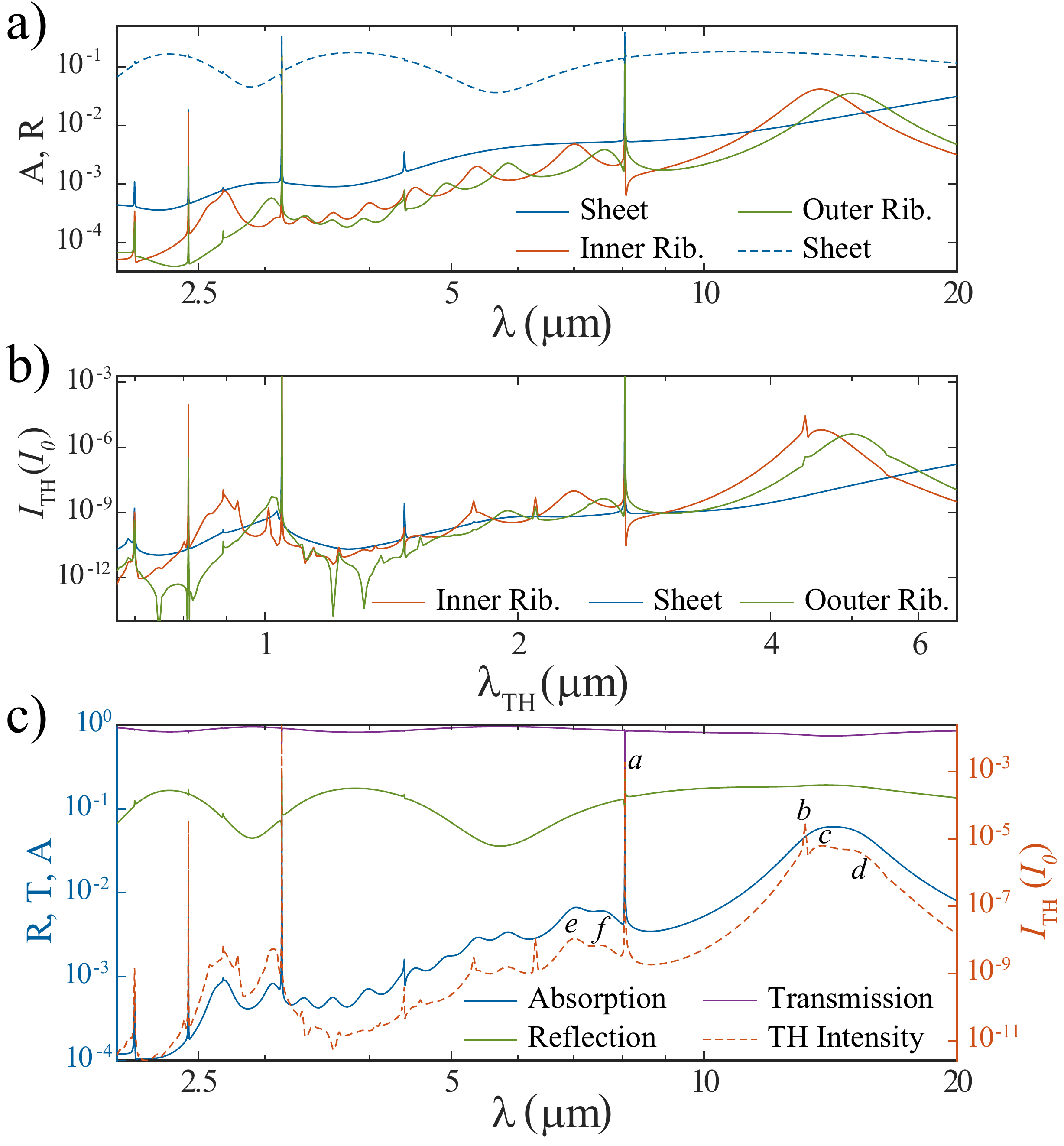}
\caption{\label{fig:WGR_graphene_spectrum}a) Absorption spectra (solid lines) of the homogeneous
graphene sheet and the graphene ribbons on top of inner and outer parts of the slab waveguide,
c.f. Figs.~\ref{fig:WG_setting}(a) through \ref{fig:WG_setting}(c), and reflection spectrum
(dashed line) of the homogeneous graphene sheet. b) The nonlinear radiation spectrum of the same
structures as in a). c) Linear spectra (reflection, transmission, and absorption) and TH spectrum
for the graphene ribbons placed on top of both parts of the waveguide, as per
\figref{fig:WG_setting}(d).}
\end{figure}

The linear absorption spectra for these three variations of the device, determined for values of
the fundamental wavelength ranging from \SIrange{1}{20}{\micro\meter}, are depicted in
\figref{fig:WGR_graphene_spectrum}(a), where normal incidence and TM-polarization is assumed. The
absorption of the waveguide with the covering graphene sheet follows a monotonously increasing
trend, upon which alternating, broad local minima and maxima are superimposed. These maxima are
due to the Fabry-Perot interference and primarily reveal themselves as maxima and minima of the
device reflectivity, also shown in \figref{fig:WGR_graphene_spectrum}(a). The absorption spectra
for the waveguide with graphene ribbons on top of the inner and outer parts of the waveguide
exhibit relatively broad spectral peaks due to the excitation of surface plasmons in the graphene
ribbons. Their excitation wavelength chiefly depends on the width of the ribbon and the
permittivity of the underlying dielectric: the lower refractive index material ($n^{\prime}=1.8$)
underneath the inner ribbons leads to excitation of plasmons at smaller wavelength than the
wavelength corresponding to the plasmons in the outer ribbons, placed on top of waveguide sections
with higher refractive index ($n=4$). In all three devices, the absorption exhibits additional,
spectrally narrow peaks with up to 30\% absorption due to the excitation of optical modes in the
slab waveguide grating.

The nonlinear radiation spectra in \figref{fig:WGR_graphene_spectrum}(b) complement these
findings. Thus, the intensity of the TH radiation generated by the unstructured graphene sheet
increases with wavelength and overlayed on it one can observe the effect of Fabry-Perot
resonances. Moreover, the local field enhancement due to the excitation of localized surface
plasmons in the inner and outer graphene ribbons leads to increased THG at lower and higher
wavelengths, respectively. The inherited waveguide modes maximize the amount of generated TH
radiation, which reaches values of up to $I_\mathrm{TH}=\num{2e-3}I_0$.

\begin{figure}[t]
    \centering
    \includegraphics[width=\linewidth]{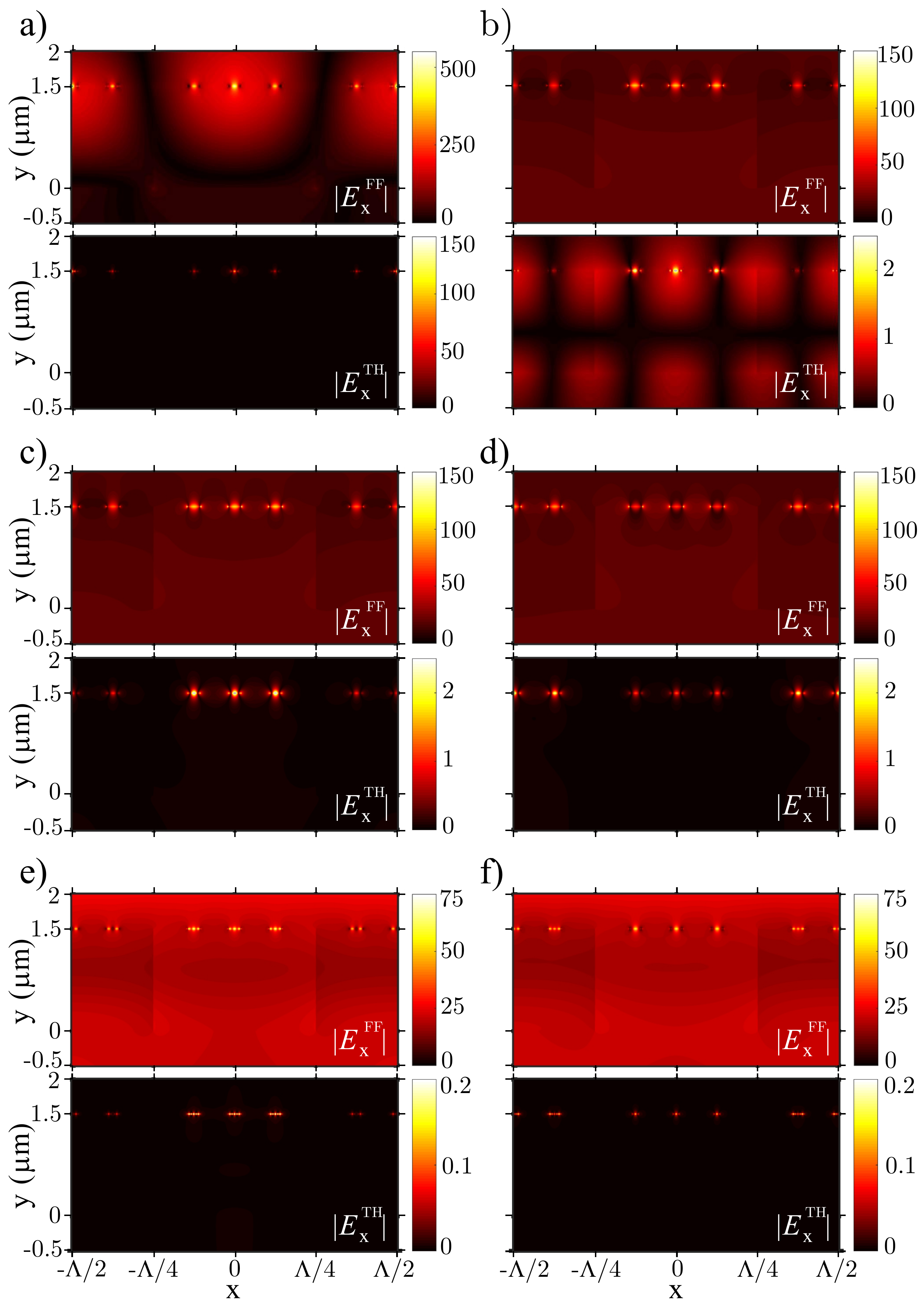}
\caption{\label{fig:WGR_graphene_NearField}Dominant component of the electric field $E_x$ at the
FF (top parts) and the TH (bottom parts) for selected fundamental wavelengths $\lambda$ and TH
wavelengths $\lambda_\mathrm{TH}=\lambda/3$: a) Field enhancement due to excitation of the TM$_0$
waveguide mode for $\lambda=\SI{8.05}{\micro\meter}$. b) Near-field profile of the TM$_0$ mode at
the TH wavelength $\lambda_\mathrm{TH} = \SI{4.4}{\micro\meter}$. c), d) Plasmonic field
enhancement in the inner and outer ribbons for $\lambda=\SI{13.82}{\micro\meter}$ and
$\lambda=\SI{15.17}{\micro\meter}$, respectively. e), f) Excitation of surface plasmons in
graphene ribbons located on top of waveguide sections with $\epsilon_r^{\prime}$ (inner ribbons,
$\lambda=\SI{7.1}{\micro\meter}$) and $\epsilon_r$ (outer ribbons,
$\lambda=\SI{7.6}{\micro\meter}$), respectively. The labels of the panels correspond to the labels
of the peaks in \figref{fig:WGR_graphene_spectrum}(c).}
\end{figure}

The interplay of these resonant effects in the combined graphene-waveguide structure with ribbons
distributed over the whole area of the waveguide is revealed by the results presented in
\figref{fig:WGR_graphene_spectrum}(c).  Note that the abscissa of this figure gives the values of
the incoming wavelength, $\lambda$, for the linear reflection, transmission, and absorption
spectra and the TH wavelength, $\lambda_\mathrm{TH}=\lambda/3$, for the intensity of THG. The absorption at
the FF displays an increasing trend in the range of \SIrange{2}{20}{\micro\meter}; however, there
are several spectrally broad and narrow absorption peaks, which are the manifestation of different
phenomena. Thus, the narrow resonances are due to the excitation of the TM$_0$ mode of the slab
waveguide, e.g. those at $\lambda = \SI{8.05}{\micro\meter}$, $\lambda = \SI{4.4}{\micro\meter}$,
and $\lambda = \SI{3.14}{\micro\meter}$. The electric field, $E_x^\mathrm{FF}$, at the FF,
$\lambda = \SI{8.05}{\micro\meter}$, shows a strong enhancement at the top of the waveguide
region, as per \figref{fig:WGR_graphene_NearField}(a, top). Similarly to the TM$_0$ mode shown in
\figref{fig:WGR_TMDC_closeupNearField}(b), this field profile has two maxima over the $x$-extent
of one unit cell of the grating and two maxima along the $z$-extent of the waveguide, where the
maximum near the top of the waveguide is much larger than the one at the bottom. This strong
fundamental field enhancement increases the absorption to $33\%$ and leads to a strong nonlinear
source current, which in turn generates a strong electric field at the TH, the near-field of which
is depicted in \figref{fig:WGR_graphene_NearField}(a, bottom). This TH field is mostly localized
around the graphene ribbons and has an evanescent nature. The total intensity of TH radiated into
the cover and substrate amounts to $I_\mathrm{TH}=\num{1.3e-3}I_0$.

The excitation of the TM$_0$ mode at $\lambda=\SI{4.4}{\micro\meter}$ is equally interesting. It
appears both as a sharp local maximum in the linear absorption spectrum at
$\lambda=\SI{4.4}{\micro\meter}$ and as an increase of TH intensity at the TH wavelength
$\lambda_\mathrm{TH}=\SI{4.4}{\micro\meter}$ in \figref{fig:WGR_graphene_spectrum}(c). The
nonlinear electric near-field profile in \figref{fig:WGR_graphene_NearField}(b, bottom) confirms
the excitation of this intrinsic nonlinear TM$_0$ mode for
$\lambda_\mathrm{TH}=\SI{4.4}{\micro\meter}$.

Another reason for increased absorption is the excitation of localized surface plasmons on the
graphene ribbons. In order to illustrate this, let us consider the two absorption peaks with
largest wavelengths, around $\lambda \approx \SI{14}{\micro\meter}$ and $\lambda \approx
\SI{7.3}{\micro\meter}$. The absorption maximum with the largest wavelength appears in the linear
spectrum as a very broad resonance around $\lambda\approx\SI{14}{\micro\meter}$, but the electric
near-field profiles in the two top panels of Figs.~\ref{fig:WGR_graphene_NearField}(c) and
\ref{fig:WGR_graphene_NearField}(d) reveal that the actually surface plasmons are excited on the
inner and outer graphene ribbons at slightly different wavelengths,
$\lambda=\SI{13.82}{\micro\meter}$ and $\lambda=\SI{15.17}{\micro\meter}$, respectively, which is
explained by the difference in the electromagnetic environment probed by the corresponding
plasmons. The excitation of these surface plasmons can also be seen in
\figref{fig:WGR_graphene_spectrum}(c), as the local maxima labeled by ``\textit{c}'' and
``\textit{d}''. Moreover, they can be viewed directly in the nonlinear far-field radiation
spectrum, too, as spectrally separated peaks. As a consequence of excitation of localized surface
plasmons, the TH near-field shows its highest values near the inner and outer ribbons at the TH
wavelengths, $\lambda_\mathrm{TH}=\SI{13.82}{\micro\meter}/3$ and
$\lambda_\mathrm{TH}=\SI{15.17}{\micro\meter}/3$, as illustrated in the bottom panels of
Figs.~\ref{fig:WGR_graphene_NearField}(c) and \ref{fig:WGR_graphene_NearField}(d), respectively.

Similar physics describe the lower-order plasmon corresponding to $\lambda \approx
\SI{7.3}{\micro\meter}$. Thus, by inspecting the profiles of the fundamental near-field in the top
panels of Figs.~\ref{fig:WGR_graphene_NearField}(e) and \ref{fig:WGR_graphene_NearField}(f), one
can see that, correspondingly, a plasmon with three peaks in the field profile is excited on the
inner ribbons at $\lambda = \SI{7.1}{\micro\meter}$ whereas this same type of plasmon is excited
on the outer ribbons at $\lambda = \SI{7.6}{\micro\meter}$. This near-field pattern is in
accordance to the surface plasmon field profiles that were found in \secref{sec:convergence:2D},
\figref{fig:2DFieldProfiles}(b). Moreover, no optical coupling between the plasmonic fields near
adjacent ribbons could be observed. The spectra of the TH radiation at the wavelengths
$\lambda_\mathrm{TH} = \SI{7.1}{\micro\meter}/3$ and $\lambda_\mathrm{TH} =
\SI{7.6}{\micro\meter}/3$ also exhibits two local maxima, due to the enhancement of the
fundamental field. The number of electric field maxima near the inner and outer graphene ribbons
in the bottom of Figs.~\ref{fig:WGR_graphene_NearField}(e) and
\ref{fig:WGR_graphene_NearField}(f), respectively, agrees with the 5 field maxima in
\figref{fig:2DFieldProfiles}(d).

The two physical mechanisms that lead to the resonant enhancement of the optical response of the
graphene-waveguide structure can partly be observed in the reflection and transmission spectra at
the fundamental wavelength, which mainly exhibit a variation between minima and maxima due to the
Fabry-Perot interference. This pattern is complemented by spectrally very narrow regions of
increased reflection and decreased transmission due to the excitation of waveguide modes. The
influence of surface plasmons on the reflection and transmission is only apparent at the largest
excitation wavelengths around $\lambda\approx\SI{14}{\micro\meter}$, where reflection and
transmission are notably increased and decreased, respectively.
\begin{figure}[t]
    \centering
    \includegraphics[width=\linewidth]{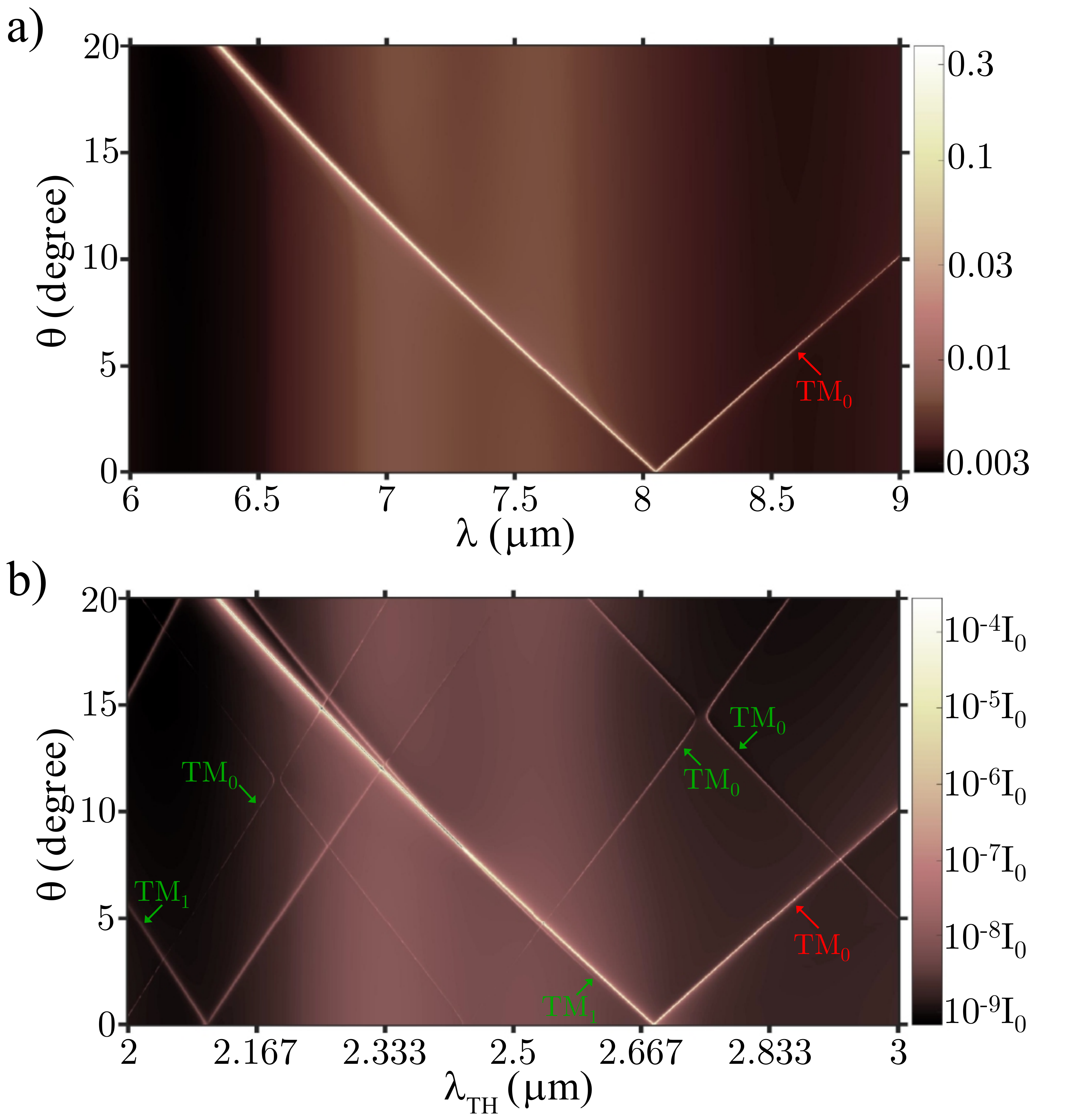}
\caption{a) Map of absorption spectra vs. angles of incidence, $\theta$, determined in the
wavelength range of the second-order surface plasmon of graphene ribbons. b) The map of nonlinear
radiation spectra shows the inherited (red labels) and intrinsic, nonlinear (green labels) modes
of the slab waveguide. The left and right branches of intrinsic modes of the same order exhibit
anti-crossing mode interaction, which leads to the formation of spectral band-gaps.}
    \label{fig:WGR_graphene_scans}
\end{figure}

An effective way of tuning the characteristics of the light radiated by our nonlinear diffraction
grating is by varying the angle of incidence of the incoming light. Not only will this showcase
the effectiveness of the proposed numerical method in the oblique-incidence configuration, but it
will also demonstrate the influence of the angle of incidence on the optical response of the
grating. For example, the interaction of the spectrally narrow TM$_0$ waveguide mode at
$\lambda=\SI{8.05}{\micro\meter}$ with the two broad continua corresponding to the excitation of
graphene plasmons at $\lambda=\SI{7.1}{\micro\meter}$ and $\lambda=\SI{7.6}{\micro\meter}$
provides a convenient optical setting for studying a tunable Fano resonance resulting from the
interaction between a discrete state and multiple continua, a phenomenon that has recently been
explored in a different plasmonic structure \cite{osley13prl}.

To investigate the interaction between a waveguide mode and the two absorption peaks, $105747$
simulations with $N=251$ harmonics have been performed for increasing angle of incidence,
$\theta$, ranging from \SIrange{0}{20}{\degree} (and constant azimuthal angle
$\varphi=\SI{0}{\degree}$) in the fundamental wavelength range of the second-order plasmon peak,
namely from \SIrange{6}{9}{\micro\meter}. Higher spectral resolution was employed near the
waveguide resonance wavelengths of the slab waveguide.

The resulting absorption map is shown in \figref{fig:WGR_graphene_scans}(a). The spectrum for
$\theta=\SI{0}{\degree}$ is the same as in \figref{fig:WGR_graphene_spectrum}(c) and shows two
broad maxima of $A\approx \num{6e-3}$, due to the excitation of surface plasmons, and a sharp
maximum of $A=0.33$ due to the excitation of the TM$_0$ waveguide mode at
$\lambda^{\mathrm{TM}_0}=\SI{8.05}{\micro\meter}$. Increasing the angle of incidence, $\theta$,
leaves the spectral location of the plasmon excitation unchanged; however, the spectral location
of the waveguide mode resonance varies. For $\theta>0$, the TM$_0$ mode is excited at two
wavelengths, $\lambda^{\mathrm{TM}_0^l}(\theta)<\lambda^{\mathrm{TM}_0}$ and
$\lambda^{\mathrm{TM}_0^r}(\theta)>\lambda^{\mathrm{TM}_0}$. Their separation from
$\lambda^{\mathrm{TM}_0}$ increases with $\theta$ and a more effective excitation on the left
branch of the TM$_0$ mode can be seen as compared to the right branch. For
$\theta=\SI{4.96}{\degree}$ and $\theta=\SI{10.71}{\degree}$ the wavelengths of the mode,
$\lambda^{\mathrm{TM}_0^l}(\SI{10.71}{\degree})=\SI{7.1}{\micro\meter}$ and
$\lambda^{\mathrm{TM}_0^l}(\SI{4.96}{\degree})=\SI{7.6}{\micro\meter}$, coincide with the central
wavelength of the plasmon absorption peaks of the inner and outer ribbons, yielding absorption of
$A=0.331$ and $A=0.238$, respectively.

Along the TM$_0$ band, i.e. the path of excitation of the TM$_0$ mode in the
($\lambda,\theta$)-space, strong enhancement of conversion efficiency due to inherited effect of
this mode can be observed in \figref{fig:WGR_graphene_scans}(b). The simultaneous excitation of
waveguide modes and graphene plasmons yields strong THG, with intensities of
$I_\mathrm{TH}=\num{5.85e-05}I_0$ at $\lambda_\mathrm{TH}=\SI{2.37}{\micro\meter}$ and
$\theta=\SI{10.71}{\degree}$ and $I_\mathrm{TH}=\num{3.89e-04}I_0$ at
$\lambda_\mathrm{TH}=\SI{2.53}{\micro\meter}$ and $\theta=\SI{4.96}{\degree}$. However, the
strongest TH intensity of $I_\mathrm{TH}=\num{0.0537}I_0$ is generated at
$\lambda_\mathrm{TH}=\SI{2.42}{\micro\meter}$ and $\theta=\SI{9}{\degree}$, namely where the
intrinsic TM$_1$ band crosses the inherited TM$_0$ band. The increase of the conversion efficiency
due to sole excitation of the intrinsic nonlinear modes is notable, but orders of magnitude lower
than in the case of inherited effects, e.g. the same intrinsic TM$_1$ mode away from the
simultaneous resonance, at $\lambda_\mathrm{TH}=\SI{2.16}{\micro\meter}$ and
$\theta=\SI{20}{\degree}$, yields TH radiation of intensity $I_\mathrm{TH}=\num{7.13e-09}I_0$.

Inspection of the bands of the intrinsic modes reveals and additional interesting feature.
Specifically, intrinsic modes of the same order show anti-crossing behavior, e.g. TM$_0$ at
$\lambda_\mathrm{TH}=\SI{2.75}{\micro\meter}$ and $\theta=\SI{14.42}{\degree}$ or TM$_1$ at
$\lambda_\mathrm{TH}=\SI{2.33}{\micro\meter}$ and $\theta=\SI{12.25}{\degree}$, whereas bands of
different intrinsic modes pass through each other, i.e. TM$_0$ and TM$_1$ at
$\lambda_\mathrm{TH}=\SI{2.26}{\micro\meter}$ and $\theta=\SI{8.375}{\degree}$. This crossing of
modes of same order occurs when the transverse component of the incident $k$ vector,
$k_\parallel$, reaches the edge of the first Brillouin zone.

\section{Concluding Remarks}\label{sec:conclusions}
In conclusion, we have derived an improved, an accurate formulation of the rigorous coupled-wave
analysis (RCWA) method to describe linear and nonlinear optical interactions between light and
periodically patterned 2D materials, such as graphene and transition-metal dichalcogenide (TMDC)
monolayers. Unlike previous approaches, our numerical formalism does not depend on the height of
the 2D material, a poorly defined physical quantity, and as such is applicable to any 2D material,
as long as their linear and nonlinear optical surface conductivities are known. A key ingredient
that markedly improves the accuracy and convergence of this numerical method is a vanishingly
small, added conductivity, which allows for correctly solving the Fourier factorization problem
and consequently a reliable computational investigation of 2D materials. In particular, this small
value of the added conductivity yields accurate results when convergence with respect to the
number of harmonics has been achieved. The proposed numerical method also allows one to describe
the nonlinear optical response of generic periodically patterned 2D materials. In this context, we
have found that correct nonlinear optical physics in these structures can only be captured when
using the accurate near-field formulation of RCWA introduced in \cite{wgp15jo}. Importantly, our
approach employing boundary conditions for the linear and nonlinear fields can be readily extended
to other methods \cite{cpo07prb,cpb09prb,bp10prb} used to describe nonlinear optical effects at
interfaces, as in the case of graphene and TMDC monolayer materials the optical higher-harmonics
are generated in a single atomic layer.

Our numerical method has been comprehensively validated by comparing its predictions to results
obtained using an alternative method. Upon successful validation, we have used it to investigate
the characteristics of various kinds of diffraction gratings comprising graphene and TMDC
monolayers. We found that these materials interact differently with light, which is explained by
their metallic or semiconductor nature. Thus, graphene exhibits THG as the lowest-order nonlinear
optical interaction, due to the inversion symmetry properties of its atomic lattice, and supports
surface plasmons. We found that the excitation of surface plasmons leads to increased linear
absorption and enhanced THG, which points to significant potential for tunable THG in graphene.
The TMDC monolayer materials, on the other hand, are semiconductors and non-centrosymmetric. As a
result, their linear optical absorption spectra show a series of exciton resonances, whereas in
this case the SHG is the lowest-order nonlinear optical process.

As an application of our numerical method, we have demonstrated that by coupling a TMDC monolayer
with a photonic structure that possesses optical resonances, namely a periodically patterned slab
waveguide, one can achieve strong, frequency selective field enhancement and consequently
increased nonlinear optical response of the TMDC monolayer. In addition, we have showed that by
coupling graphene with a similar waveguiding device, the interplay between plasmon resonances in
graphene and leaky waveguide resonances of the slab waveguide leads to rich physics explained by
intriguing phenomena, such as multi-continua Fano resonances and enhanced SHG via simultaneous
excitation and efficient coupling of optical modes at the FF and SH.

The formulation of our method is general enough to describe most nonlinear optical processes of
practical interest in the undepleted pump approximation, which is valid in essentially all
experimental settings. In order to tackle those cases where this approximation might be less
accurate, such as the optical Kerr effect, one can easily extend our method beyond the undepleted
pump approximation by employing an iterative, self-consistent solution process, similar to the
approach introduced in \cite{pbo03ol,pbo04josab,bej14josab} for describing optical Kerr effects in
periodic bulk media. Equally important, it is also possible to investigate important carriers
related optical effects in 2D materials with the proposed method, namely the influence of charge
doping on the optical properties of photonic structures containing 2D materials. Whereas a
rigorous description of transient effects in such systems would require the incorporation of
charge dynamics in our numerical algorithm, a nontrivial but tractable task, the optical response
in the steady-state can be determined by simply modifying the electric permittivity of the 2D
materials so as to take into account the dependence of the permittivity on the charge density.

\begin{acknowledgments}
This work was supported by the European Research Council, Grant Agreement no. ERC-2014-CoG-648328.
The work of M. W. was partly supported through a UCL Impact Award graduate studentship funded by
UCL and Photon Design, Ltd. The authors thank Y. S. Kivshar and D. F. G. Gallagher for many
insightful discussions and wish to acknowledge support from the Royal Society's International
Exchanges Scheme and the hospitality of the Nonlinear Physics Centre of the Australian National
University. The authors acknowledge the use of the UCL Legion High Performance Computing Facility
(Legion@UCL) and associated support services in the completion of this work.
\end{acknowledgments}

\appendix*
\section{Outline of the rigorous coupled-wave analysis}\label{sec:A:RCWA}
For the sake of completeness, we outline in this Appendix the mathematical description of the RCWA
based approach used to find the layer-wise solution of the diffraction grating problem. For now,
let us drop the superscript $\omega$, as the description of the modal form of the electromagnetic
field is independent on whether a pump or generated frequency is considered.

Known as the Bloch theorem, the solution of MEs for a periodic structure is pseudo-periodic, i.e.
periodic with an additional transverse phase shift. Therefore, the permittivity
$\epsilon_r(x,y,z)$ of the structure in the $(x,y)$-plane can be expressed as a 2D Fourier series:
\begin{align}
    \epsilon_r(x,y,z) = \sum_{n=-\infty}^\infty  \epsilon_{r,n}(z) e^{2\pi i \left( \frac{n_1}{\Lambda_1}x +
    \frac{n_2}{\Lambda_2}y\right)},
    \label{eq:defPermittivityFS}
\end{align}
with $z$-dependent coefficients $\epsilon_{r,n}(z)$. The sum over $n=(n_1,n_2)$ is to be
understood as the double infinite sum over the integers $n_i\in \mathbb{Z}$, $i=1,2$. Similarly,
the electromagnetic field quantities, $f=\E, \H, ...$, can be expressed as a Fourier series with
phase shift,
\begin{align}
    &f(x,y,z) = \sum_{n=-\infty}^\infty  f_{n}(z) e^{i\left( k_{nx}x + k_{ny}y \right)}
    := \rec{\fvec{f(z)}}(x,y),   \label{eq:defFS}
\end{align}
where $k_{nx/y} = k_{0x/y} + 2\pi n_{1/2}/\Lambda_{1/2}$ is the $x/y$ component of the $n$th
diffraction order. The intrinsic phase shift is determined by the transverse component of the
$k$-vector of the incident plane wave, $k_{0x/y} = k_{x/y}$. The sequence of Fourier coefficients
shall be denoted by $\fvec{f(z)}$, and the evaluation of a Fourier by means of the sum in
Eq.~\eqref{eq:defFS} is denoted by the reconstruction operator, $\rec{\fvec{f(z)}}$.

In actual calculations, the infinite sums in Eqs.~\eqref{eq:defPermittivityFS} and
\eqref{eq:defFS} have to be truncated. A rectangular truncation approach, namely
$n=(n_1,n_2)\in\left\{-N_1,\ldots,N_1\right\}\times \left\{-N_2,\ldots,N_2\right\}$, will be used
throughout this derivation, yielding a total of $N_0=(2N_1+1)(2N_2+1)$ Fourier series coefficients
(or harmonics).

In the remaining part of the section we will provide the general mathematical formulation of the
modal field expansion of the electromagnetic field. Thus, the underlying assumption of the modal
field expansion in a periodic bulk layer of the grating is that the permittivity function,
$\epsilon_r(x,y)$, in layer $z\in\left[z^+,z^-\right]$ is $z$-invariant, where $z^+$ and $z^-$
denote the bottom and the top of the periodic bulk layer. This implies that the Fourier
coefficients, $\epsilon_{r,n}$, of $\epsilon_r(x,y)$ are $z$-independent, too.

Since the field solution is periodic according to Bloch theorem, the electromagnetic fields inside
this periodic bulk layer are pseudo-periodic, i.e. they can be expressed as Fourier series with
phase shift using the reconstruction operator, $\mathcal{R}$:
\begin{subequations}\label{eq:PeriodicAnsatz}
    \begin{align}
        \E(\br) =& \ \rec{\fvec{E_{x}(z)}\e_x + \fvec{E_{y}(z)}\e_y + \fvec{E_{z}(z)}\e_z},\\
        \H(\br) =& \ \rec{\fvec{H_{x}(z)}\e_x + \fvec{H_{y}(z)}\e_y + \fvec{H_{z}(z)}\e_z}.
    \end{align}
\end{subequations}

Before using these ansatz functions to determine the electromagnetic fields, the correct Fourier
factorization rules \cite{lm96josaa,l97josaa,li03josaa,srk07josaa} have to be used in order to
factorize the product,
\begin{align}
    \D(x,y,z) = \epsilon_0\epsilon_r(x,y) \E(x,y,z),\label{eq:ConstRelationSpace}
\end{align}
as they ensure accuracy and convergence of the RCWA modal expansion even for low number of
harmonics.

In this work, the normal vector field approach \cite{srk07josaa,wgp15jo} in bulk layers is used to
accurately solve the Fourier factorization problem in 2D. To this end, let
$\nvfv=(\nvf_x,\nvf_y,\nvf_z)^T$ be a continuation of the surface normal vectors of the grating
structure, i.e. a normal vector field. The Fourier series factorization of the constitutive
relation Eq.~\eqref{eq:ConstRelationSpace} then reads,
\begin{align}
    \fvec{D_\alpha(z)} &= \epsilon_0\sum_{\beta=1}^3 \left(\delta_{\alpha,\beta}\fmatrix{\epsilon_r} -
    \Delta\textsf{\nvf}_{\alpha\beta}\right)\fvec{E_\beta(z)}, \label{eq:ConstRelationNVF}
\end{align}
where $\delta_{\alpha\beta}$ is the Kronecker delta. Here, $\fmatrix{g}$ denotes the Toeplitz
matrix of Fourier coefficients of a function $g$, and the matrix
$\Delta\textsf{\nvf}_{\alpha\beta}$ is given by $\Delta\textsf{\nvf}_{\alpha\beta} = \frac
12\left(\Delta \fmatrix{\nvf_\alpha \nvf_\beta} +\fmatrix{\nvf_\alpha \nvf_\beta}\Delta\right)$,
with $\Delta = \fmatrix{\epsilon_r} - \fmatrix{1/\epsilon_r}^{-1}$ and $\nvf_\alpha$ being the
$\alpha$-component of the normal vector field, $\nvfv$, at the material boundary.

Inserting Eq.~\eqref{eq:PeriodicAnsatz}, the permittivity given by
Eq.~\eqref{eq:defPermittivityFS}, and the correct factorization provided by
Eq.~\eqref{eq:ConstRelationNVF} into the MEs one obtains a linear system of ordinary differential
equations for the $z$-dependent amplitudes of the modal fields. This system is solved assuming
exponential dependency for modal Fourier coefficients, $\fvec{E_\alpha(z)} =
\fvec{E_\alpha}e^{ik_0\propconst z}$ and $\fvec{H_\alpha(z)} = \fvec{H_\alpha}e^{ik_0\propconst
z}$, with the complex propagation constant, $\propconst$. Then, the system of ordinary
differential equations can be rewritten as an algebraic eigenvalue problem for $\fvec{E_{xy}}$,
\begin{align}
    \textsf{M}_1\textsf{M}_2 \left(\begin{array}{c} \fvec{E_x}\\\fvec{E_y}\end{array}\right) = \propconst^2 \left(\begin{array}{c} \fvec{E_x}\\\fvec{E_y}\end{array}\right), \label{eq:defEVproblem}
\end{align}
and an additional relation for $\fvec{H_{xy}}$,
\begin{align}
    \sqrt{\propconst^2} \left(\begin{array}{c} \fvec{H_x}\\\fvec{H_y}\end{array}\right) = \textsf{M}_1 \left(\begin{array}{c} \fvec{E_x}\\\fvec{E_y}\end{array}\right). \label{eq:defEVH}
\end{align}
In these relations, $\textsf{M}_{1,2}$ are $2N_0\times 2N_0$ matrices of block-matrix form:
\begin{align*}
    \textsf{M}_1 = & \left(\begin{array}{cc}
        \textsf{K}_x \fmatrix{\epsilon_r}^{-1}\textsf{K}_y &\textsf{I}- \textsf{K}_x \fmatrix{\epsilon_r}^{-1}\textsf{K}_x \\
        \textsf{K}_y \fmatrix{\epsilon_r}^{-1}\textsf{K}_y - \textsf{I} & - \textsf{K}_y\fmatrix{\epsilon_r}^{-1}\textsf{K}_x
    \end{array}\right),\\
    \textsf{M}_2 = & \left(\begin{array}{cc}
        \Delta\textsf{\nvf}_{yx} - \textsf{K}_x \textsf{K}_y & \textsf{K}_x \textsf{K}_x - \textsf{C}_y \\
        \textsf{C}_x - \textsf{K}_y \textsf{K}_y & \textsf{K}_y \textsf{K}_x - \Delta\textsf{\nvf}_{xy}
    \end{array}\right).
\end{align*}
Here, $\textsf{C}_\alpha=\fmatrix{\epsilon_r}-\Delta\textsf{\nvf}_{\alpha\alpha}$, the matrices
$\textsf{K}_\alpha=\operatorname{diag}\left(k_{\alpha n}\right)$, $\alpha=x,y$, are diagonal
matrices of the in-plane propagation constants, $k_{\alpha n}$, of the diffraction orders, and
$\textsf{I}$ is the identity matrix of size $N_0\times N_0$.

The eigenvalue problem defined by Eq.~\eqref{eq:defEVproblem} has $2N_0$ solutions consisting of
the eigenvalues $\propconst_m^2$ and eigenvectors $\left(\fvec{E^{m}_x},\fvec{E^{m}_y}\right)$,
$m=1,\ldots,2N_0$. Defining the positive and negative roots of $\propconst_m^2$ as
\begin{align*}
    {\propconst_m^+}^2 &:= \propconst_m^2, \text{ if } \Re{\propconst_m^+} + \Im{\propconst_m^+} >    0, \\
    {\propconst_m^-}^2 &:= \propconst_m^2, \text{ if } \Re{\propconst_m^-} + \Im{\propconst_m^-} \leq 0,
\end{align*}
respectively, one obtains a total of $2N_0$ upward and $2N_0$ downward propagating modes of the
grating. The upward (downward) mode with index $m$ is defined by its propagation constant,
$\propconst_m^+$ ($\propconst_m^-$), and its transverse mode profile given by the Fourier vector
coefficients, $\fvec{E^{(m,+)}_{x/y}}$ and $\fvec{H^{(m,+)}_{x/y}}$ ($\fvec{E^{(m,-)}_{x/y}}$ and
$\fvec{H^{(m,-)}_{x/y}}$), where $\fvec{H^{(m,+)}_{x/y}}$ ($\fvec{H^{(m,-)}_{x/y}}$) are obtained
from Eq.~\eqref{eq:defEVH} by setting $\propconst=\propconst_m^+$ ($\propconst=\propconst_m^-$).

Since the bulk grating layer is considered to be made of linear optical materials, the linear
superposition of modes is a solution to the MEs, too. Therefore, the total electric field in the
grating is given by
\begin{align}
    E_\alpha(\br) =&\sum_{m=1}^{2N_0} c^+_m\rec{\fvec{E^{(m,+)}}}(x,y)e^{ik_0\propconst_m^+(z-z^{-})} \nonumber \\
    &+ c^-_m\rec{\fvec{E^{(m,-)}}}(x,y)e^{ik_0\propconst_m^-(z-z^{+})}, \label{eq:defRCWAmodalExpansion}
\end{align}
where the complex mode coefficient $c^{+}_m$ ($c^{-}_m$) determines the contribution of each
upward (downward) propagating mode to the total grating field and $z^{+}$ ($z^{-}$) denotes the
$z$-coordinate of the bottom (top) of the considered grating layer. The components of the magnetic
field, $H_\alpha(\br)$, can be found from a similar equation. Given this structure of the modes,
the electromagnetic fields in the grating are hence fully determined by $4N_0$ mode coefficients,
$c^{\pm}_m$. Their values are obtained by means of the electromagnetic boundary conditions
described in \secref{sec:inhomSMatrix}.

For reasons related to the practical implementation of RCWA, it is useful to rewrite
Eq.~\eqref{eq:defRCWAmodalExpansion} in terms of $z$-dependent Fourier coefficients, similar to
Eqs.~\eqref{eq:PeriodicAnsatz}, but interchanging the order of summation of modes and Fourier
components in Eq.~\eqref{eq:defRCWAmodalExpansion}:
\begin{subequations}\label{eq:defRCWAmodalField}
    \begin{align}
        &E_\alpha(\br) = \ \rec{\fvec{E_\alpha^{+}(z)}}(x,y) + \rec{\fvec{E_\alpha^{-}(z)}}(x,y), \\
        &H_\alpha(\br) = \ \rec{\fvec{H_\alpha^{+}(z)}}(x,y) + \rec{\fvec{H_\alpha^{-}(z)}}(x,y),
    \end{align}
\end{subequations}
where $\fvec{E^{\pm}(z)}$ and $\fvec{H^{\pm}(z)}$ are given by:
\begin{subequations}\label{eq:defRCWAmodalMatrix}
    \begin{align}
        \fvec{E^{\pm}_\alpha(z)} = & \sum_{m=1}^{2N_0} \fvec{E^{(m, \pm)}_\alpha(z)}
        = \sum_{m=1}^{2N_0} \fvec{ E^{(m,\pm)}_\alpha} e^{ik_0\propconst^\mp_m (z-z^\mp)} c_m^\pm \nonumber \\
        &=  \textsf{E}^\pm_\alpha \propmat^\pm(z)\bc^\pm,\\
        \fvec{H^{\pm}_\alpha(z)} = & \textsf{H}^\pm_\alpha \propmat^\pm(z)\bc^\pm.
    \end{align}
\end{subequations}
Here, the $2N_0\times 2N_0$ mode-shape matrix $\textsf{E}^\pm_\alpha$ ($\textsf{H}^\pm_\alpha$)
contains the vector of Fourier coefficients, $\fvec{ E^{(m,\pm)}_\alpha}$ ($\fvec{
H^{(m,\pm)}_\alpha}$), in its $m$th column. Moreover, the propagation matrix, $\propmat^\pm(z)$,
is a diagonal matrix containing the $z$-dependence of each mode on its diagonal,
$\propmat_{mm}^\pm(z) = e^{ik_0\propconst_m^\pm (z-z^\mp)}$, with
$m\in\left\{1,\ldots,2N_0\right\}$, and $\bc^\pm$ denotes the vector of upward (``$+$'') and
downward (``$-$'') propagating mode coefficients.

Using Eqs.~\eqref{eq:defRCWAmodalField} and \eqref{eq:defRCWAmodalMatrix} one can determine the
electromagnetic field everywhere in and around the grating, namely in the periodic or homogeneous
bulk layers forming the grating as well as in the cover and substrate. However, for homogeneous
layers a Rayleigh expansion \cite{l97josaa} with the diffraction orders as modes is preferable to
the solution of the RCWA eigenproblem defined by Eq.~\eqref{eq:defEVproblem} as it is
computationally less demanding to calculate $\textsf{E}^\pm_\alpha$, $\propmat^\pm$, and $\bc^\pm$
for the Rayleigh expansion.

As a concluding remark, we note that the importance of Eqs.~\eqref{eq:defRCWAmodalMatrix} resides
in that it translates the electromagnetic field from its modal representation to its direct
representation as Fourier series. The modal representation in terms of $\textsf{E}^\pm_\alpha$,
$\propmat^\pm$, and $\bc^\pm$ in the r.h.s. of Eqs.~\eqref{eq:defRCWAmodalMatrix} is restricted to
a computational layer, but very advantageous therein, as it separates mode-shape quantities,
$\fvec{E^{(m, \pm)}_\alpha}$, propagation and decay constants $\propconst_m^\pm$, and excitation
strength coefficients $c_m^\pm$. On the other hand, the direct field representation as a Fourier
series with coefficients $\fvec{E^{\pm}_\alpha(z)}$ in the l.h.s. of
Eqs.~\eqref{eq:defRCWAmodalMatrix} has the same complex exponentials, $e^{ik_{nx}x+ik_{ny}y}$, as
basis functions everywhere in the grating, and hence it facilitates the comparison of quantities
in different computational layers.

% Bibliography
% \bibliography{mybib}

%

\end{document}